\documentclass[a4paper,12pt,oneside]{article}
\usepackage[showframe,paper=a4paper,margin=0.2in]{geometry}
\usepackage{lipsum}
\usepackage{graphicx}
\usepackage{sidecap}
\usepackage{amsmath}
\usepackage{textcomp}
\usepackage{wrapfig} 
\usepackage{color}
\usepackage{setspace}
\usepackage{subfigure}
\usepackage{placeins}
\usepackage{verbatim}
\usepackage{siunitx}
\usepackage{gensymb}
\usepackage{floatrow}
\usepackage{tabu}
\usepackage{float}
\usepackage{graphicx}
\usepackage{pagecolor}
\usepackage{afterpage}
\usepackage{amssymb}
\usepackage{color}

\definecolor{EPD_blue}{RGB}{40,75,200} 

\begin{document}

\pagecolor{white}
\newgeometry{top=0.6in,bottom=0.0in,right=0.2in,left=0.0in}
\begin{titlepage}
\newpagecolor{EPD_blue}\afterpage{\restorepagecolor}

\begin{figure}[p]
    \vspace*{-2.3cm}
    \hspace*{0.2cm}
    \makebox[\linewidth]{
        \includegraphics[width=1.18\linewidth]{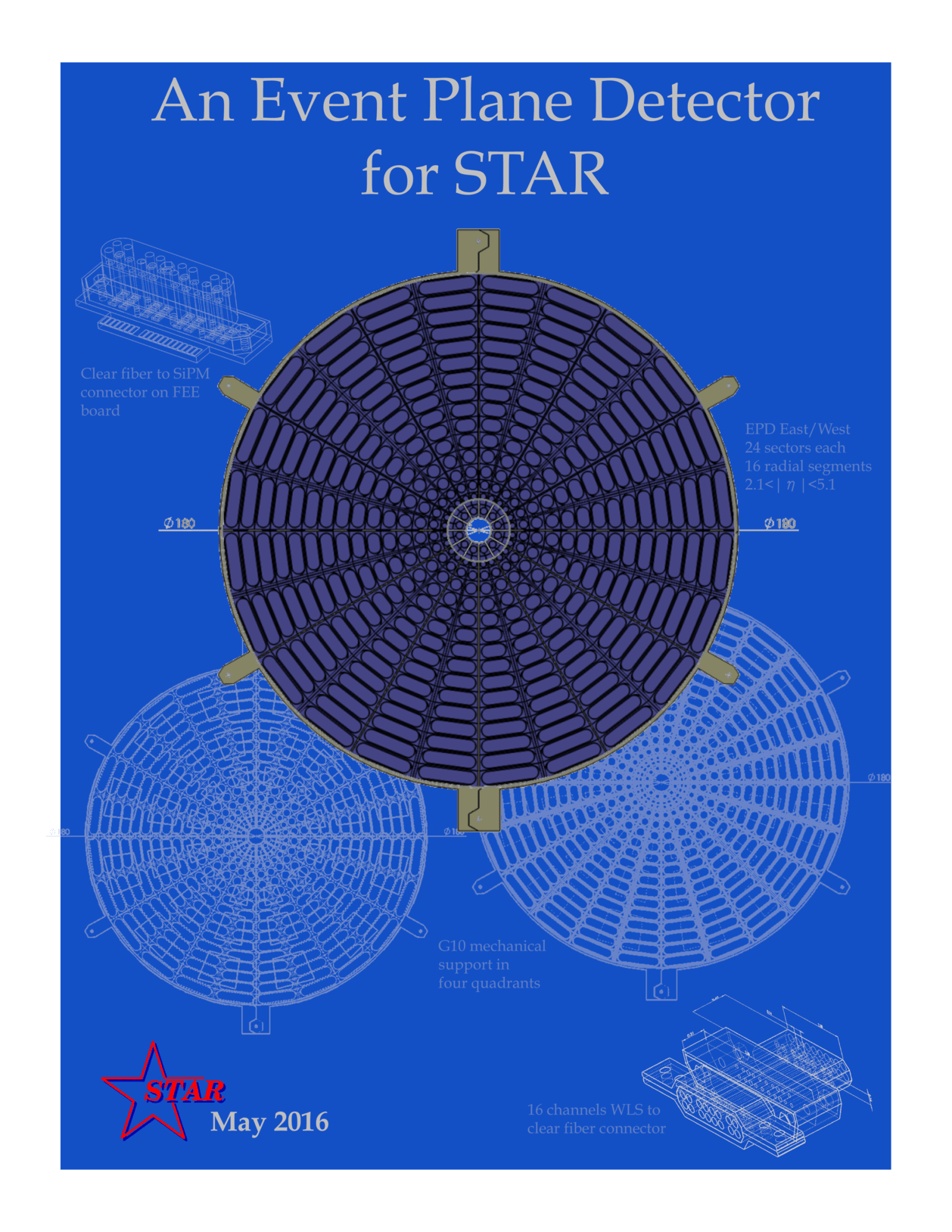}
    }
\end{figure}

\end{titlepage}
\newpage
\newgeometry{top=1.2in,bottom=1.2in,right=1.0in,left=1.2in}

\tableofcontents
\newpage
\listoffigures
\newpage

\vspace{-1cm}
\section{Executive Summary}
\label{sec_ex_summary}
We propose to construct a new, dedicated {\bf E}vent {\bf P}lane, centrality, and trigger {\bf D}etector (EPD) in the forward direction of STAR for the Beam Energy Scan (BES) phase II, anticipated for years 2018-2020. The new detector will cover the pseudo-rapidity range between 2.1 and 5.1, with high radial and azimuthal segmentation.  The EPD will allow the centrality and the event plane to be measured in the forward region, reducing the systematics due to autocorrelations from mid-rapidity analyses. The baseline detector design utilizes scintillator plastic, wavelength-shifting fibers and silicon photomultipliers (SiPMs). 

The construction budget request is \$242k plus \$45k for new QT boards.  These figures include overhead.  There have been several prototypes already built and tested for the EPD.  The "pre-prototype" was installed for run 15, with the goal of insuring that we could integrate with the STAR electronics and to test the basic design principals.  This test was very successful.  Our "prototype" is currently in STAR, taking data during run 16.  The goal of the prototype was to build one sector of the EPD and to test it, in order to insure we have the performance necessary to meet the detector requirements.  The details of the procedure for the construction of the prototype were also tested, and this informed many of the decisions made that are outlined in this proposal.

The EPD consists of two disks that will be placed on either side of the STAR interaction region, in the current location of the Beam-Beam Counter (BBC).  We will install 1/4 of one of the disks for run 17 for detector commissioning.  The remaining 7/8 of the detector will be built and installed for run 18. 

The construction proposal is structured in the following way: in section~\ref{sec_intro} we elaborate on physics motivations and the need for the proposed detector, section~\ref{sec_design} summarizes the design and the simulation results which informed this design, section~\ref{sec_design} we elucidate the construction needs, goals and design, and in section~\ref{sec_costnschedule} we list the construction cost and schedule.



\newcommand {\SdotL} {\mbox{$\vec{S}\cdot\hat{L}$}~}
\newcommand {\SdotLhat} {\SdotL}

\newcommand {\SdotLAve} {\mbox{$\left\langle\overline{\SdotL}\right\rangle$}~}

\newcommand {\Lhat} {\mbox{$\hat{L}$}~}
\newcommand {\LhatEst} {\mbox{$\Lhat^{\rm est}$}~}

\newcommand {\RPres} {\mbox{$R^{(1)}_{\rm EP}$}~}

\newcommand{\lam} {\mbox{$\Lambda$}~}
\newcommand{\lamBar} {\mbox{$\overline{\Lambda}$}~}
\newcommand{\polVec} {\mbox{$\vec{P}$}~}
\newcommand{\rootsnn} {$\sqrt{{s}_{\textrm{NN}}}$~}
\newcommand{\roots} {$\sqrt{{s}_{\textrm{NN}}}$~}
\newcommand{\snn}  {$\sqrt{{s}_{\textrm{NN}}}$~}
\newcommand{\npart} {${N}_{part}$}
\newcommand{\cijk}    {${c}_{ijk}$}

\newcommand {\PH} {\mbox{$\overline{P}_{\rm H}$}~}

\newcommand {\Plam} {\mbox{$\overline{P}_{\Lambda}$}~}
\newcommand {\Palam} {\mbox{$\overline{P}_{\overline{\Lambda}}$}~}

\newcommand {\Sig} {\mbox{$\Sigma^0$}~}
\newcommand {\SigBar} {\mbox{$\overline{\Sigma}^0$}~}

\newcommand{\red} {\textcolor{red}}
\newcommand {\CN} {\red{CN}~}  
\newcommand{\caab}     {${c}_{aab}$}
\newcommand{\cabc}     {${c}_{abc}$}
\newcommand{\cbbd}     {${c}_{bbd}$}
\newcommand{\cbce}     {${c}_{bce}$}

\section{Physics of the EPD}
\label{sec_intro}

\begin{figure}[htbp]
\begin{center}
{
\mbox{\includegraphics[width=0.60\textwidth]{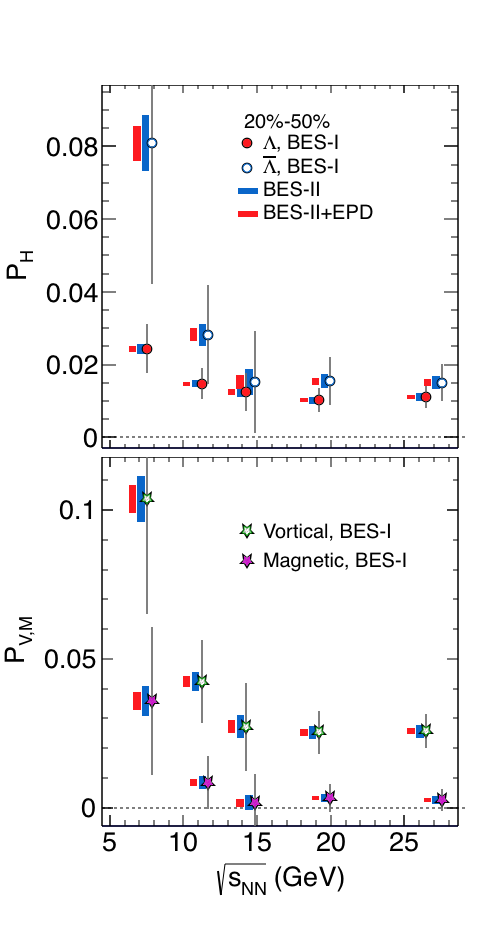}}
}
\end{center}
\caption{Upper panel: The average polarization of $\Lambda$ and $\bar{\Lambda}$ from 20-50\% central $Au+Au$ 
collisions is plotted as a function of collision energy. The  effects  of  feed-down  have  
not  been  removed, and only statistical errors are plotted. Lower panel: The vortical and magnetic contributions to $\Lambda$ and $\bar{\Lambda}$ emitted 
directly from the hot zone created in a heavy ion collision. Only statistical uncertainties are 
shown, but the scale of $P_V$ has an uncertainty of +60\% and -5\% due to uncertainty in the amount 
of feed-down from $\Sigma^0$ decays. The corresponding systematic uncertainty for $P_B$, however, is 
negligible. 
}
\label{fig:LambdaPol}
\end{figure}

\begin{figure}[htbp]
\begin{center}
{
\mbox{\includegraphics[width=0.85\textwidth]{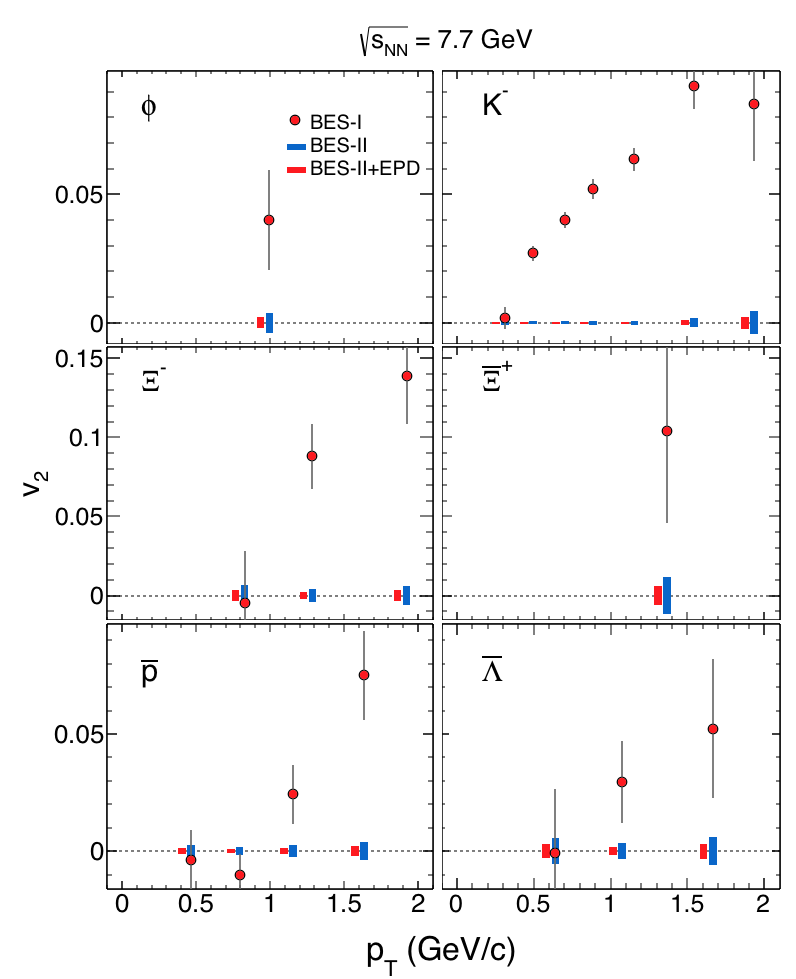}}
}
\end{center}
\caption{These six figures show ${v}_{2}$ versus ${p}_{T}$ for six different species: $\phi$ in the upper left, ${K}^{-}$ in the upper right, ${\Xi}^{-}$ in the middle left, ${\bar{\Xi}}^{+}$ in the middle right, $\bar{p}$ in the lower left and ${\bar{\Lambda}}$ in the lower right.  The red points are the data collected by STAR during the first Beam Energy Scan.  The blue vertical bars indicate how the error bars will improve from BESI to BESII.  The red vertical bars show the improvement that will be seen including the EPD.
}
\label{fig:BESII_v2}
\end{figure}

\begin{figure}[htbp]
\begin{center}
{
\mbox{\includegraphics[width=0.70\textwidth]{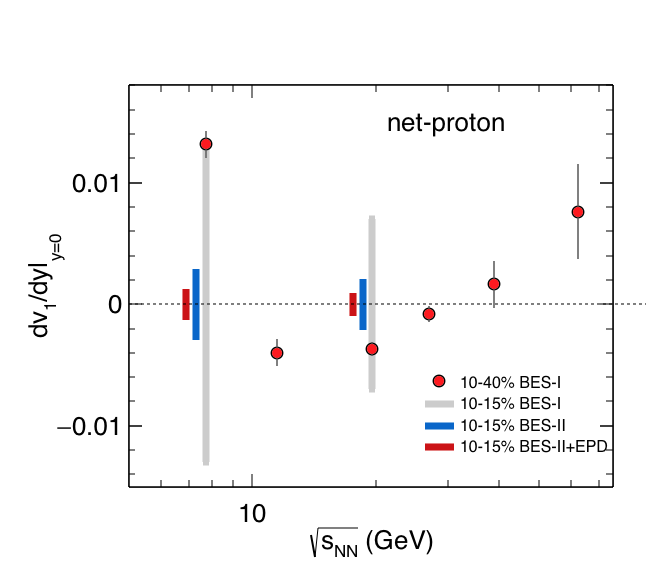}}
}
\end{center}
\caption{This shows the net proton ${v}_{1}$ versus the $\sqrt{{s}_{NN}}$ at mid-rapidity.  The red points show the BESI data from semi-central collisions (10-40\%).  The grey bars indicate what the error bars would have been with a narrow centrality selection from 10-15\%.  The blue bars indicate what the error bars will be for BES II.  The red bars indicate what the uncertainty will be in BESII with the inclusion of the EPD.
}
\label{fig:BESII_v1}
\end{figure}

\subsection{Introduction}

The RHIC Beam Energy Scan (BES) has been identified as a top priority by Brookhaven Lab, and 
  considerable beam time, personnel and resources for accelerator and detector upgrades have
  been dedicated to it.
The program's importance has been recognized by the broader U.S. nuclear physics community, 
  as evidenced by its endorsement and highlight in the Nuclear Science Advisory Council's
  (NSAC) 2015 Long Range Plan~\cite{Geesaman:2015fha}.
The importance of a systematic scan of energies in the BES region  \roots$\approx5-30$~GeV)
 is also recognized by the larger worldwide heavy ion community, as large, new facilities
  such as FAIR and NICA are being constructed to perform such scans.
These facilities are years off, however, and do not span the complete range of energies
  over which a study-- which should include regions on the high and low-\roots
  side of any ``transition'' energy range-- needed for a compelling systematic study.

A collider (with unchanging kinematics as \roots changes) expressly built flexible enough
  to study different species and energies, RHIC is an existing facility ideal to perform BES.
A detector expressly built for multi-purpose, correlated studies with broad acceptance,
  STAR is an existing detector likewise ideal for BES.
The RHIC scientific community-- collaborations, collider staff, and lab management-- recognize
  the opportunities and importance of BES and are currently assembled and dedicated to it.
This is a unique moment with unique physics opportunities.
It is, furthermore, a moment that will pass; all recognize that future developments imply
  that a BES cannot be an incremental, decade-long program.
We must exploit the physics potential of this unique moment; as sketched here,
  the Event Plane Detector is an integral component of this mandate.

The intense community-wide focus on the BES energy range is based on theoretical expectations
  that QCD-- a theory with an intrinsic scale associated with a fundamental phase transition--
  will display nontrivial, observable behaviour shedding light on the nature of that transition.
More than theoretical expectations, however, the community's interest has been prompted by STAR's
  experimental observations~\cite{Aggarwal:2010wy,Adamczyk:2012ku,Adamczyk:2013gw,Adamczyk:2013gv,Adamczyk:2013dal,Abdelwahab:2014yha,Adamczyk:2014fia,Adamczyk:2014ipa,Adamczyk:2014mzf,Adamczyk:2014mxp,Sarkar:2014wva,Adamczyk:2015eqo,Adamczyk:2015lvo,Adamczyk:2016exq}.
  in the first phase of the RHIC scan (BES-I).


BES-I was a principle proof of two things.
First is a clear confirmation of the expectation that nontrivial features in the data-- some
  long-anticipated and some not-- were lurking at \roots$\approx10-20$~GeV.
Second was that a dedicated experiment-- STAR-- could plan in detail~\cite{Aggarwal:2010cw} and
  execute successfully a program to find and study those features.
BES-I also revealed the statistics, \roots granularity, and detector upgrades required
  to probe this physics more deeply.
The EPD described here meets those requirements.

The EPD serves several purposes.
Firstly, it provides a trigger for the experiment, identifying a ``good'' collision that
  needs to be recorded; this aspect of the device is covered elsewhere.
In the remainder of this section, we discuss the importance of the EPD on extracting physics
  via an analysis of recorded events.
We break out various analyses according to their dependence on the event plane.

\subsection{Event-plane-independent measurements}

Fluctuation studies in BES-II require the EPD not for determination of the event plane, but rather
  for centrality determination in a region well-separated (in rapidity) from the midrapidity
  region under study.
Others, such as forward-backward correlations and balance functions, require the extended rapidity
  coverage as part of the analysis itself.

\subsubsection{Moments of distribution of conserved quantities}

In an infinite, static system in thermal equilibrium, the scale of local 
  fluctuations of conserved quantities may be associated with chemical susceptibilities.
If the state of the system is near a critical point in its phase diagram, these susceptibilities--
  and the scale of the fluctuations-- diverge.
The phenomenon of critical opalescence in some liquids is a favorite visual example.

\begin{figure}
\includegraphics[width=0.45\textwidth]{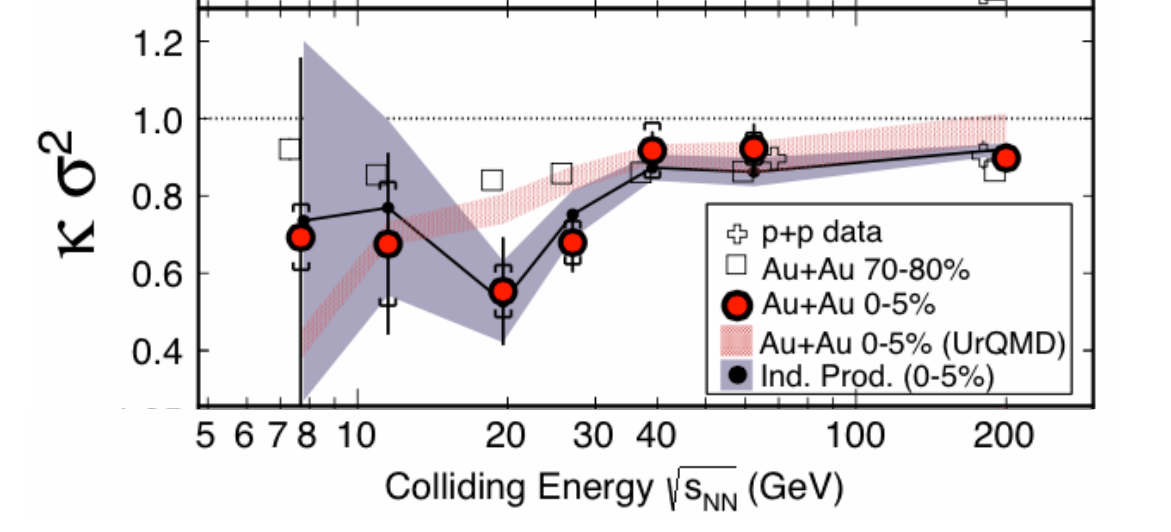}
\hfill
\includegraphics[width=00.45\textwidth]{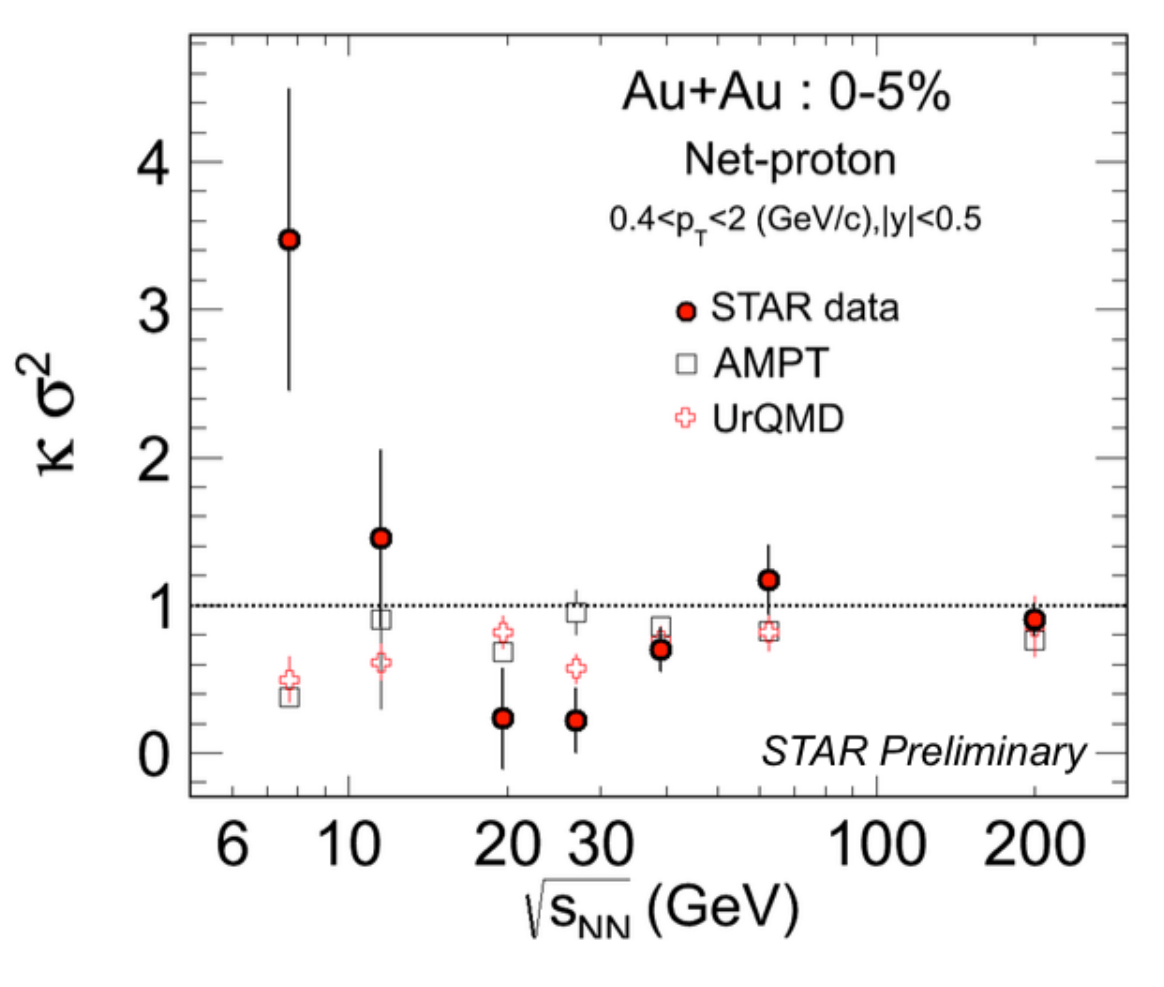}
\caption{
\label{fig:Phys:NetProton1}
The product of $4^{\rm th}$ and $2^{\rm nd}$ moments of the net proton distribution from central
  collisions depends on \roots in a non-trivial.
Left: The published values from STAR's 2014 PRL~\cite{Adamczyk:2013dal}.
Right: Preliminary STAR data presented at CPOD 2014~\cite{Luo:2015ewa}.
}
\end{figure}

STAR has published results on the moments of conserved quantities in QCD--
  net electrical charge~\cite{Adamczyk:2014fia},
  net strangeness (restricted to kaons)~\cite{Sarkar:2014wva},
  and net baryon number (restricted to protons)~\cite{Adamczyk:2013dal,Luo:2015ewa}.
The measurement of higher-order moments of the net proton distributions have
  attracted the most attention, as they display an energy dependence
  that is predicted for a system near a critical point.
Of key focus is the product of the fourth to the second moment of the net-proton
  distribution, which is expected to evolve from above unity
  (super-Poissonian) to below unity, and then return to unity, as the energy
  of the collision increases.
STAR's results, which have attracted intense attention, are shown in figure~\ref{fig:Phys:NetProton1}.

The first message from figure~\ref{fig:Phys:NetProton1} is that the analysis is extremely
  sensitive to the phasespace studied.
The right panel was shown at a conference in 2014, the same year that the panel on the left
  was published in PRL, but the qualitative sense one gets is quite different.
The increased $p_T$ limit increases the particle yield by less than a factor of 2.

\begin{figure}
\includegraphics[width=0.45\textwidth]{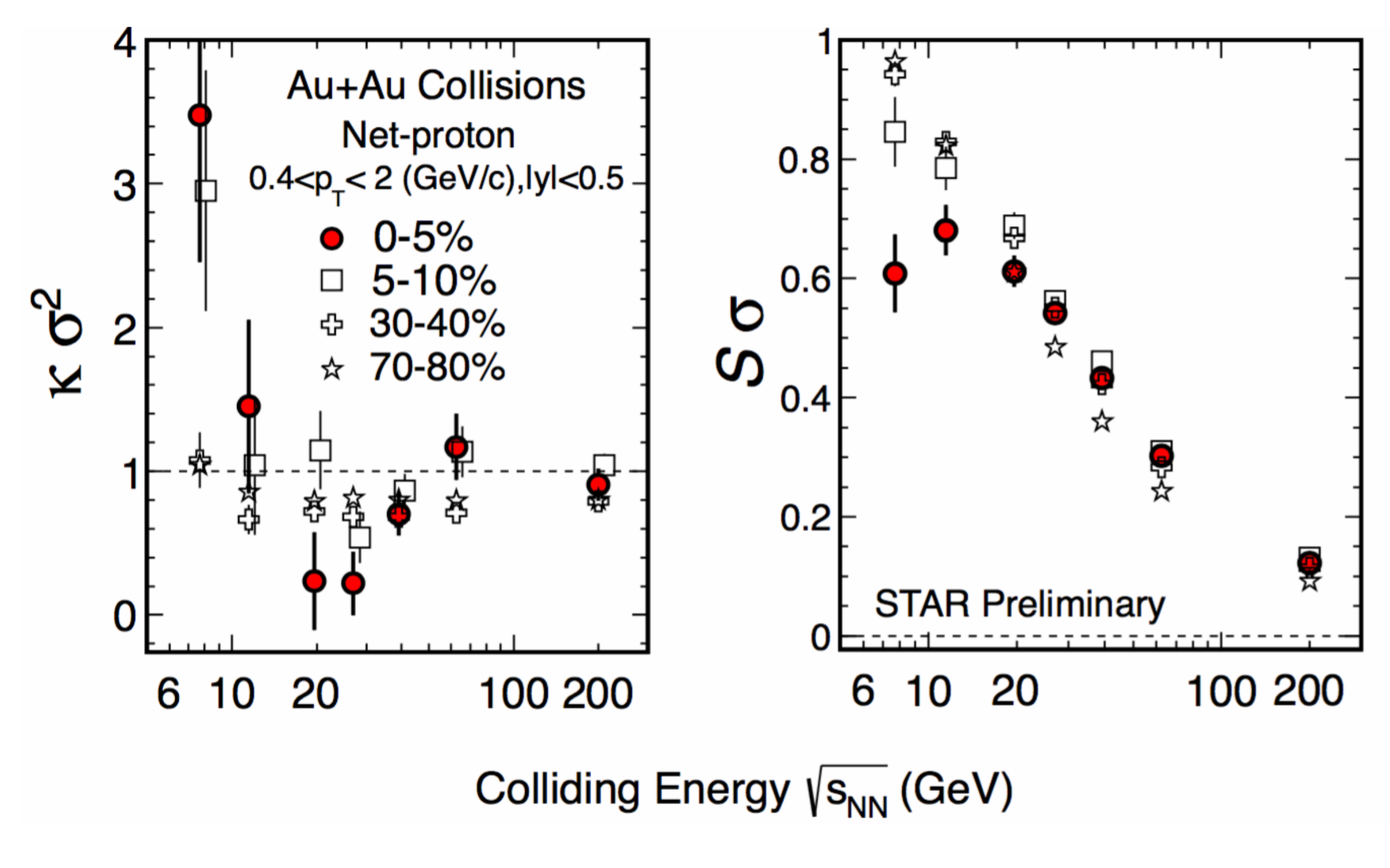}
\hfill
\includegraphics[width=0.45\textwidth]{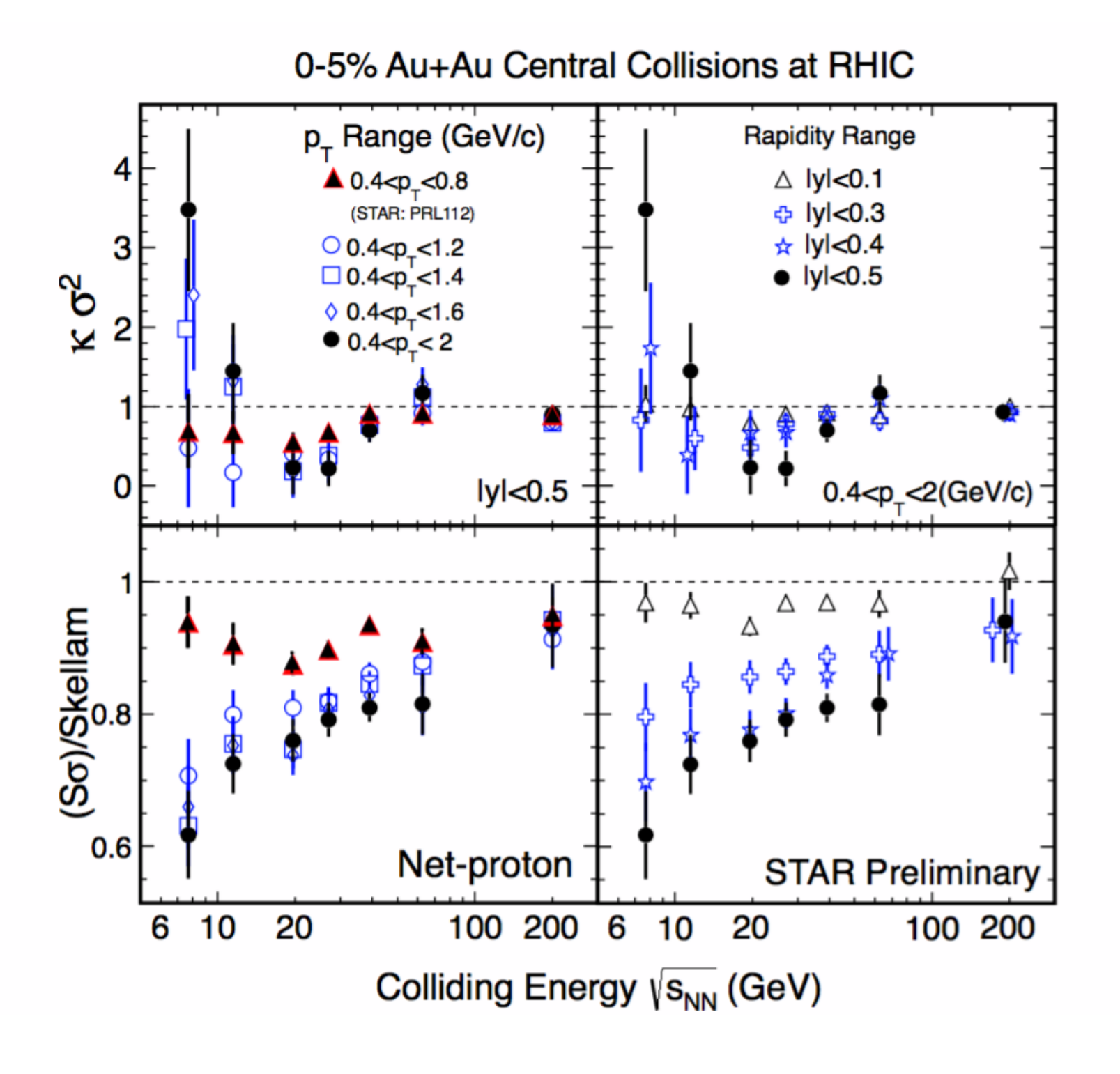}
\caption{
\label{fig:Phys:NetProton2}
The kurtosis signal is strongly dependent on the rapidity range used, 
  as well as the centrality of the collision, as estimated from {\it midrapidity}
  charged particle yeild
Preliminary STAR data presented at CPOD 2014~\cite{Luo:2015ewa}.
}
\end{figure}

Figure~\ref{fig:Phys:NetProton2} emphasizes the strong sensitivity of the signal on collision centrality
  and rapidity range.
These are particularly relevant for BES-II, for which the iTPC upgrade is partly motivated by the need
  to widen the rapidity range of the net-proton analysis; it is increasingly recognized that only
  by studying the signal as a function of rapidity window, will the competing effects of
  critical fluctuations and finite system size be disentangled.
As disussed above, a determination of event centrality {\it at forward rapidity} will
  be crucial to make a compelling and lasting message from this intriguing data.

Neither the TPC itself, nor the small BBC tiles can serve this function; the EPD will be crucial.

Charged-particle multiplicity remains the primary experimental means of estimating
  the centrality of a collision.
Several high-profile analyses~\cite{Aggarwal:2010wy,Adamczyk:2013dal,Adamczyk:2014fia,Luo:2015ewa}
  focus on event-by-event multiplicity distributions of particles in the midrapidity region.
Obviously, one doesn't want to use the same particles to {\it characterize} the global
  characteristic of the event (centrality) and
  to look for {\it changes} in the event as a function of its global characteristics.

The small tiles of the BBC allow for multiplicity determination away from midrapidity.
However, collisions at BES energies neither (1) produce as many particles nor (2) focus them
  so far forward.
Therefore, the most interesting observables from BES have been restricted to small
  rapidity range (0.5), in order to allow the edges of the TPC acceptance (0.5-1)
  for a multiplicity measure (RefMult2).
Even this marginal situation is not sufficient.

The net-proton kurtosis analysis uses for centrality estimation the charged-particle
  multiplicity at $|\eta|<1.0$-- overlapping the same region ($|y|<0.5$) from which the analyzed protons and antiprotons
  are drawn-- excluding the protons and antiprotons themselves.
This is hardly ideal.
After all, pions make up the bulk of the midrapidity charged particles, and most of these arise from 
  midrapidity $\Delta$ baryons, especially at low \roots (which is where the most interesting signal lies).
The other decay product of a midrapidity $\Delta$ is a midrapidity proton.

The net-charge analysis~\cite{Adamczyk:2014fia} and net-kaon analysis~\cite{Sarkar:2014wva} are hardly much
  better: each draws the particles of interest from $|\eta|<0.5$ and uses
  charged-particle multiplicity in $0.5<|\eta|<1.0$ for centrality estimation.

Especially given the very strong sensitivity to centrality estimation and rapidity range shown below,
  a separation of at least one unit of rapidity between the region used for the proton analysis,
  and the region used for centrality estimation, is absolutely required if STAR's intriguing
  but somewhat unstable fluctuations signals will result in a convincing and lasting message of the
  RHIC BES program.~\footnote{Volker Koch, private communication}

\subsubsection{Forward-backward correlations}

\begin{figure}[t]
\includegraphics[width=0.8\textwidth]{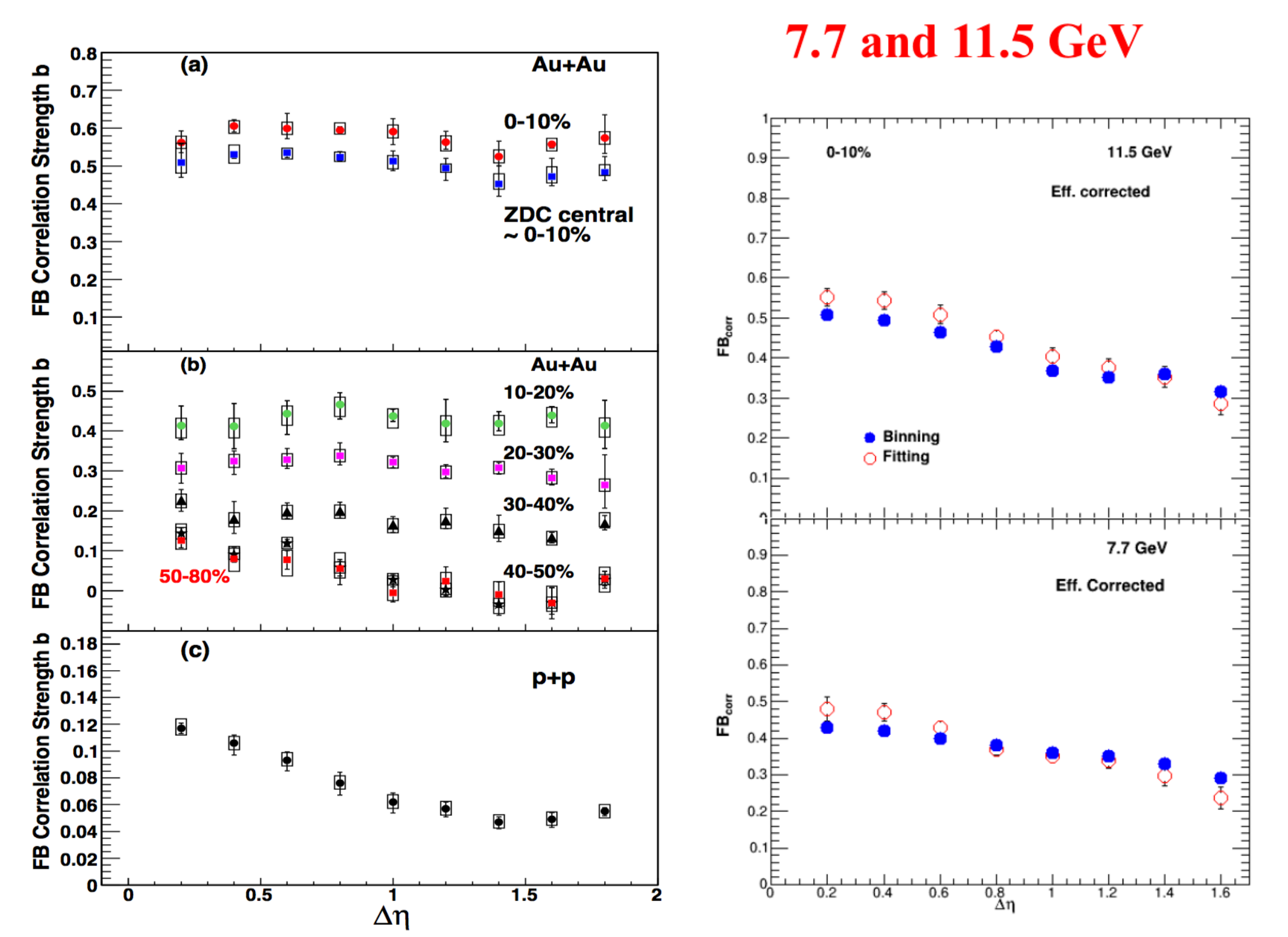}
\caption{
\label{fig:FBcorrelations1}
Published~\cite{Abelev:2009ag} forward-backward $\eta$ correlations from Au+Au
  and p+p collisions at 200~GeV exhibit significantly different structures.
At lower energy (preliminary results by Brijesh Srivastava), the correlation width
  decreases; this may be due to simple rapidity scaling, or it may indicate a more
  fundamental shift in the degrees of freedom in the intitial state.
}
\end{figure}

In 2009, STAR published~\cite{Abelev:2009ag} a first study of forward-backward correlations in \roots = 200~GeV 
  Au+Au and p+p collisions.
The results, shown in the left panel of figure~\ref{fig:FBcorrelations1}, are predicted~\cite{Walker:2004tx}
  to appear when multiple parton scattering becomes dominant and qualitatively confirmed predictions
  of the dual-parton model~\cite{Capella:1992yb} and --perhaps-- of the colored-glass condensate~\cite{Kovchegov:1999ep}.
As seen in the figure, the scale of the correlation in p+p collisions is on the order of a unit in $\eta$,
  but extends far beyond the TPC range for central Au+Au collisions.

Capturing more of the structure would allow for quantitative comparison to models-- the overall scale of the
  correlation carries less information.
In STAR, there is renewed interest in continuing these studies in BES; see right panels of figure~\ref{fig:FBcorrelations1},
  shown in a recent bulkcorr working group meeting in April, 2016.
In a spirit similar to the
  break-down of constituent-quark-number scaling of $v_2$~\cite{Adamczyk:2012ku,Adamczyk:2013gw},
  reduction of the width (beyond simple scaling with the beam rapidity) as \roots is lowered,
  would provide evidence that partonic degrees of freedom do not dominate even the intial state.
``Disappearance of QGP signature'' measures are important, to bolster our confidence that we are
  probing both sides of the phase transition in BES.

The right panels of the figure indeed suggest that the width of the correlation structure is
  reduced at low \roots, but whether it is simple scaling with beam rapidity, we are unable
  to determine.
It is important to reveal the entire structure of the correlation, as is possible with p+p
  collisions shown for the 200~GeV data.

The smooth $\eta$ coverage of the EPD is essential to extract the long-range structure of
  the intitial state at low \roots.

\subsubsection{Balance functions}

Complementary to the forward-backward correlations of the previous section, which 
  are charge-independent, through conditional probabilities,
  balance fuctions~\cite{Bass:2000az} probe the rapidity separation in the final
  state, of a pair of charges created at the same
  space-time point in the initial state.
The ``charge'' may be baryon number, strangeness or electric charge.

Due to strong longitudinal flow and coupling of space-time to momentum-space
  rapidity, a pair of charged particles initially created together will become increasingly
  separated as the collision evolves.
If hadronization is delayed due to QGP formation, the charges have less time to separate,
  and the balance function narrows~\cite{Bass:2000az}.
Comparing Au+Au collisions with p+p collisions at \roots=200~GeV, STAR reported~\cite{Adams:2003kg}
  first narrowing of the balance functions, supporting the concept of delayed hadronization.

\begin{figure}
\includegraphics[width=0.45\textwidth]{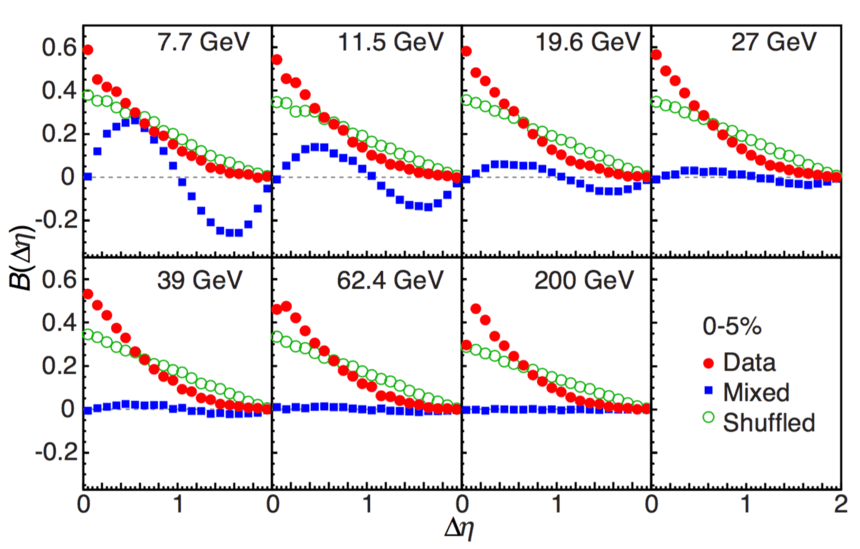}
\hfill
\includegraphics[width=0.45\textwidth]{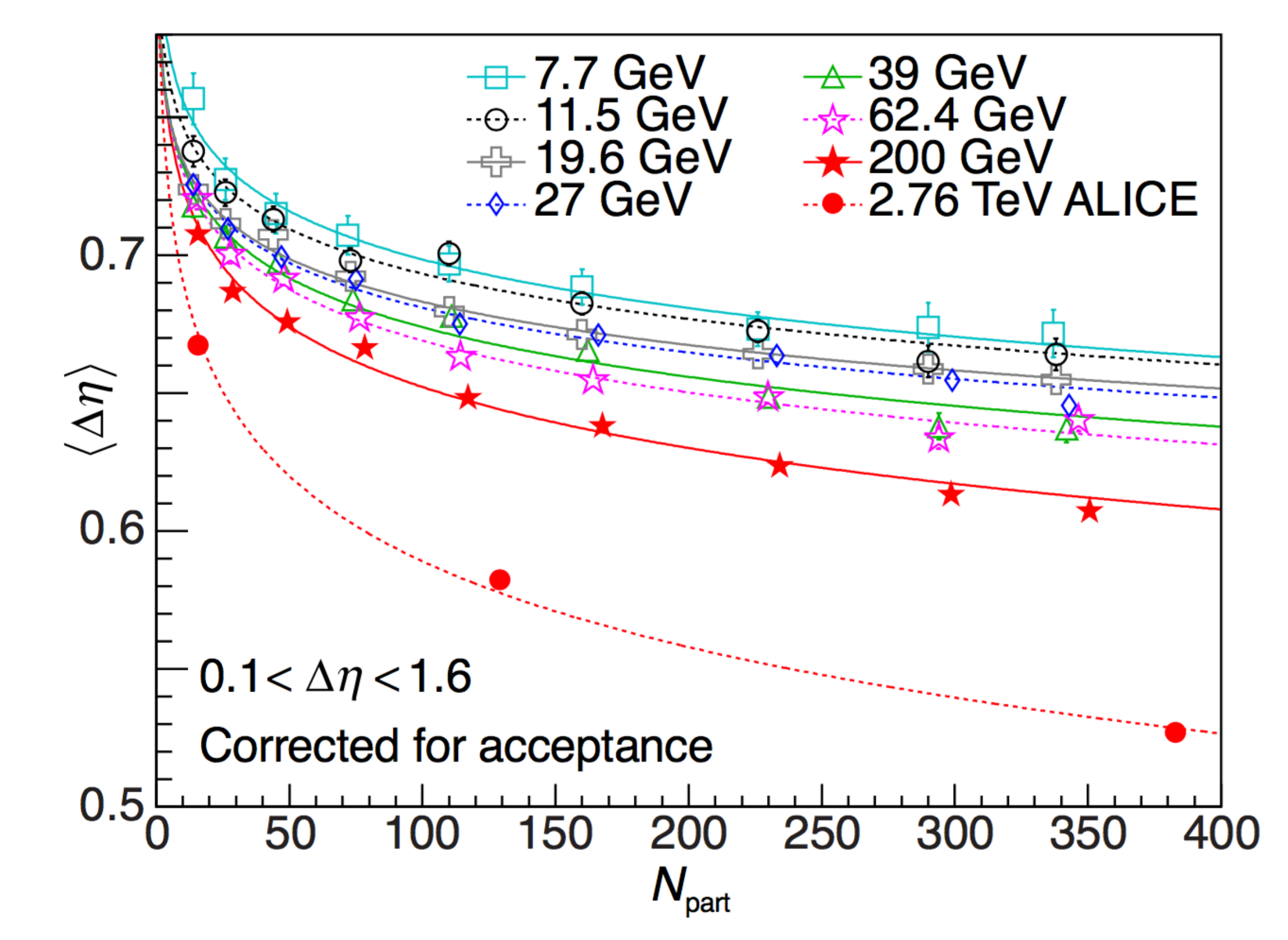}
\caption{
\label{fig:BalanceFunctions1}
Electrical-charge balance functions published by STAR~\cite{Adamczyk:2015yga}.
Left panel: the balance functions for the most central collisions.
Right panel: the narrowing of the balance function, relative to the most peripheral
  collisions, is strongest at high \roots and $N_{\rm part}$.
}
\end{figure}

Figure~\ref{fig:BalanceFunctions1} shows STAR's measurement of the balance function
  in BES-I.
From the left panel, it appears that the entire structure of the balance function is captured
  for the most central collisions, in the TPC alone.
However, it is the centrality {\it evolution} of the function that provides information 
  on the QGP evolution time; this is shown in the right panel.
For all centralities, the balance function is wider for more peripheral collisions; STAR
  cannot capture the entire structure of the effect, and efficiency effects become very
  large at large $\Delta\eta$ using the TPC alone.
Indeed, ALICE claims a ``universal'' scaling of the narrowing of the balance function,
  which may contradict a narrative in which the qualitative nature of the degrees of 
  freedom differ at very high and low \roots.

The balance function has the potential to provide convincing evidence of the dissapearance
  of a QGP at the lowest BES-II energies.
However, quantitative fits to the balance function depend crucially on the corrections at large
  $\Delta\eta$, and few have been attempted.
A convincing and lasting case of the evolution of the underlying physics will only become 
  possible with $\Delta\eta$ coverage that exposes the entire balance function structure.
The EPD is crucial to expose this underlying physics.

\subsection{Physics with the first-order event plane}

\subsubsection{Directed flow}
The directed flow is expected to be especially sensitive to the early evolution of a heavy-ion
collision. Three-fluid hydrodynamic models with an explicit first-order phase transition in the 
equation of state predict a dip in the slope of the (net)proton $v_1$ at mid-rapidity 
\cite{Rischke:1995ir}, \cite{Brachmann:1999xt}, \cite{Stoecker:2004qu}.
This non-monotonic behavior is not seen in transport models, such as UrQMD \cite{Bass:1998ca}\cite{Bleicher:1999xi}, nor is it seen in 
hybrid\cite{Steinheimer:2014pfa} or PHSD\cite{Konchakovski:2014gda} calculations. A double sign change in the mid-rapidity net-proton $v_1$ slope is 
seen in the STAR BES for mid-central (10-40\%) collisions (figure \ref{fig:BESII_v1}) 
\cite{Adamczyk:2014ipa}. 
Net-protons are used as a proxy of transported particles pushed to mid-rapidity, as anti-protons 
may only be produced. Near a critical point one would expect a mixed partonic-hadronic phase with
small transverse pressure gradients. This softening of the EOS would allow transported particles to 
move through the collision region with low resistance leading to a minimum in the net-baryon 
$v_1$ slope. At collision energies much lower than the phase transition point the system is dominated
by transported particles, whereas at high collision energies produced particles (with no preferred
direction) dominate at mid-rapidity leading to a flattening in the slope. 

$v_1$ results require forward instrumentation to measure the first order event plane. For 
$7.7-39GeV$ this analysis is performed using the BBC, while at $62.4GeV$ and $200GeV$ the ZDC is
used. Exactly how the ratio of transported to produced particles changes as a function of rapidity
for different beam energies is poorly understood and can lead to bias in this result. A detector, 
such as the EPD, with high $\eta$ segmentation increases our understanding of this dependence. More
obviously high segmentation in $\phi$ reduces statistical error by providing a better event-plane
resolution. The statistical error here scales as $1/(R_1*\sqrt{N})$, where $R_1$ is the first order
event plane resolution. Error reductions from increased EPD $R_1$ and BES II statistics allow for a 
more finely centrality binned analysis which will lead to better understanding of the relevance of 
baryon stopping at low beam energies.

\subsection{Global $\Lambda$ polarization}

Non-central heavy-ion collisions provide a system with non-zero total angular momentum which can be 
transferred, in part, to the fireball via baryon stopping. It has been predicted that this angular 
momentum will lead to a net spin of emitted particles through coupling with the bulk material. Due 
to its parity violating decay the Lambda baryon is self-analyzing, which allows us to associate the 
daughter proton decay direction with spin. Therefore we can use Lambda spin as a probe of 
net-hadron spin. The global polarization is measured as a correlation between the system angular momentum direction (from the first order event 
plane) and the daughter proton's momentum in the $\Lambda$ rest frame (denoted by a *) 
$P_H \sim \sin(\Psi_1 - \phi^*_{p})$ and is shown for $\Lambda$s and $\bar{\Lambda}$s in the BES
figure \ref{fig:LambdaPol} (panel 1).
Global polarization arises principally from either QCD spin-orbit coupling of spins with system
angular momentum (or vortices in a fluid) or from magnetic moment coupling with the 
(extremely large - though short lived) magnetic field left at mid-rapidity as charged spectators
are sheared off and pass by the fireball. In a vortical coupling picture one expects both 
$\Lambda$s and $\bar{\Lambda}$s to have identical, positive, global polarization. Since the 
magnetic moment of the $\Lambda$ is negative, and $\mu_{\Lambda} = -\mu_{\bar{\Lambda}}$, one expects
the global polarization due to the magnetic coupling to be negative for $\Lambda$s and positive 
(though same magnitude) for the $\bar{\Lambda}$s. In the limit of low polarization these effects
are simply additive. Therefore by averaging the $\Lambda$ and $\bar{\Lambda}$ polarization one can 
get the contribution due to vorticity and by subtracting one from the other and dividing by 2 one
can get the magnetic coupling. These results (corrected for feed down from $\Sigma^0$s) are shown 
in figure \ref{fig:LambdaPol} (panel 2).

Though individual data points are not hugely significant the overall positive trend with 
$\sqrt{s_{NN}}$ gives us confidence that we are seeing, for the first time in any experiment, a 
positive signal of global polarization. Vorticity is a brand new observable and can, perhaps, be a 
new lever with which to evaluate hydrodynamic and transport models. The polarization is 
corrected for the resolution in $\Psi_1$ the same way as a $v_1$ measurement so, just like in 
$v_1$, the reported statistical errors are scale by $1/(R_1*\sqrt{N})$ (where $N$ is the number of 
Lambdas). At present the analysis is too statistically challenged for real differential studies to
be performed. It is not clear, and is of interest, how this variable depends on 
$E_{\Lambda}$, $\phi_{\Lambda}$, and centrality. There appears to be a decreasing trend with 
$\sqrt{s_{NN}}$ (STAR has reported a null result for $62.4GeV$ and $200GeV$) which is not universally
predicted or very significant in the data. It is furthermore of interest what, if any, spin coupling with
spectator magnetic field may exist and what $\sqrt{s_{NN}}$ dependence this magnetic field may have. At present it
is not possible to claim a positive result for magnetic polarization, but with the increased 
statistics of BES II and the increased resolution of the EPD it may be possible to make such a
positive measurement in the future.

\subsubsection{Spatial tilt and twist and first-order azimuthally-sensitive HBT}

STAR published extensive systematics on azimuthally-sensitive pion intensity interferometry (so-called ``HBT'')
  radii in the BES-I~\cite{Adamczyk:2014mxp}.
These measurements, which were relative to the {\it second}-order event plane, provided coordinate-space
  information analogous to elliptic flow.
In particular, they revealed a source that remained out-of-plane extended, even while elliptic flow is characterized
  by an in-plane extension in momentum space.
This six-dimensional freezeout configuration provides strong contraints for dynamic models of the system evolution.

\begin{figure}
\includegraphics[width=0.3\textwidth]{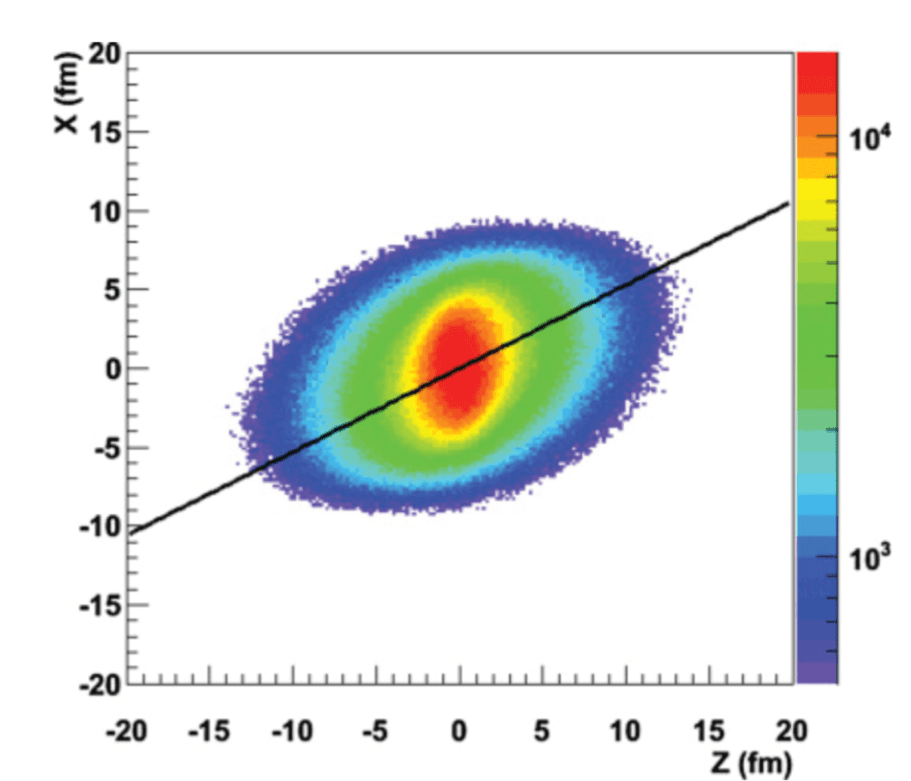}
\hfill
\includegraphics[width=0.3\textwidth]{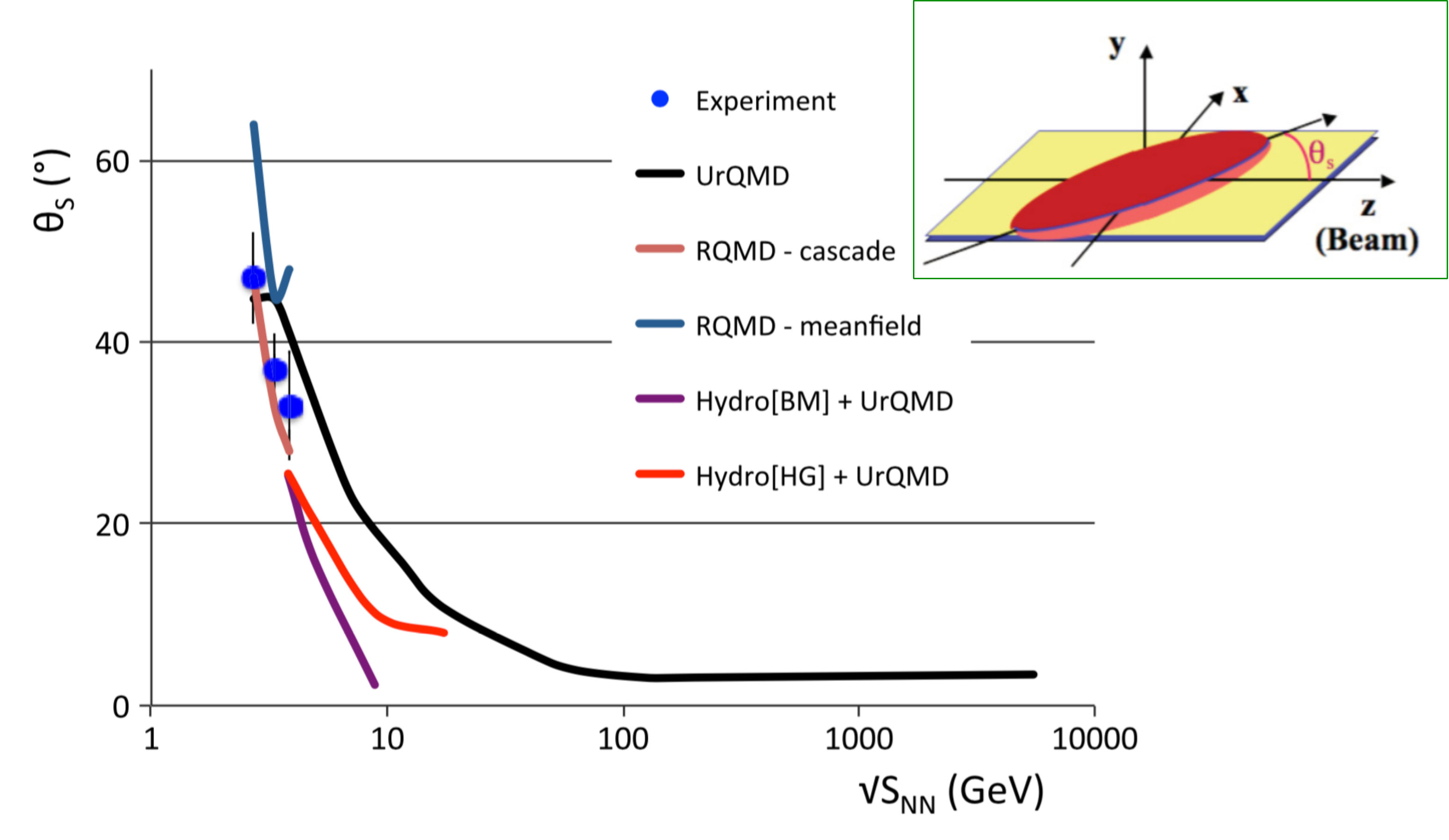}
\hfill
\includegraphics[width=0.3\textwidth]{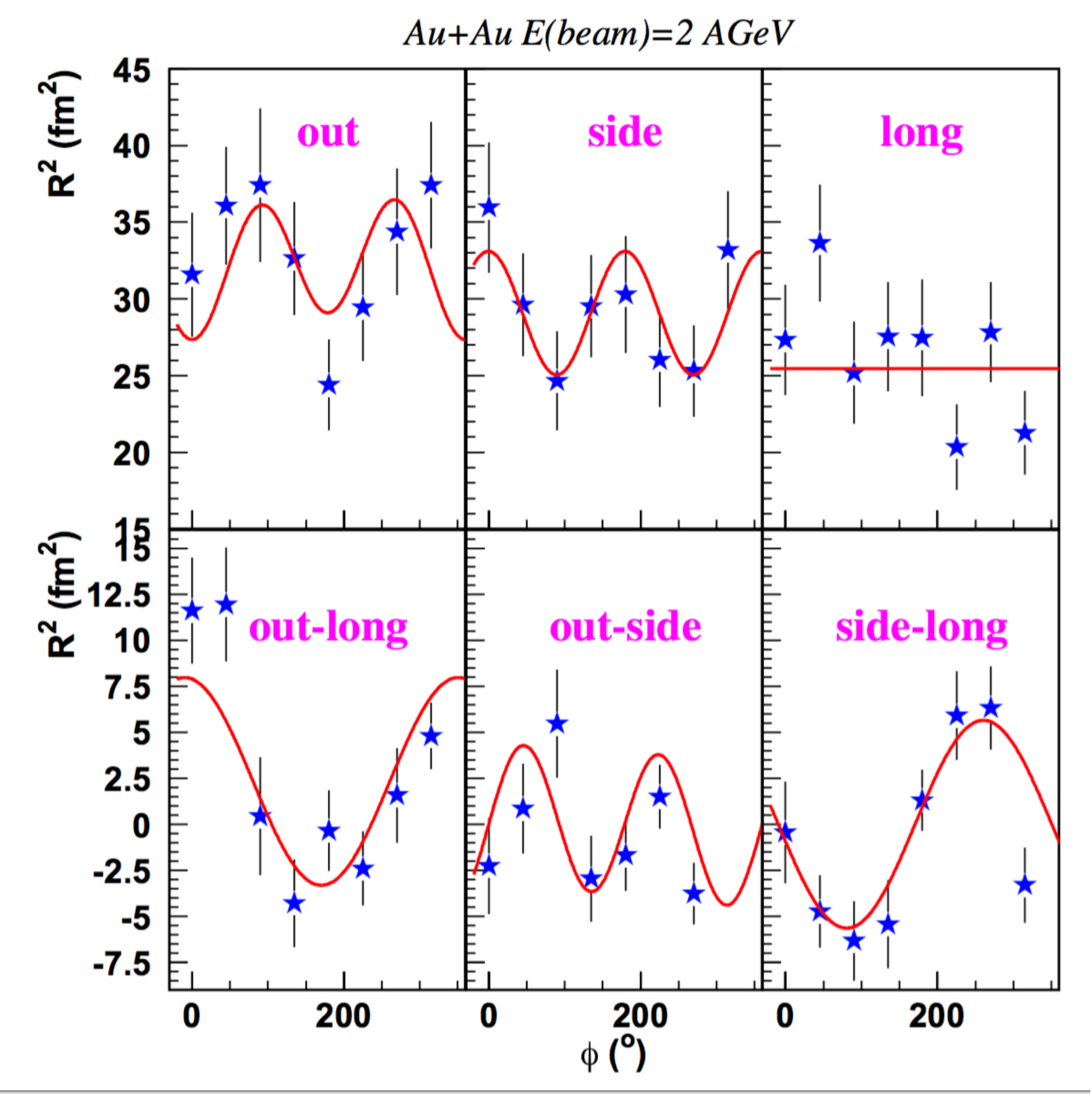}
\caption{
\label{fig:asHBT1}
Left: UrQMD prediction for the freezeout configuration in coordinate space, projected onto the reaction plane,
  for Au+Au collisions at \roots=3.84~GeV.~\cite{Mount:2010ey}\newline
Center: Predictions for the tilt angle from mid-central Au+Au collisions as a function of \roots, for
  transport calculations.~\cite{Lisa:2011na}\newline
Right: Measurement of the first-order azimuthal dependence of the six ``HBT radii'' from Au+Au
  collisions in the fixed-target experiment E895 at the AGS; beam energy was 2~AGeV.~\cite{Lisa:2000xj}
}
\end{figure}

Much more information on the femtoscopic substructure of the system may be revealed through {\it first}-order
  anisotropies of the HBT radii.
In particular, the spatial tilt~\cite{Lisa:2000ip,Lisa:2011na,Csernai:2013vda} and ``twist''~\cite{Mount:2010ey,Graef:2013wta}
  of the hot emission zone may be extracted.
The left panel of figure~\ref{fig:asHBT1} shows the freezeout configuration in coordinate space in the
  reaction plane ($\hat{z}$ is the beam direction and the impact parameter is directed along $\hat{x}$).

The primary feature of the configuration is a strong tilt away from the beam direction.
Importantly, this feature is not to be confused with directed flow (though it frequently has been):
  for example, the spatial tilt angles at AGS energies have been measured to be $\sim50^\circ$, as
  compared to momentum-space angles $\lesssim2^\circ$; to make the point even more strongly,
  the spatial tilt angle has the opposite sign to the momentum tilt angle~\cite{Lisa:2000xj,Lisa:2000ip}.
The spatial tilt encodes independent information from momentum-space tilt (i.e. directed flow $v_1$), but does
  reveal some of the physics behind directed flow~\cite{Lisa:2000ip}.
Given the prominence of the directed flow measurements~\cite{Adamczyk:2014ipa} to the BES program,
  the importance of a tilt measurement is elevated to high priority.

A measure of the spatial tilt has only successfully been performed at the Brookhaven AGS; results from
  fixed-target experiment E895 are shown in the left panel of figure~\ref{fig:asHBT1}, together
  with fits that allowed extraction of the ellipticity and tilt of the source (leading to the datapoints
  in the center panel of the figure).
In that experiment, the first-order event-plane resolution varied from 0.72 at 6~AGeV, to 0.94 at 2~AGeV;
  these are tremendously good resolutions.
As with the hyperon polarization, somewhat worse resolution may be accommodated by increased pion pair
  statistics.
In particular, again as with the hyperon polarization, the figure of merit is 
  $\sqrt{N_{\rm pair}}\times RP_{\rm EP}^{(1)}$.

E895 recorded $\sim10^6$ pion pairs in a wide centrality range, whereas in BES-II, for a narrow range (say 10\%) in centrality,
  we may expect $~10^7$ pairs (at low $q_{\rm inv}$ in the low-\roots collisions.
Hence, a decent measure of the tilt should be attainable with event plane resolutions on order
  0.3.

Chris Anson attempted this analysis with BES-I data, and the BBC resolution was (barely, by our
  estimates) insufficient to extract this interesting and unique feature of the geometric substructure
  of the plasma.
With the combined statistics of BES-II and the improved event-plane resolution of the EPD, we will
  be able to measure it well.

(We point out that a further analysis of the substructure, corresponding to the ``twist'' of the source
  (essentially corresponding to the obvious fact that the ``tilt'' of the source in the left panel of
  figure~\ref{fig:asHBT1} depends on scale), has been proposed~\cite{Mount:2010ey,Graef:2013wta}
  a the formalism developed.
This sub-femtoscopic feature encodes the rotational hisory of the evolving QGP~\cite{Graef:2013wta}.
We believe that with BES-II statistics and EPD resolution, we should be able to extract the
  twist, but we have no firm estimates to present at the moment.)

\subsection{Physics with the higher-order event planes}
\begin{figure*}[hbtp]
\centering\mbox{
\includegraphics[width=0.99\textwidth]{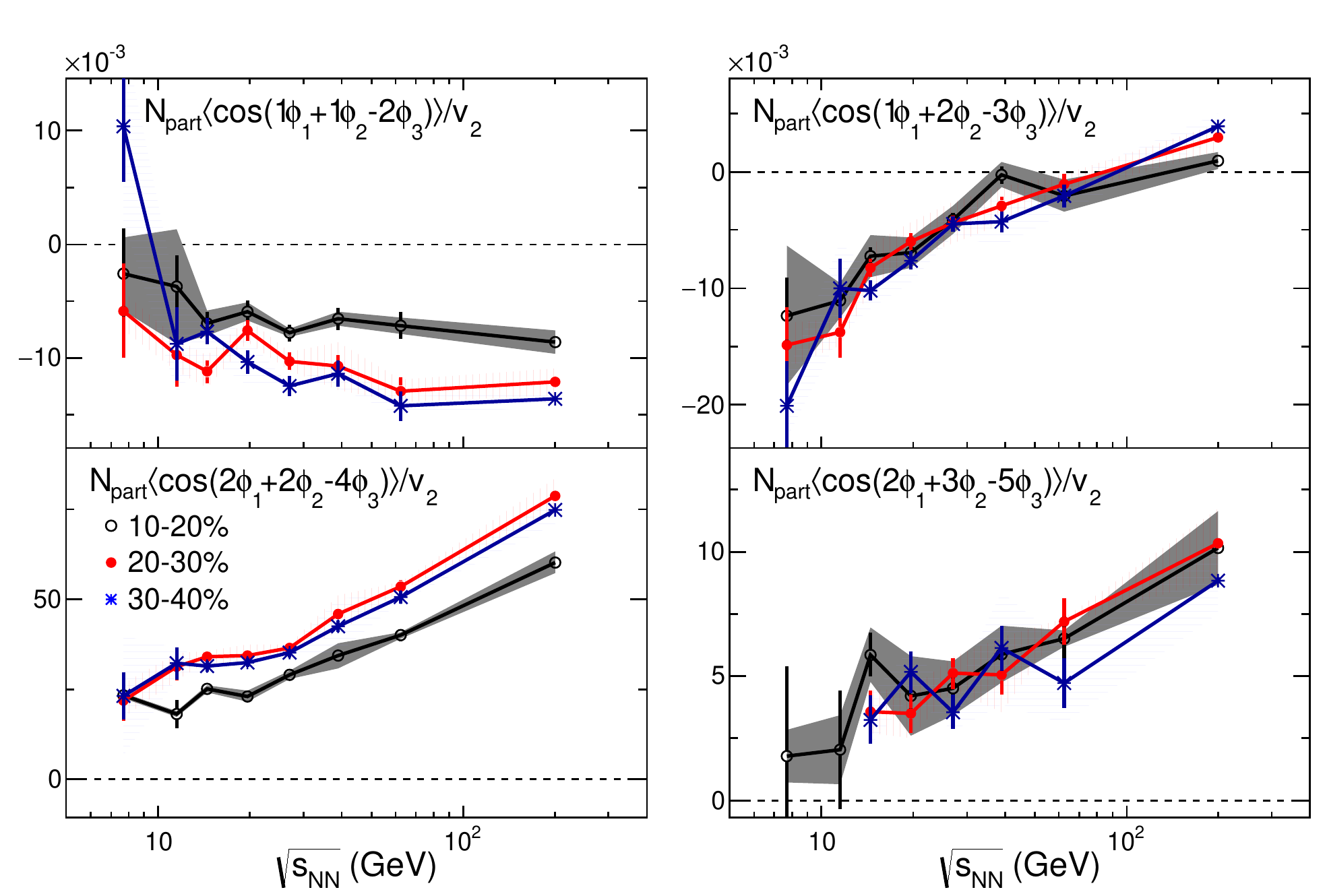}}
\caption{ (color online) The {\snn} dependence of
  {\npart}{\cijk}{/$v_2$} for $(m,n)=(1,1)$ (top left), $(1,2)$ (top
  right), $(2,2)$ (bottom left) and $(2,3)$ (bottom right) for three
  selected centrality intervals. In the bottom right panel, the lowest
  energy points for the 20-30\% and 30-40\% centrality intervals,
  having large uncertainties, are omitted for clarity. Statistical
  uncertainties are shown as vertical error bars while the statistical
  errors are shown as shaded regions. }
 \label{edep}
\end{figure*}

Three particle azimuthal correlation measurements like {\cijk} scaled
by the appropriate products of $v_n$ are equivalent to reaction plane
correlations. STAR has measured {\cijk} for energies from 7.7 GeV up
to 200 GeV but only within the STAR TPC acceptance of
$|\eta|<1$. Fig.~\ref{edep} shows {\npart}{\cijk}$/v_2$ as a
function of {\snn} for three centrality intervals: 10-20\%, 20-30\%,
and 30-40\%. The correlations {\caab}, {\cabc}, {\cbbd}, and {\cbce}
have either $m=2$, $n=2$, or $m+n=2$. When $v_2$ is large, as it is
for the 10-20\%, 20-30\% and 30-40\% centrality intervals, then
$\langle\cos(m\phi_1+n\phi_2+2\phi_3)\rangle/v_2 \approx
\langle\cos(m\phi_1+n\phi_2+2\Psi_{\mathrm RP})\rangle$ where
$\Psi_{RP}$ is the reaction plane angle. Correlations including a
second harmonic should then provide information about two-particle
correlations with respect to the second harmonic reaction plane:
\begin{eqnarray*}
  \langle\cos(\phi_1+\phi_3-2\phi_2)\rangle/v_2 &\approx& \langle\cos(\phi_1'+\phi_2')\rangle, \\
  \langle\cos(1\phi_1+2\phi_3-3\phi_2)\rangle/v_2 &\approx& \langle\cos(\phi_1'-3\phi_2')\rangle, \\
  \langle\cos(2\phi_1+2\phi_3-4\phi_2)\rangle/v_{2} &\approx& \langle\cos(2\phi_1'-4\phi_2')\rangle, \\
  \langle\cos(2\phi_3+3\phi_1-5\phi_2)\rangle/v_{2} &\approx& \langle\cos(3\phi_1'-5\phi_2')\rangle,
\end{eqnarray*}
where $\phi'=\phi-\Psi_{\mathrm RP}$. Since we are integrating over
all particles in these correlations, the subscript label for the
particles is arbitrary so we've reassigned them for convenience. At
200 GeV, all correlations except $\langle\cos(\phi_1'+\phi_2')\rangle$
are positive. This suggests an enhanced probability for a pair of
particles in one of two possible configurations: either 1)
$\phi_1'\approx \pi/3$ and $\phi_2'\approx 2\pi/3$ or 2) $\phi_1'
\approx -\pi/3$ and $\phi_2' \approx -2\pi/3$. This result is
surprising since it implies a preference for correlated particles to
either both be in the upper hemisphere, or both in the lower
hemisphere. For energies below 200 GeV, {\cabc} changes sign so that
$\langle\cos(\phi_1'+\phi_2')\rangle$ and
$\langle\cos(1\phi_1'-3\phi_2')\rangle$ are both negative while
$\langle\cos(2\phi_1'-4\phi_2')\rangle$ and
$\langle\cos(3\phi_1'-5\phi_2')\rangle$ are both positive. This
condition indicates a preference for particle pairs with $\phi_1'
\approx 0$ and $\phi_2' \approx \pi$. This preference for back-to-back
particle pairs aligned with the reaction plane is consistent with an
increased importance for momentum conservation at lower
energies. These conclusions are stable for the second order event
plane because the correlations {\cijk} are mostly independent of
$\Delta\eta$ between particles as long as one of the particles is
associated with the second harmonic. There is however a strong
$\Delta\eta$ dependence for correlations like {\cabc} when looking at
the $\Delta\eta$ between particles one and three. This makes it
impossible to interpret the correlations in terms of reaction
planes. It is unclear whether the $\Delta\eta$ dependence arises from
short-range two-particle correlations or from reaction plane
decoherence. The addition of the EPD with it's high segmentation will
allow us to extend these measurements to large $\eta$ to eliminate any
interference between short and long range effects. Without this
capability, it will not be possible to to reliably determine event
planes for harmonics above n=2.

\subsection{Chiral Magnetic/Vortical Effects in the Strong Interaction}

STAR has been searching for evidence of chiral magnetic/vortical effects for more than a decade, and so far the 
experimental observables support the pictures of the chiral magnetic effect, the chiral magnetic wave and the chiral 
vortical effect. To draw a firm conclusion, an effective way is needed to disentangle the signal and the background 
contributions, the latter of which is interwined with collective flow. Collisions of isobaric nuclei, i.e. 
$^{96}_{44} Ru + ^{96}_{44} Ru$ and $^{96}_{40} Zr + ^{96}_{40} Zr$ , present a unique opportunity to vary the initial 
magnetic field by a significant amount while keeping everything else almost the same.

A three-point correlator, $\gamma = \left\langle \cos(\phi_{\alpha} + \phi_{\beta} - 2\Psi_{RP}) \right\rangle$, 
sensitive to the CME was proposed as a tool for disentangling the flow background, where $\phi$ is the azimuthal 
angle, the subscripts $\alpha$ and $\beta$ denote the particle charge (positive or negative), and $\Psi_{RP}$
is the angle of the reaction plane of a given event.

In addition to the chiral magnetic effect, a chiral vortical effect (CVE) has also been predicted. In this case,
the role of the magnetic field is played by vorticity coupled with a baryon chemical potential ($\mu_B$), and the 
resultant electric charge separation will be replaced with the baryonic charge separation. Vorticity in heavy-ion 
collisions is a natural consequence of angular momentum conservation, and there is always a finite $\mu_B$ at RHIC 
energies. The CVE can be searched for via $\gamma$ correlations between two baryons, one of which is preferably 
charge neutral to avoid the complication by the CME.

\begin{figure}[htbp]
\begin{center}
{
 \mbox{ \includegraphics[width=0.45\textwidth]{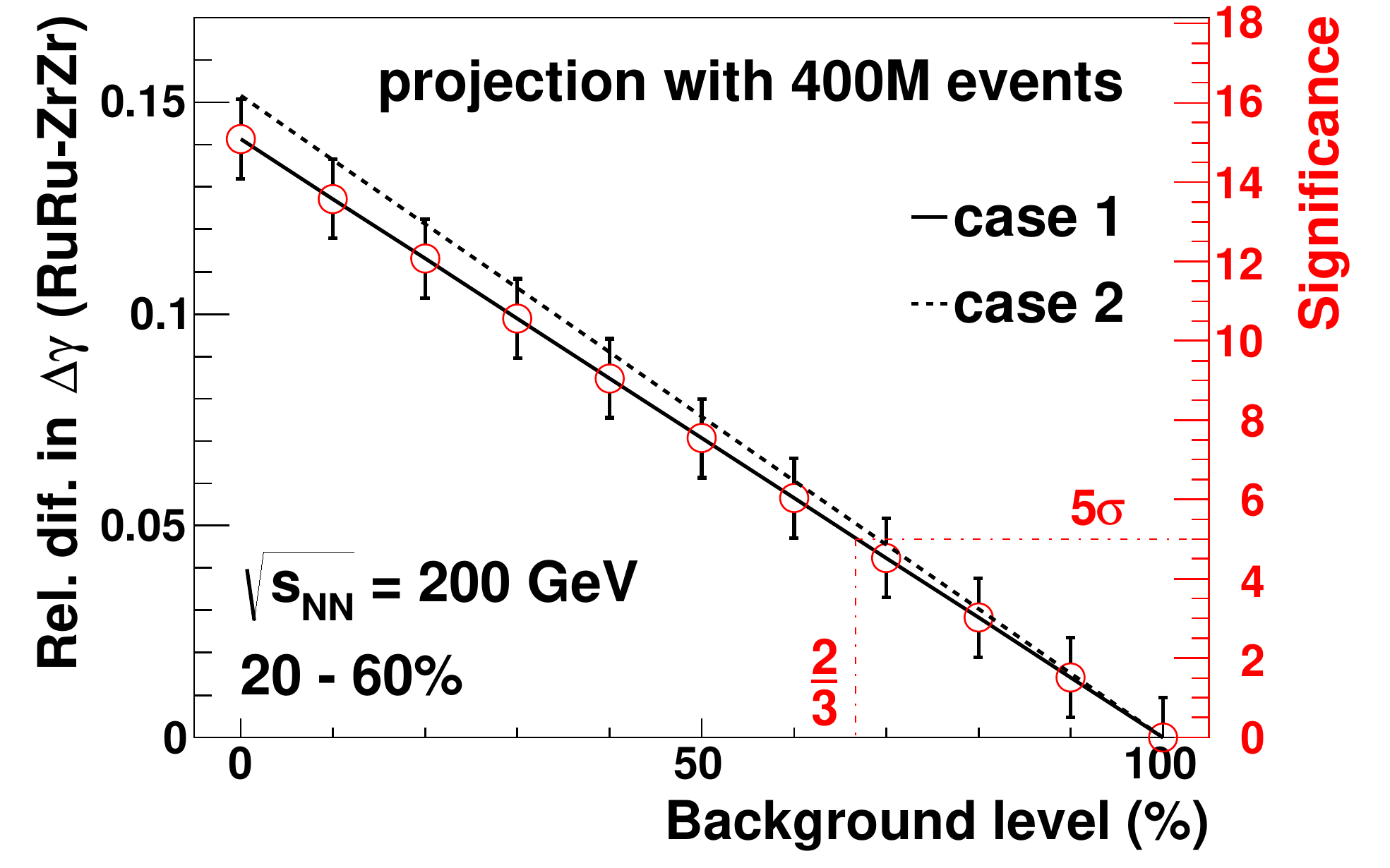}}
  \mbox{\includegraphics[width=0.45\textwidth]{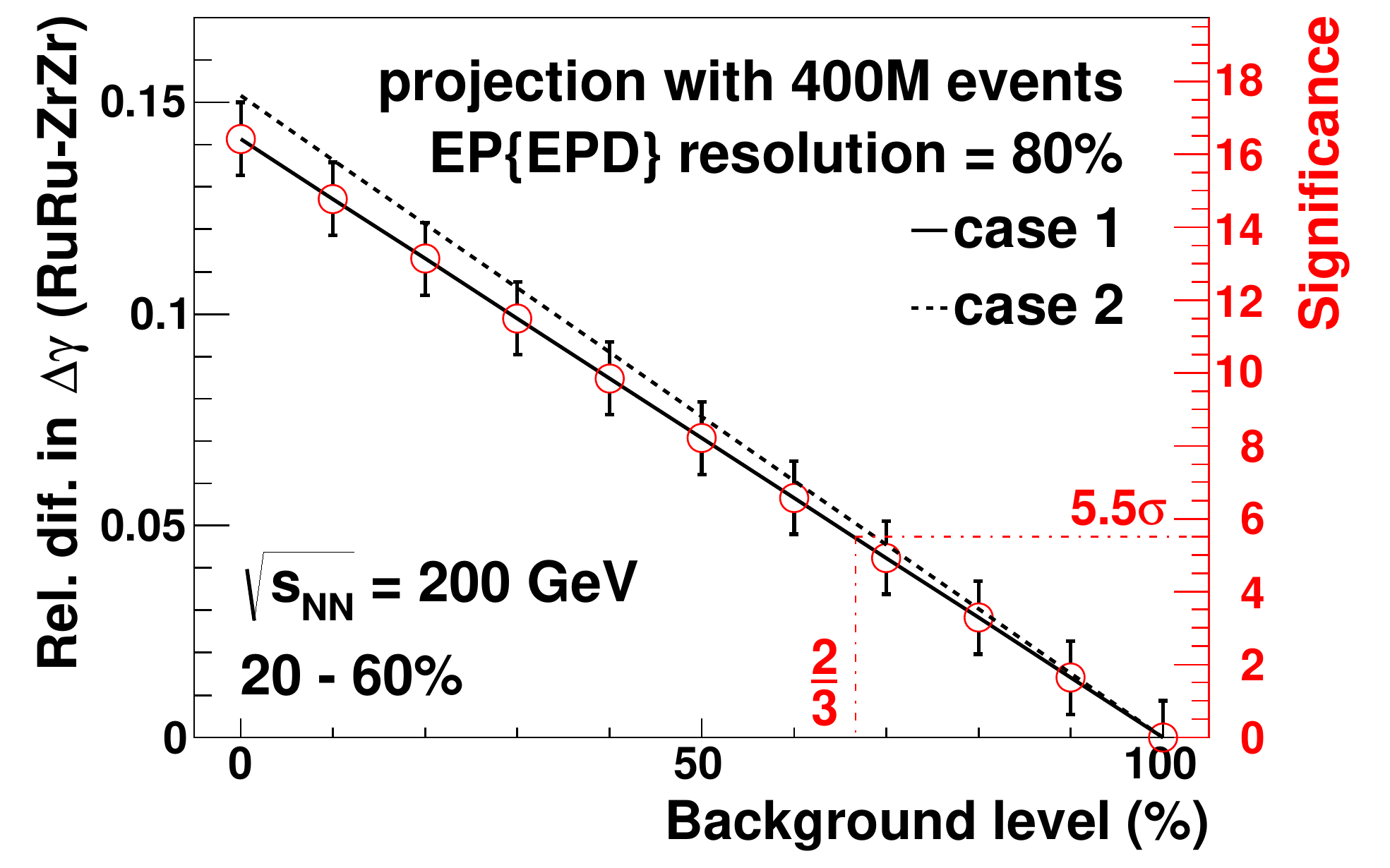}}
  }
  \end{center}
  \caption{Magnitude (left axis) and significance (right
axis) of the relative difference in the projected $\Delta\gamma\times N_{part}$ between $_{44}^{96}Ru+_{44}^{96}Ru$ and $_{40}^{96}Zr+_{40}^{96}Zr$ at 200 GeV. The right figure is the same as the left, except with the event plane from the EPD instead of the TPC. The EPD event plane resolution has been simulated to be 80\%.}
  \label{fig:sigTPCvEPD}
\end{figure}

When a different background level is assumed, the magnitude and significance of the projected difference in $\delta\gamma$ between $Ru+Ru$ and $Zr+Zr$ change accordingly, as shown in figure \ref{fig:sigTPCvEPD}. The future measurements of the isobaric collision data will determine whether there is a finite CME signal observed in the $\gamma$ correlator, and if the answer is ``yes'', will ascertain the background contribution when compared with this figure. Up to now, the projections have been made with the event plane reconstructed with the STAR TPC. By 2018 when the isobaric collisions are supposed to occur, the EPD would provide an independent event plane in the forward/backward rapidity regions. The right side of figure \ref{fig:sigTPCvEPD} shows the magnitude and significance of the projected difference in $\delta\gamma$ with the event plane reconstructed from the EPD with 400M events for each collision system at 200 GeV. Using the simultated EPD event plane resolution of 80\% over the centrality range of interest, we will be able to obtain the CME signal with a significance better than 5.5$\sigma$ if the background contribution is less than two thirds. This increase from 5.0 to 5.5$\sigma$ is due to a better $2^{nd}$ harmonic event plane resolution by the EPD due to statistical effects (though there are fewer particles per rapidity window in the EPD acceptance, the $\eta$ acceptance of the EPD is larger than that of the TPC).

Currently, the systematic uncertainties due to auto-correlations are checked with the ZDC event plane ($R_2 \sim 0.3$). The results are in agreement with the TPC event plane method within large statistical errors. Because of the large $\eta$ gap afforded by the EPD, systematic errors will be reduced similar to the ZDC event plane. The much higher event plane resolution of the EPD compared to the ZDC will allow us to study those systematic effects in detail.

\subsubsection{Fixed Target Analyses}
\label{sec_FXT}

The implementation of an internal fixed-target at STAR will allow the BES-II physics program to be
extended to lower energies and to reach the higher $\mu_B$ regions below the expected onset of
deconfinement. Many of the physics analyses proposed for the fixed-target (FXT) will be extensions of the
BES-II analyses to lower energies and will place similar requirements on the detector upgrades. Primarily
the requirements will be for accurate and independent measurements of reaction plane and centrality.
However it should be noted that the effective $\eta$ coverage of the EPD is significantly different for
fixed-target events than for collider events.

For fixed-target events, the EPD will cover the psuedorapidity range from $2.4 < |\eta| < 4.8$. The
single beam rapidities will range from $y = 5.5$ for the highest energies proposed, to $y = 2.1$ for the
lowest energies. This indicates that the EPD will cover the projectile fragmentation regime, which means
that its role will be similar the role played by the ZDC in 200 GeV Au+Au collisions. The centrality
measurement in the EPD will essentially be the number of spectator protons, while the event will be
determined by the recoil of these spectator protons off of the compressed interaction region. These
methodologies for centrality and event-plane determination have been well-established during the AGS
fixed-target program in the 1990s.

The physics program of the fixed target will require event-plane determination for both directed and
elliptic flow studies and for azimuthally dependent HBT measurements. Across the energy range of the
FXT program, quantifying the rise and then the fall of the directed flow will determine the energy of
maximum compression and will provide constraints of models describing a softening of the equation of
state at the onset of deconfinement. The elliptic flow will be negative (out-of- plane squeeze-out) for the
lowest energies for the FXT program. As the FXT energies are increased, the elliptic flow values will rise,
cross zero, and become positive. The precise point where the elliptic flow crosses zero will determine
the balance between the projectile transit time and the generation of repulsive compression. The
azimuthal HBT will provide key insights to the development of the spatially elongated interaction region.
The EPD will allow the FXT program to have essentially full $\eta$ coverage from target to projectile
rapidity which will allow the $dN/d\eta$ to be mapped out across the FXT energies. This
helps determine the decrease in stopping, which is key to the variation of $\mu_B$
with collision energy.

\newpage

\section{EPD Overview and Simulation}

The EPD will be the the same size as the Beam-Beam Counter (BBC), as it is required to sit in the same space within the STAR experiment, which gives is the same acceptance in $\eta-\phi$  The EPD will extend from a radius of 4.5 cm (1.77 in) to 90 cm (35.4 in) and will be located at  $z = \pm 375$ cm, which corresponds to $5.1 < \eta < 2.1$.  The design allows for the EPD to be installed behind the BBC, or for the EPD to be installed instead of the BBC.  The EPD scintillator is 1.2 cm thick, the same thickness as the BBC.  

\begin{figure}[ht]
\centering
\mbox{\includegraphics[width=0.45\textwidth]{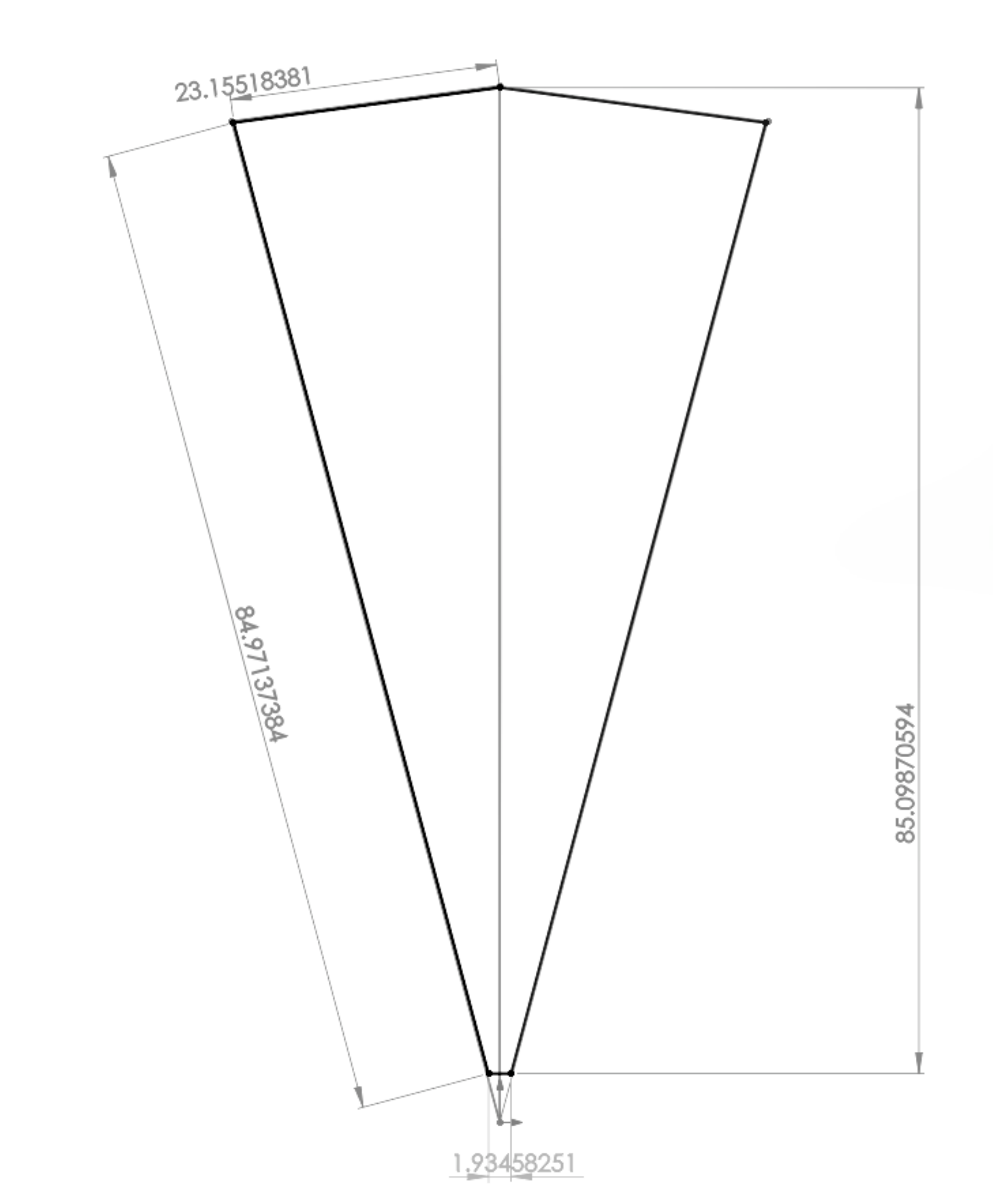}}
\caption{The frame outline of the EPD super sector.  There are 12 of these in each wheel (or disk) of the EPD, which amounts to 24 total super sectors for the whole design.}
\label{fEPD_SuperSector_Frame}
\end{figure}

\subsection{Segmentation}
\begin{figure}[ht]
\centering
\mbox{\includegraphics[width=0.45\textwidth]{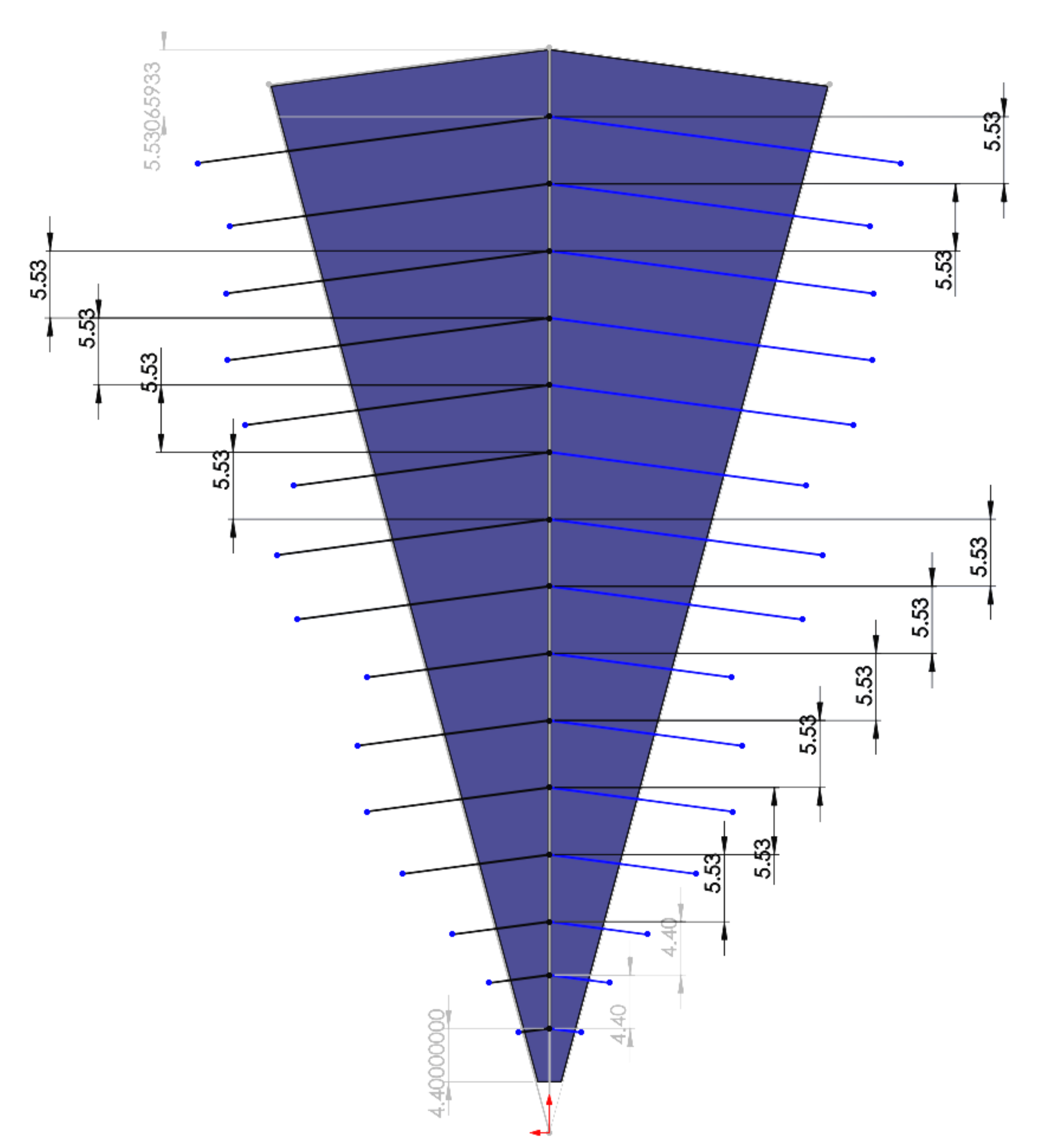}}
\mbox{\includegraphics[width=0.35\textwidth]{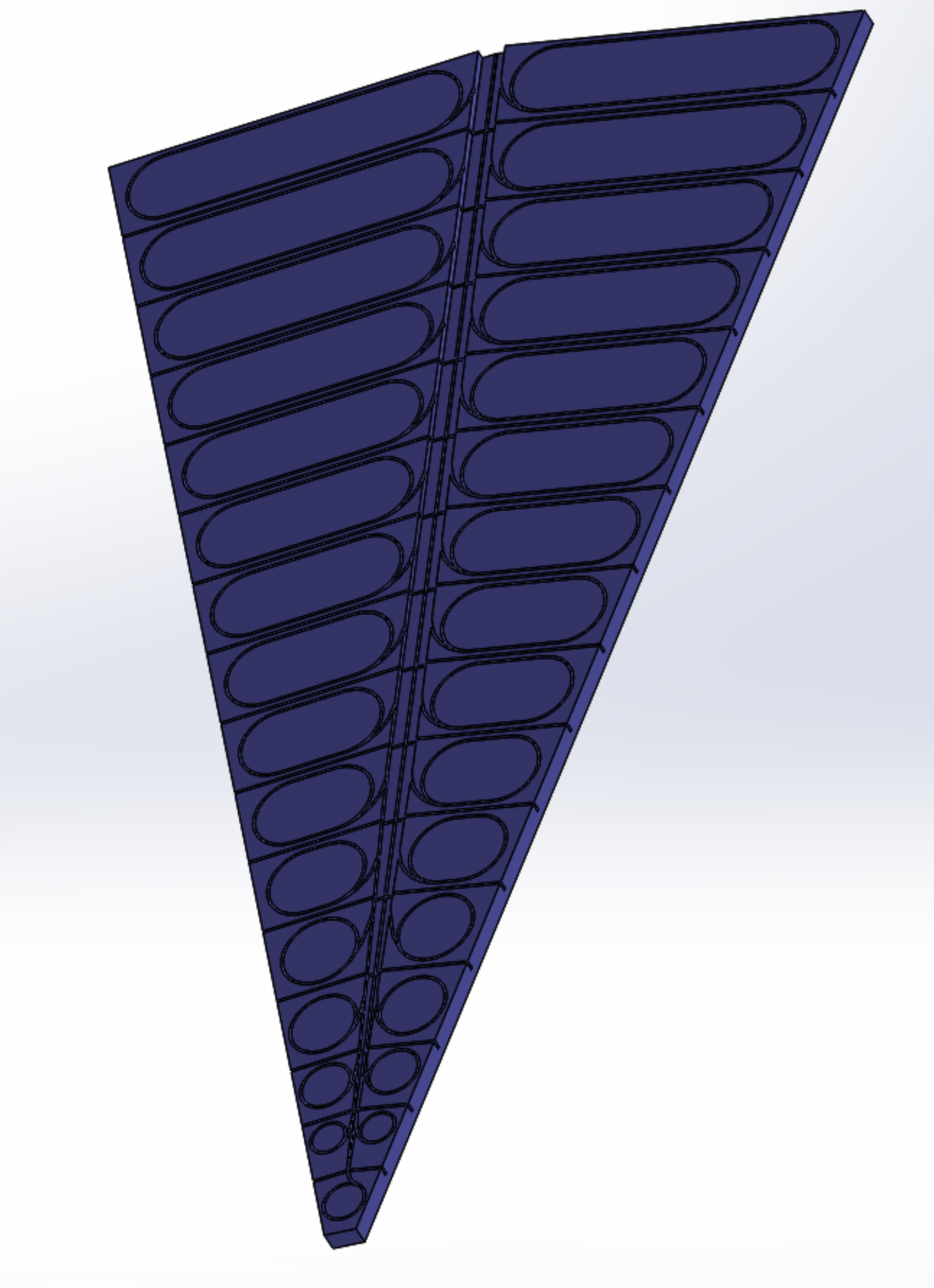}}
\caption{The super sector on the left shows the widths of all of the tiles in a single super sector.   They are slightly narrow towards the center.  On the right is the section shown at an angle with the fiber grooves visible.  The channel in the center is where the WLS fibers will be routed to the end where the fiber connectors are located.}
\label{fEPD_SuperSector_Tiles}
\end{figure}

The EPD will have 24 azimuthal segments, spanning an angle of 15\degree~which we give the label "sector".  A super sector will contain two sectors and will be machined out of a single piece of scintillator, the outer dimensions are shown in Figure \ref{fEPD_SuperSector_Frame}.  There will be 16 segments radially, however the innermost tile will span an entire super sector.  Each tile other than the innermost will span 15\degree, and the innermost tile will spans the full 30\degree of the super-sector.  The segmentation can be seen in Figure \ref{fEPD_SuperSector_Tiles}, with the details in Table \ref{tEPD_Segmentation}.   This results in 744 channels for the two EPD disks.   The tile size was chosen based on dN/dy measurements from $\sqrt{{s}_{NN}}$ = 19.6 GeV by PHOBOS, the details are shown in \ref{sec_sim}.  The EPD is designed such that the probability of multiple particle hits in the same tile would be less than 10\%.  This increases to 65\% for the top energy of 200 GeV with the same segmentation.

\begin{table}[]
\centering
\caption{EPD Segmentation by row}
\label{tEPD_Segmentation}
\begin{tabular}{|c|c|c|c|c|}
\hline
Row & ${r}_{i}$ (cm) & ${r}_{f}$ (cm) & ${\eta}_{i}$ & ${\eta}_{f}$ \\
\hline
1   & 4.5            & 8.5            & 5.10         & 4.48         \\
\hline
2   & 8.5            & 12.9           & 4.48         & 4.06         \\
\hline
3   & 12.9           & 17.3           & 4.06         & 3.77         \\
\hline
4   & 17.3           & 22.83          & 3.77         & 3.49         \\
\hline
5   & 22.83          & 28.36          & 3.49         & 3.28         \\
\hline
6   & 28.36          & 33.89          & 3.28         & 3.10         \\
\hline
7   & 33.89          & 39.42          & 3.10         & 2.95         \\
\hline
8   & 39.42          & 44.95          & 2.95         & 2.82         \\
\hline
9   & 44.95          & 50.48          & 2.82         & 2.70         \\
\hline
10  & 50.48          & 56.01          & 2.70         & 2.60         \\
\hline
11  & 56.01          & 61.54          & 2.60         & 2.51         \\
\hline
12  & 61.54          & 67.07          & 2.51         & 2.42         \\
\hline
13  & 67.07          & 72.6           & 2.42         & 2.34         \\
\hline
14  & 72.6           & 78.13          & 2.34         & 2.27         \\
\hline
15  & 78.13          & 83.66          & 2.27         & 2.21         \\
\hline
16  & 83.66          & 90             & 2.21         & 2.13      \\  
\hline
\end{tabular}
\end{table}

\subsection{Centrality Resolution}
Figure \ref{fEP_res1} shows the hit count versus the impact parameter, which drives the centrality resolution of the detector.  For this plot, it was assumed that multiple hits could not be distinguished, which decreases the centrality resolution.  This is discussed extensively in Section \ref{sec_sim}.  The prototype data has shown us that while it is not possible to distinguish individual hits, we can use ADC weighting to better determine the centrality.

\subsection{Event Plane Resolution}

\begin{figure}[ht]
\centering
\mbox{\includegraphics[width=0.45\textwidth]{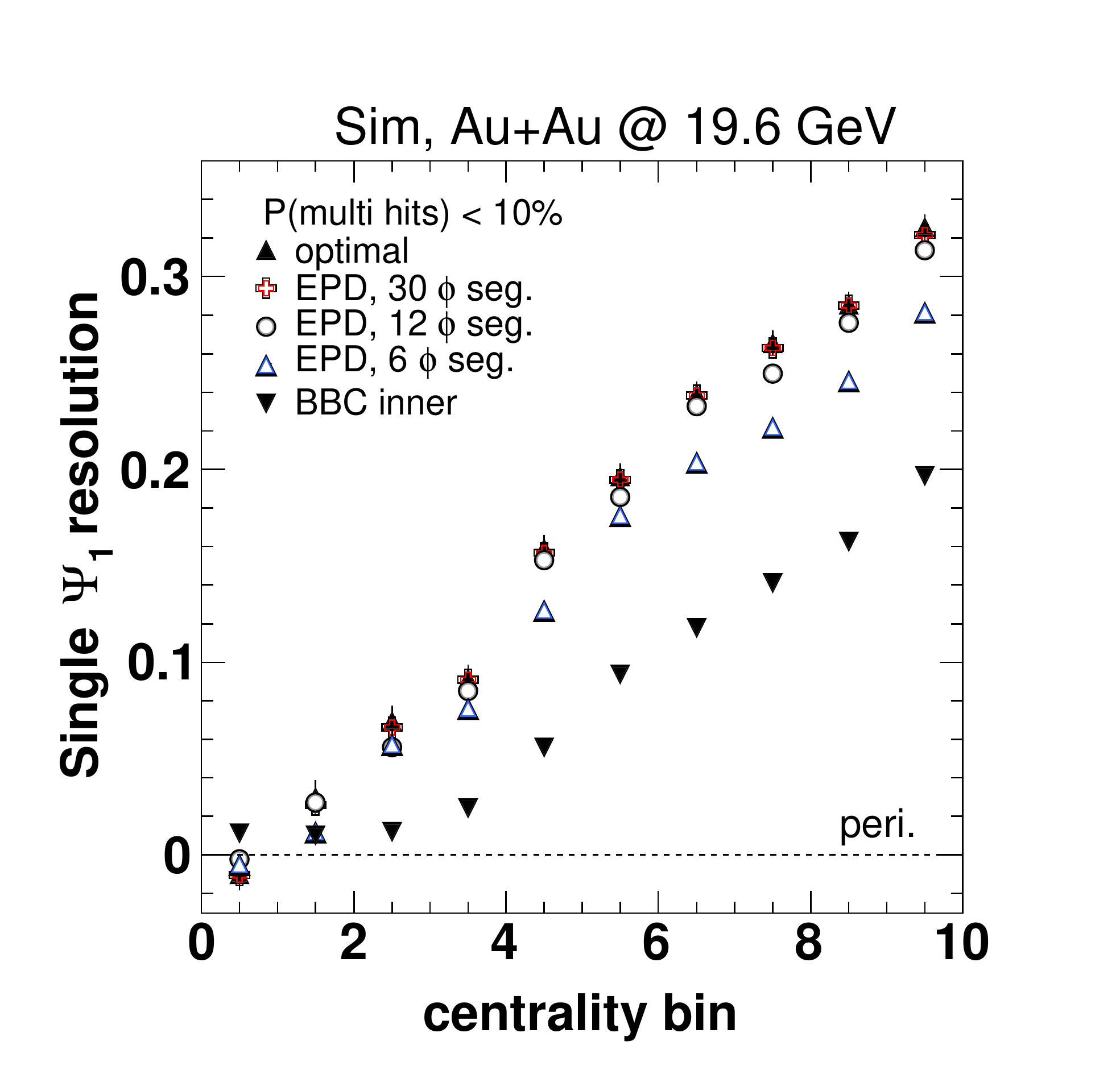}}
\mbox{\includegraphics[width=0.45\textwidth]{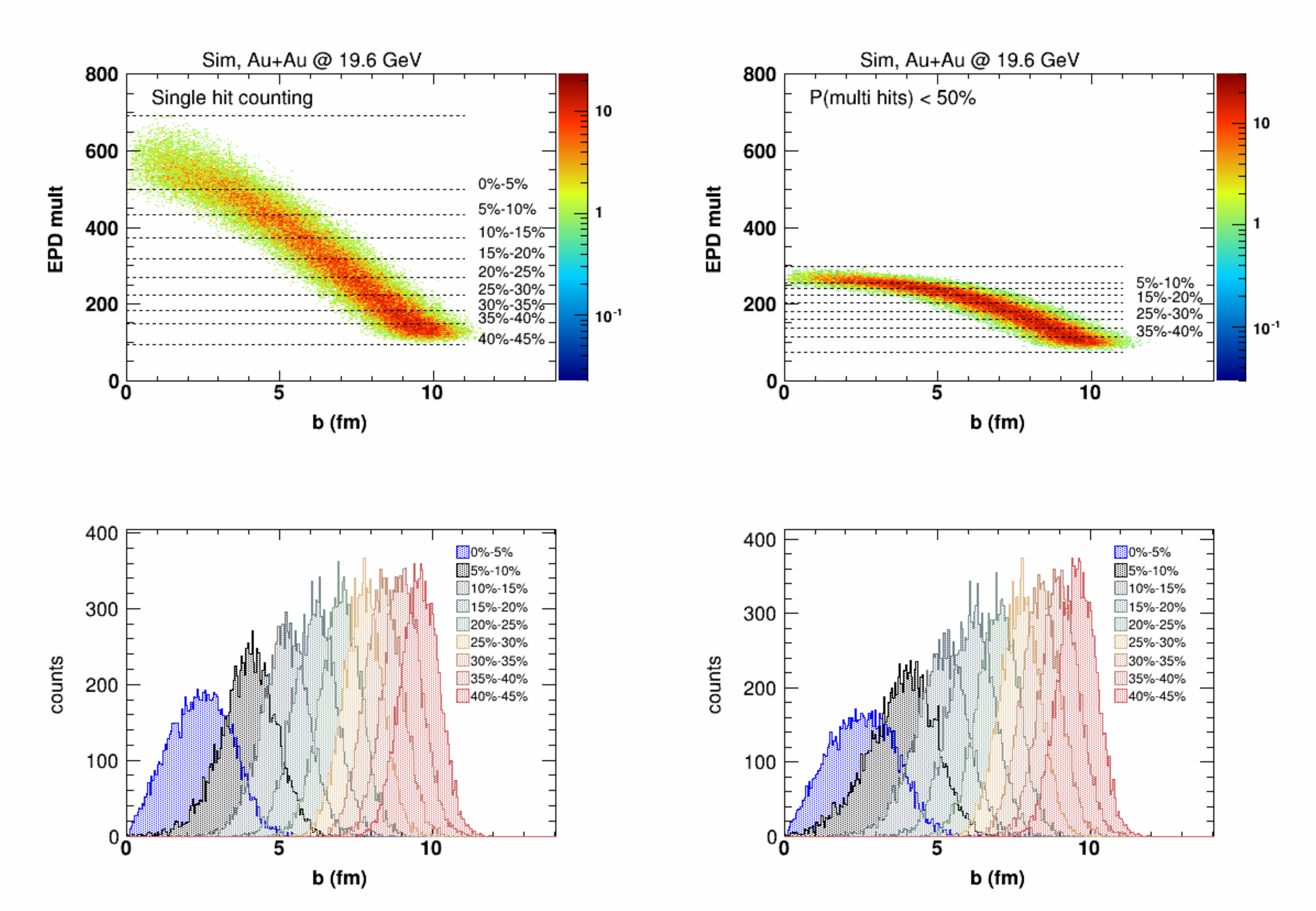}}
\caption{Left figure: $\Psi_{1}$ (single plane) event plane resolution as a function of centrality for different detector setups. For this simulation every single hit was counted without any ADC implementation. Most central events are on the left, the most peripheral bin shown corresponds to 40\%-45\%. The resolution for the EPD setup with 30 azimuthal segments coincides with the optimal resolution.  On the right is the multiplicity in the EPD acceptance as a function of the impact parameter b for multiple hits.}
\label{fEP_res1}
\end{figure}

 Figure~\ref{fEP_res1} shows the event plane resolutions for different detector setups as a function of the centrality bin.   The resolution for all configurations chosen is better than the BBC resolution.  Above 12 sectors the gain in the event plan resolution is minimal for the first order event plane.  However, as discussed in Section \ref{sec_sim}, a higher segmentation is needed for good higher order event plane resolution.  For the first through 5th order plane, the resolution starts to saturate around 20 $\phi$ segments, which brings us to our choice of 24.  
 
\subsection{Simulations}
\label{sec_sim}

\begin{figure}[ht]
\centering
\subfigure[Centrality: 0\%-3\%]{%
\resizebox{6.5cm}{!}{%
\includegraphics{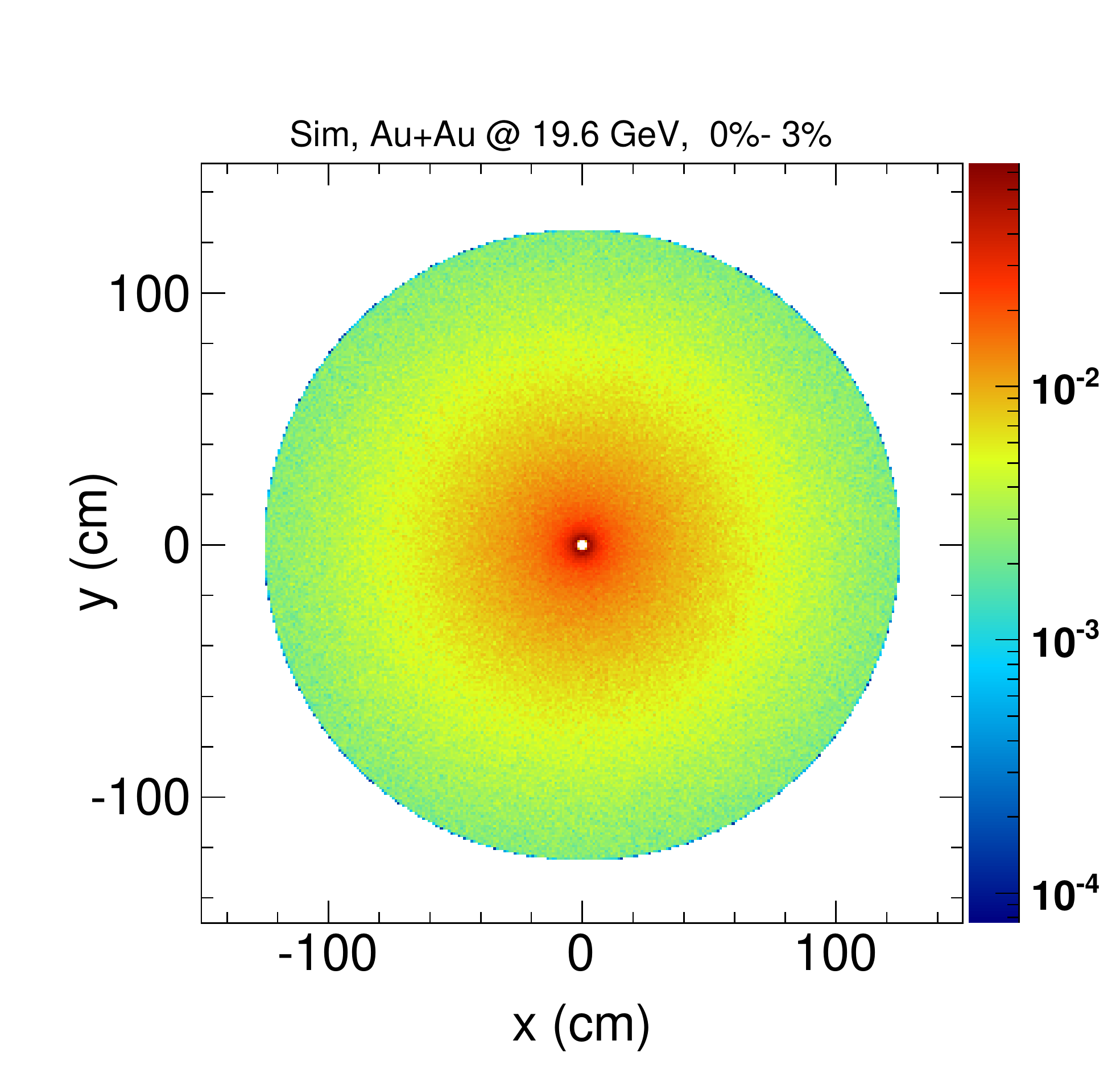}}
\label{fig:subfigure1}}
\quad
\subfigure[Centrality: 15\%-20\%]{%
\resizebox{6.5cm}{!}{%
\includegraphics{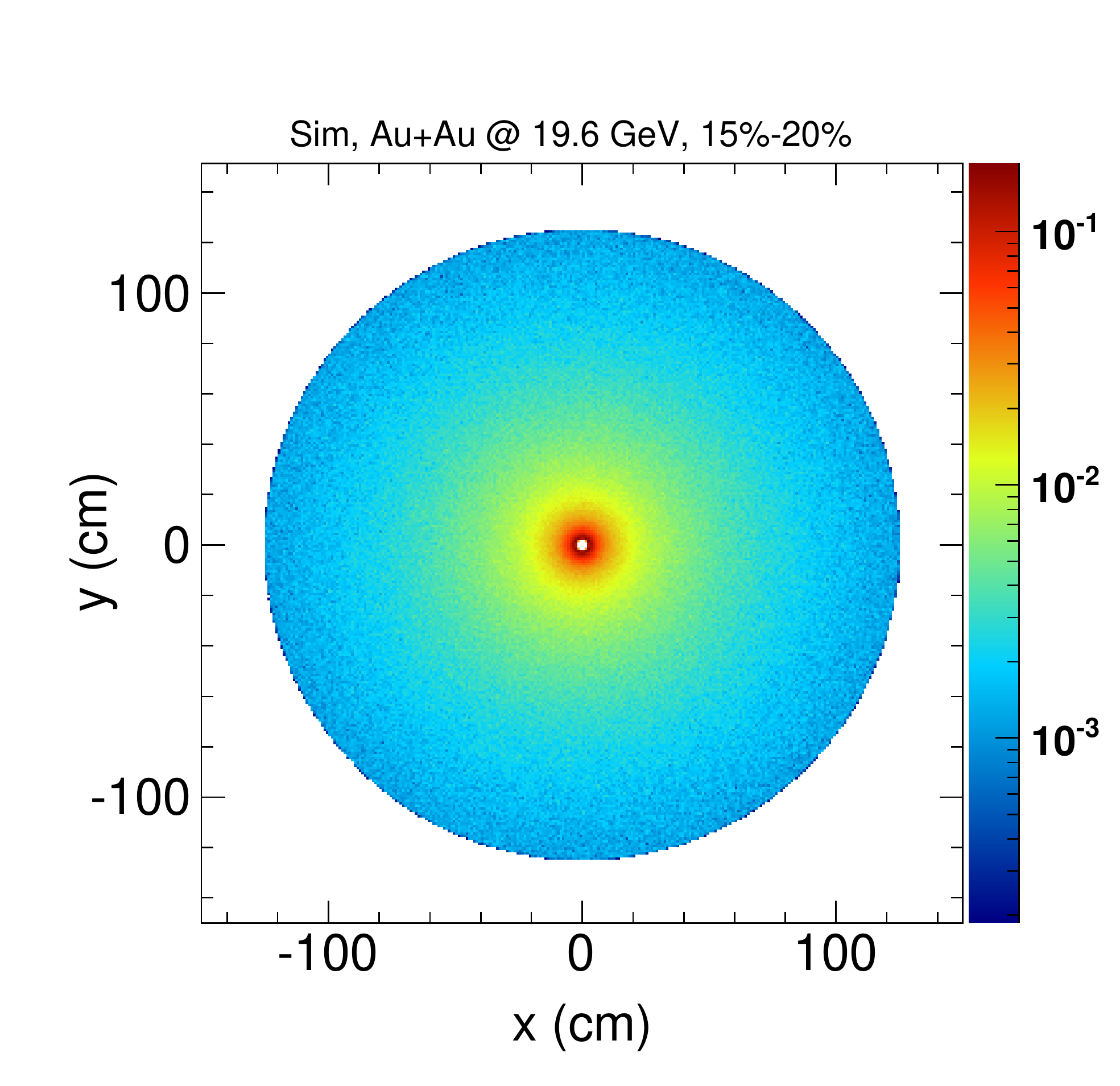}}
\label{fig:subfigure2}}

\subfigure[Centrality: 40\%-45\%]{%
\resizebox{6.5cm}{!}{%
\includegraphics{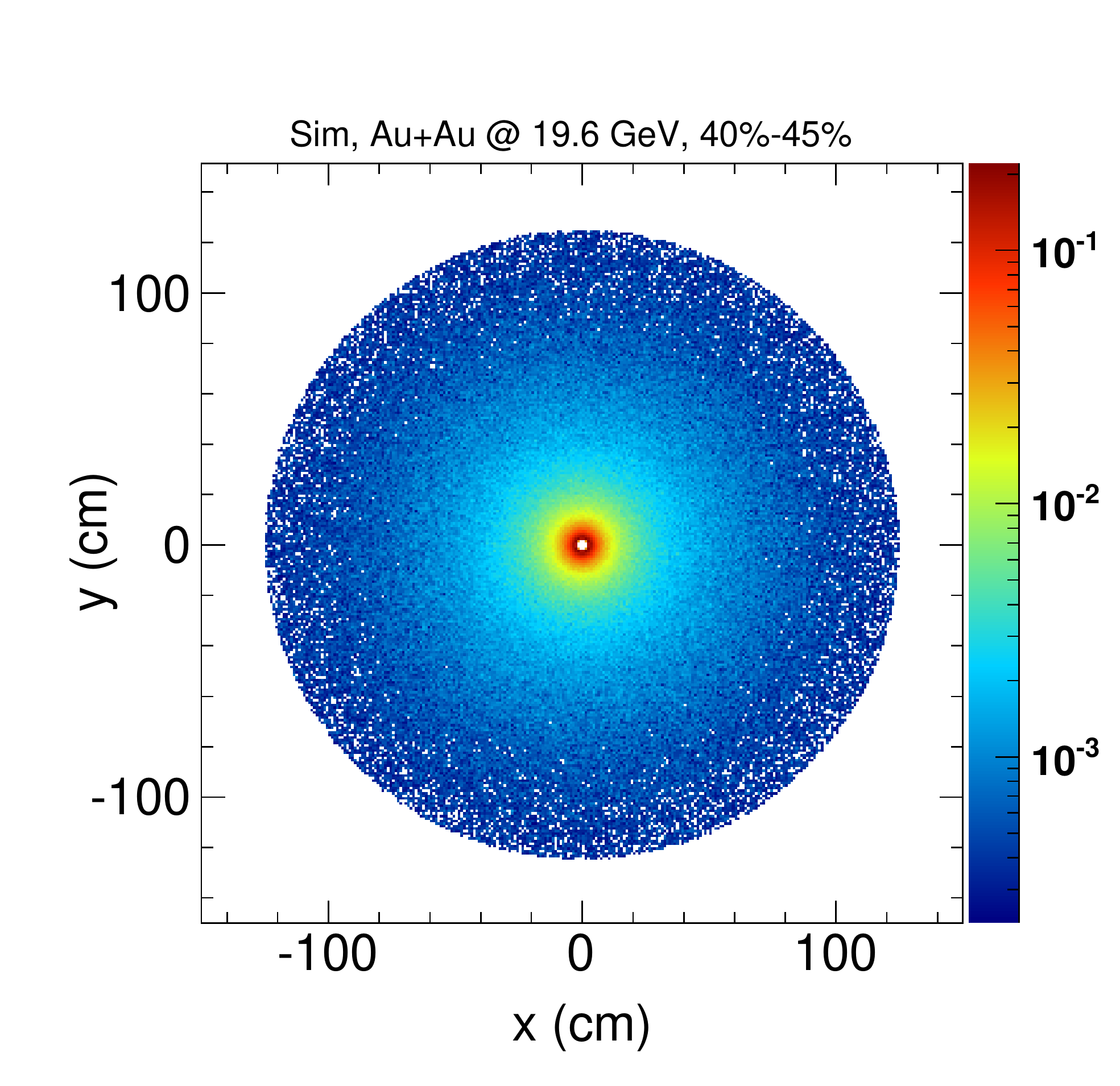}}
\label{fig:subfigure3}}
\quad
\subfigure[Centrality: 0\%-45\%]{%
\resizebox{6.5cm}{!}{%
\includegraphics{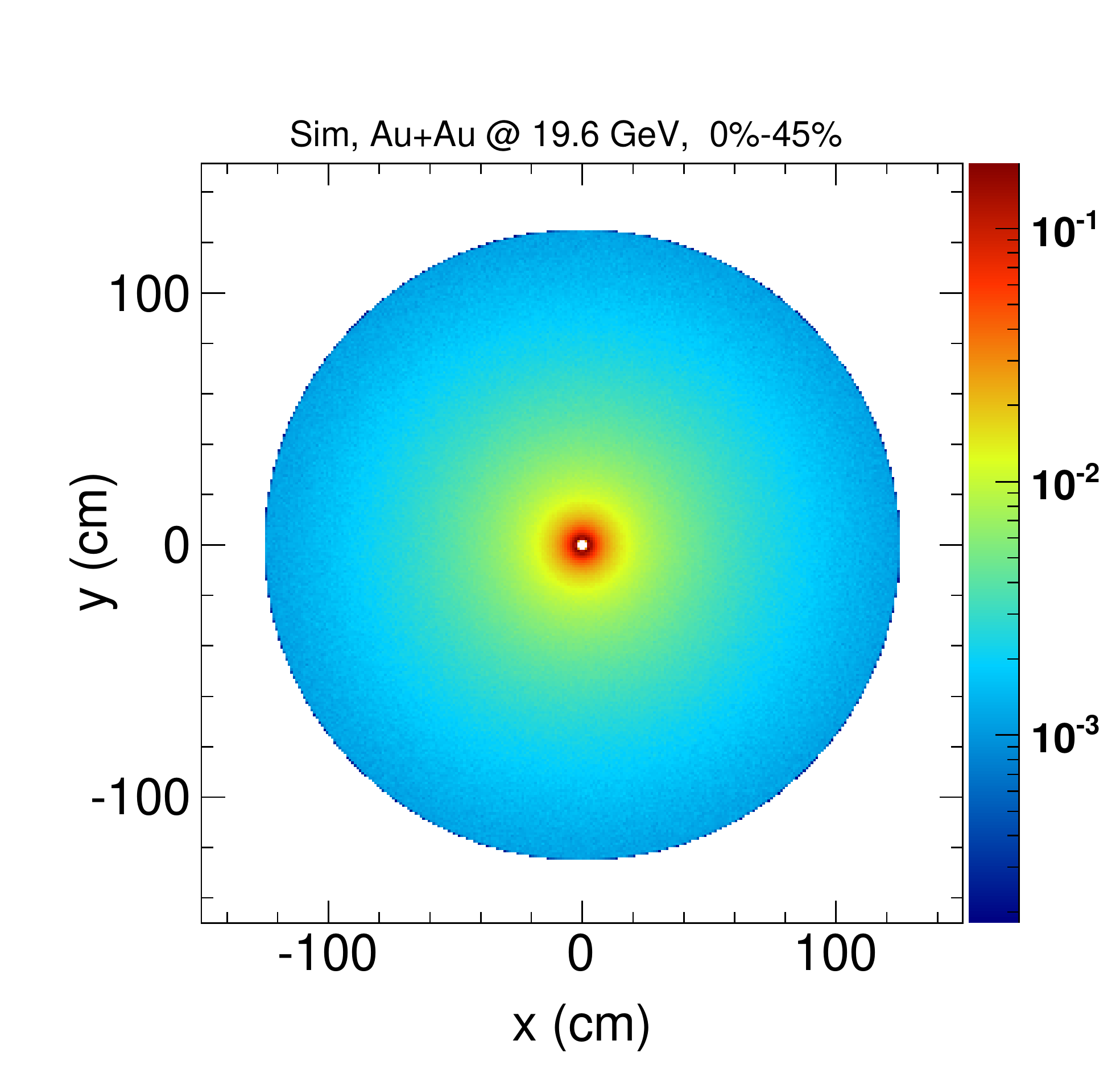}}
\label{fig:subfigure4}}
\caption{Hit densities for simulated Au+Au events at $\sqrt{s_{NN}}$ = 19.6 GeV events at z = 375 cm for different centralities.}
\label{fHit_density_2D}
\end{figure}

We performed a series of Monte Carlo (MC) simulations to optimize the geometry of the proposed detector, based on the physics requirements. To minimize the overall size of the detector for a given acceptance, it would be good to place it as close as possible to the center of the TPC. The proposed detector will replace the BBC~\cite{STAR_BBC}, which is currently located at z = $\pm$375 cm. This position in z was also used for the simulations, since the available space in the forward direction even closer to the TPC is limited. The simulated detector had an inner radius of 4 cm and an outer radius of 125 cm in the following event plane resolution studies.

The MC simulation input is based on PHOBOS $dN/d\eta$~\cite{Alver:2010ck} and both PHOBOS and STAR $v_{1}$~\cite{Yadav_v1} measurements. We first sample a number of tracks based on STAR reference multiplicity distributions. Those are scaled to the PHOBOS $dN/d\eta$ distributions in the STAR acceptance. For the measured PHOBOS centralities from 0\%-45\% we sample the $\eta$ values for each track. The $p_{T}$ values are sampled from a Boltzmann distribution, which are adjusted to the mid-rapidity slopes from STAR. The directed flow ($v_{1}$) for each track was assumed to scale linearly with pseudo-rapidity and the overall scale was also adjusted to measured data from STAR. We also included an elliptic flow $v_{2}$ component based on published STAR data from the BES~\cite{Adamczyk:2013gw}. A 5\% Gaussian smearing for $v_{1}$ and $v_{2}$ was applied to account for fluctuations. The relative angle $\phi - \Psi$ between a particle and a randomly oriented event plane was finally sampled from the following distribution:

\begin{equation}
\frac{dN}{d(\phi-\Psi)} \sim 1 + 2v_{1}\cos(\phi-\Psi) + 2v_{2}\cos(2\phi-2\Psi).
\end{equation}

\begin{figure}[htbp]
  \begin{center}
\mbox{\includegraphics[width=0.45\textwidth]{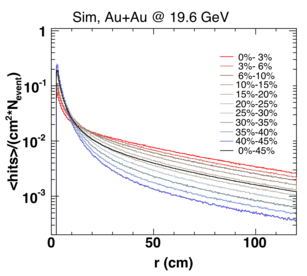}}
\mbox{\includegraphics[width=0.45\textwidth]{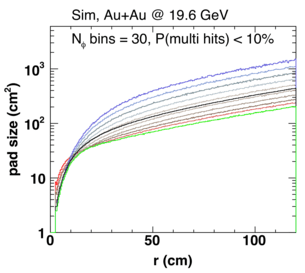}}
  \end{center}
  \caption{Left panel: Charged particle hit density from simulated Au+Au events at $\sqrt{s_{NN}}$ = 19.6 GeV events at z = 375 cm as a function of the radius to the beam axis for various centralities. Right panel: Maximum pad size as a function of the radius to the beam axis for a multi-hit probability of $\leq$ 10\% and 30 segments in the azimuthal direction.}
\label{fhit_density}       
\end{figure}

A total of 1M events were simulated for $\sqrt{s_{NN}}$ = 19.6 at z = 0 cm. The simulated particles were tracked through the (full) STAR magnetic field. Simulations for the STAR forward tracker in the same acceptance have shown that multiple scattering is a negligible effect for the event plane reconstruction. The transverse hit density per cm$^{2}$ was calculated based on the intersection points of the tracks with the detector planes. The two dimensional hit densities for various centralities for $\sqrt{s_{NN}}$ = 19.6 GeV are shown in Fig.~\ref{fHit_density_2D}. In the left panel of Fig.~\ref{fhit_density}, we show the hit densities as a function of the radius. An interesting and important feature of the distributions is that the hit density is higher for peripheral events at small radii compared to central events. The pattern switches with increasing radius due to the changed ratio of produced particles to sheared off spectators.

Based on the azimuthal symmetry of the system, we used as a starting point a pie sliced detector layout. Such a layout is optimal to select different pseudo-rapidity regions. The geometry is defined by a number of equally sized azimuthal segments and radial segments which can vary in $\Delta r$. For a given energy the size of the pads is fully determined for any radius by choosing a number of azimuthal segments and a maximum multi-hit probability per pad. With those two parameters and the known hit density distribution, one can calculate the optimal pad size as a function of the radius. An example is shown for various centralities for $\sqrt{s_{NN}}$ = 19.6 GeV in the lower panel of Fig.~\ref{fhit_density}. 


\begin{figure}[htbp]
  \begin{center}
\mbox{\includegraphics[width=0.55\textwidth]{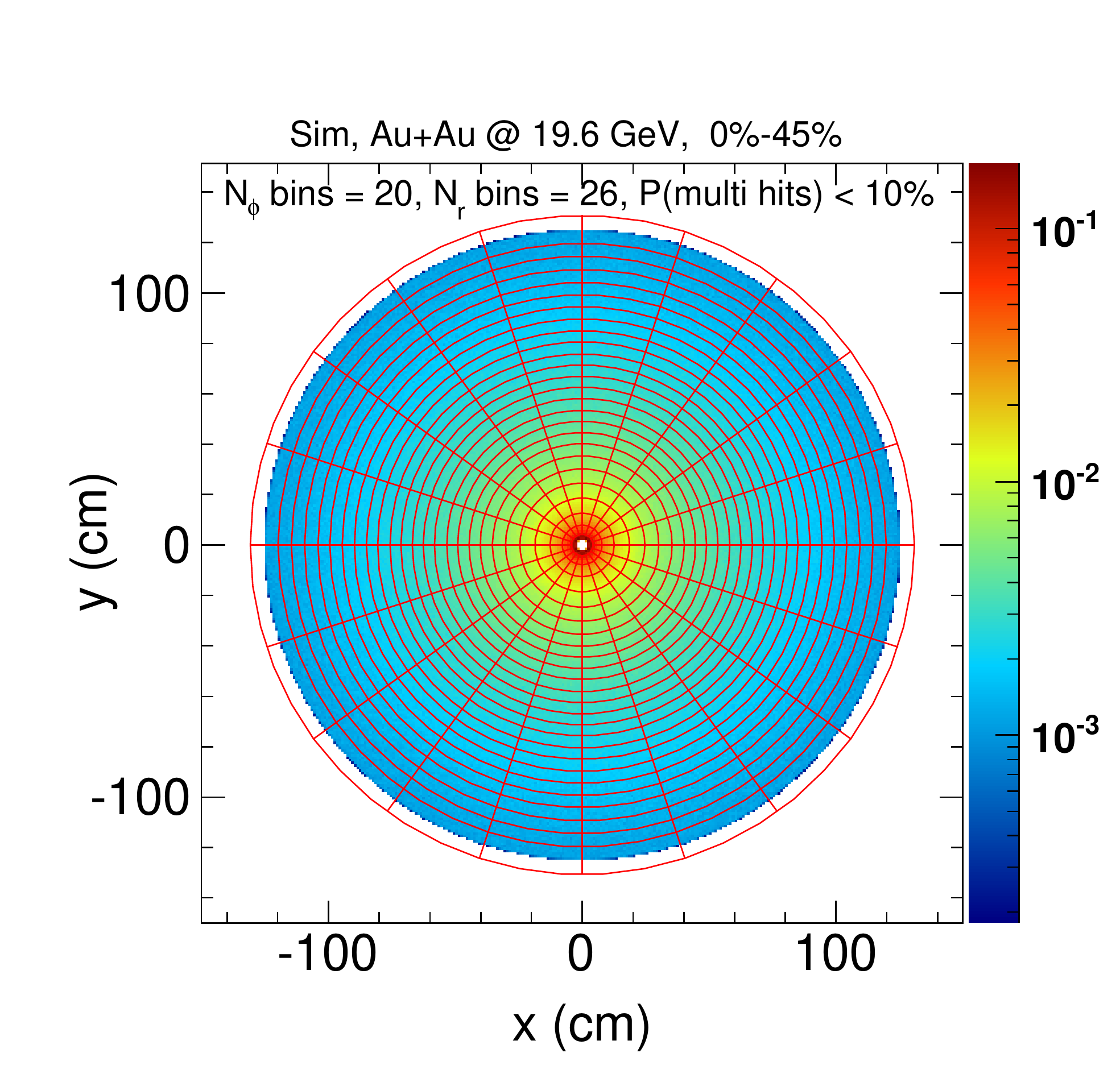}}
  \end{center}
 \caption{Detector setup with 20 azimuthal segments and for a multi-hit probability $\leq$ 10\%.}
\label{fdet_0_45_phi_20}       
\end{figure}

The minimum over all curves in Fig.~\ref{fhit_density} defines the optimal pad size for any centrality, as shown by the green curve. A possible geometry based on the calculations is shown in Fig.~\ref{fdet_0_45_phi_20} with 20 radial segments and a multi-hit probability $\leq$ 10\%. This kind of setup would result in about 500 tiles per detector plane. For the final detector we will minimize the number of different tile geometries.

\begin{figure}[htbp]
\begin{center}
{
\mbox{\includegraphics[width=0.48\textwidth]{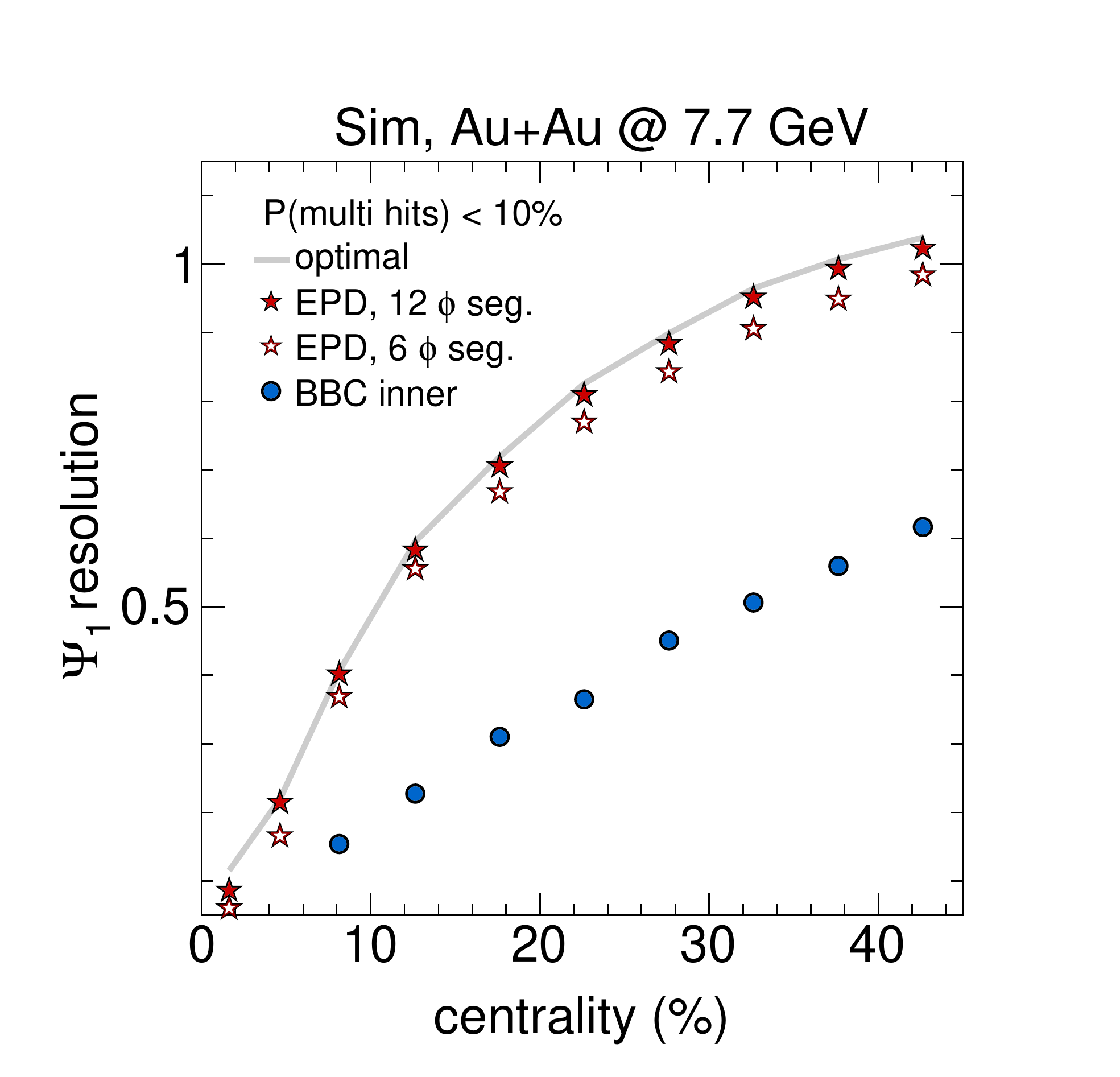}}
\mbox{\includegraphics[width=0.48\textwidth]{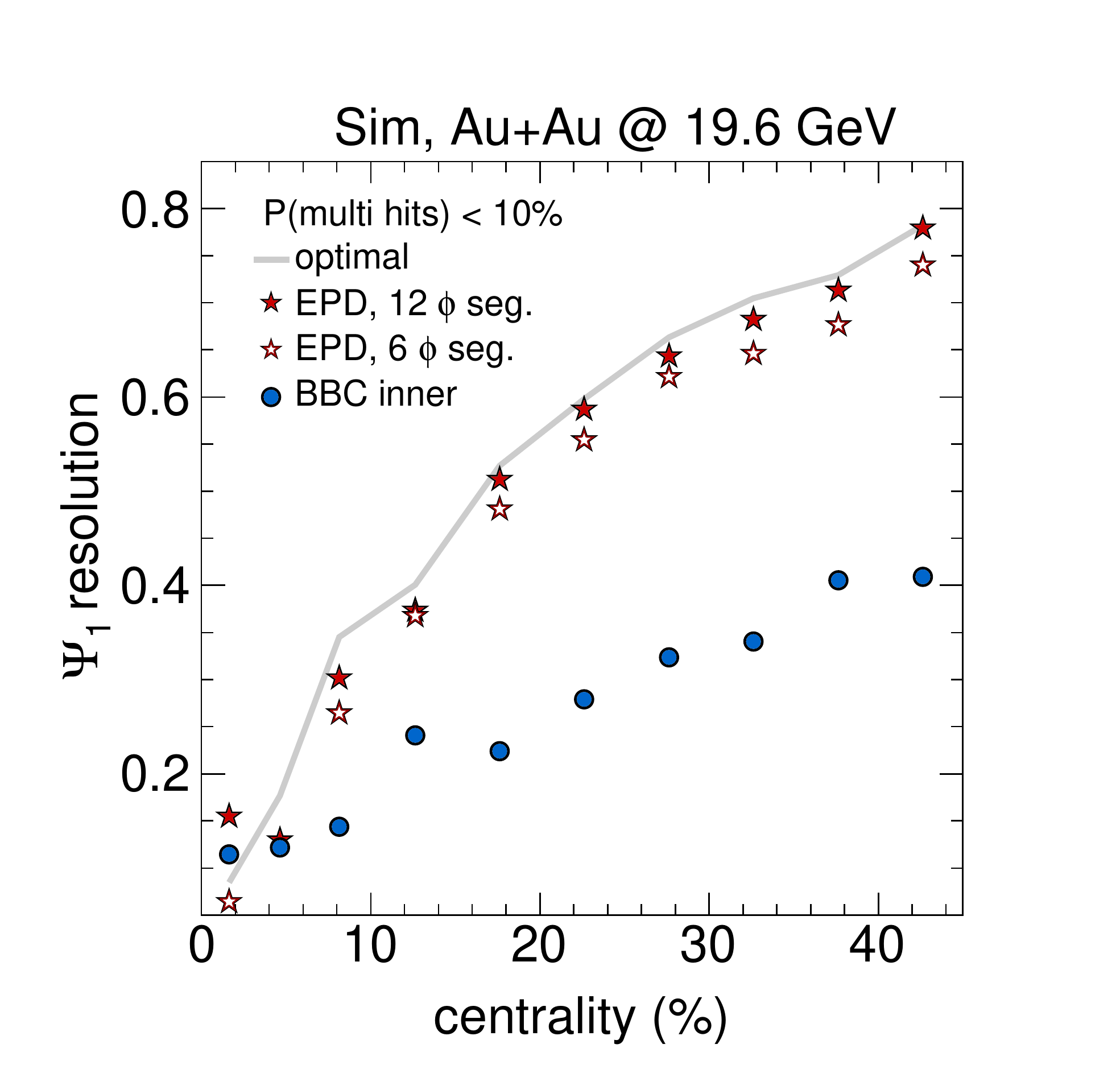}}
\mbox{\includegraphics[width=0.48\textwidth]{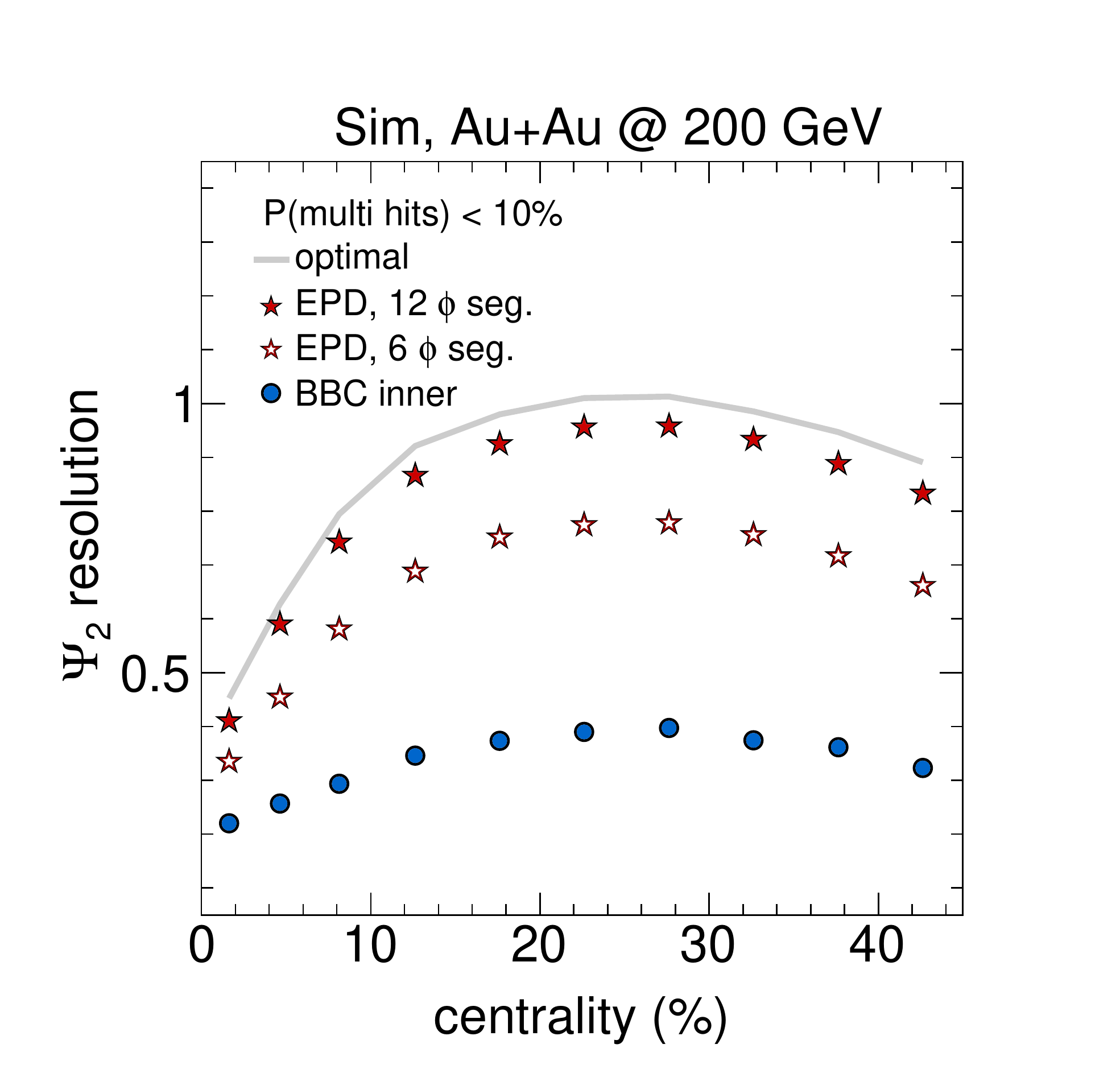}}
}
\end{center}
\caption{First harmonic ($\Psi_{1}$) event plane resolution (both detector planes) as a function of centrality for different detector setups. Each hit in this simulation was smeared based on realistic ADC distributions.
 Most central events are on the left, the most peripheral bin shown corresponds to 40\%-45\%. The plot on the top left is from $\sqrt{{s}_{NN}} = 7.7$ GeV and the plot on the top right is from $\sqrt{{s}_{NN}} = 19.6$ GeV.  The lower plot shows the event plane resolution for the second harmonic ($\Psi_{2}$). The BBC points are in blue, and we can see that for all centralities and energies the EPD improves the event plane resolution.
}
\label{fEP_res}
\end{figure}

We evaluated the event plane and centrality resolution for various detector geometries. The optimal pad sizes were calculated for 6, 8, 10, 12, 20, and 30 $\phi$ segments  and for multi-hit probabilities per pad smaller than 10\%, 20\%, 30\%, 40\%, or 50\%. As references we use the optimal event plane resolution within the detector acceptance and the BBC (inner tiles) event plane resolution. Each hit on the detector cells was smeared based on measured ADC distributions.  Figure~\ref{fEP_res} shows the event plane resolutions for different detector setups as a function of the centrality bin. 


There is a significant difference ($\sim$20\%) in the EP resolution between the EPD detector layout with 6 and 12 azimuthal segments, whereas more than 12 azimuthal segments do not contribute much more to the EP resolution. For 30 azimuthal segments we reach the optimal resolution. The $r$-segmentation has a much smaller impact on the EP resolution. The improvement compared to an optimal (see above) inner BBC setup is up to a factor 5 for the most central events and still 60\% for the centrality bin 40\%-45\% . The corresponding improvement for an elliptic flow analysis using the first harmonic event plane would be even larger.

\begin{figure*}[]
\centering
\resizebox{15cm}{!}{%
\includegraphics{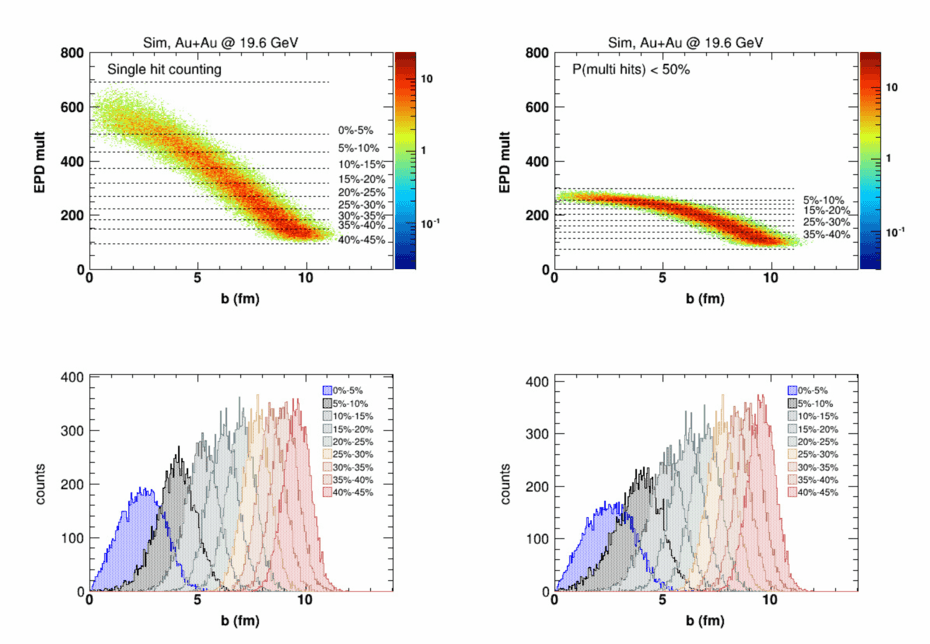}}
\caption{Upper panels: Multiplicity in the EPD acceptance as a function of the impact parameter b for single hit counting (left) and a multi-hit probability per detector tile of 50\% (right). Lower panels: Projections to the impact parameter axis for different centrality selections.}
\label{fmuls_vs_b}       
\end{figure*}

\begin{figure}[htbp]
  \begin{center}
   \includegraphics[width=0.5\textwidth]{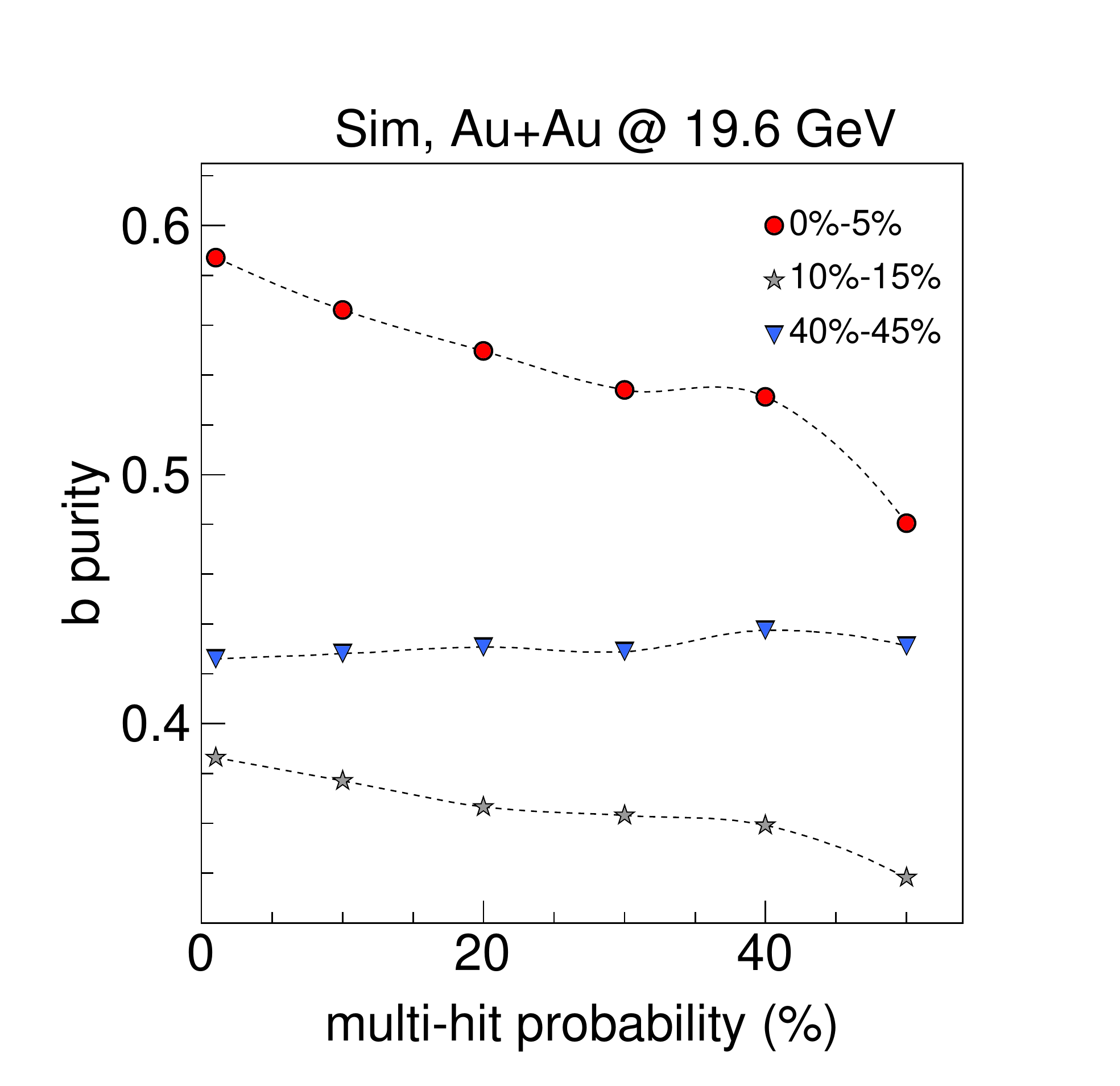}
     \end{center}
\caption{Impact parameter purity for simulated Au+Au events at $\sqrt{s_{NN}}$ = 19.6 GeV as a function of the multi-hit probability per detector tile, for three different centrality selections.}
\label{fb_purity}       
\end{figure}

For our centrality studies, we used Glauber calculations, based on measured STAR data at $\sqrt{s_{NN}}$ = 19.6 GeV, to get the correlation between the number of produced charged particles and the impact parameter b. Figure~\ref{fmuls_vs_b} shows the correlation for single hit counting (left) and for a multi-hit probability per detector tile of 50\% (right). In this calculation it was assumed that multiple hits per tile cannot be distinguished. The projections for different centrality selections to the impact parameter axis are shown in the lower plots. A clear saturation/flattening effect is observed for increased multi-hit probabilities (larger tile sizes). Based on the $b$-projections we calculated the $b$-purity (90\% confidence interval) for different centrality selections as a function of the multi-hit probability, as shown in Fig.~\ref{fb_purity}. The purity for the most central collisions significantly drops with increased mulit-hit probability, whereas the purity is almost constant for peripheral centrality selections due to the lower saturation probability. For multi-hit probabilities $\leq$ 10\% we achieve almost the optimal $b$-purity. 

It is also obvious that a limited granularity will be more sensitive to any kind of multiplicity fluctuations in the saturation region of high multiplicity events. We further want to point out that a significant amount of sheared off spectator particles mixes with the produced particles in the forward region at lower energies. It is unclear how this will affect the determination of the centrality, but we think it will be crucial to have the capability to distinguish different $\eta$ regions. Therefore, a large number of radial segments will be important. This would be another advantage compared to the BBC detector.

\subsubsection{UrQMD simulation}

\begin{figure}[htbp]
\begin{center}
\mbox{\includegraphics[width=0.48\textwidth]{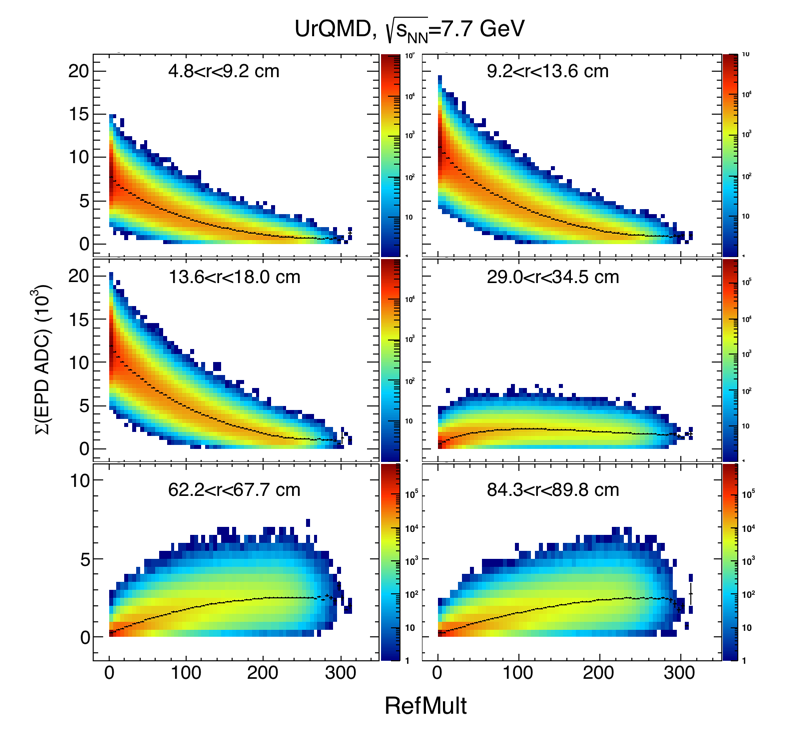}}
\mbox{\includegraphics[width=0.48\textwidth]{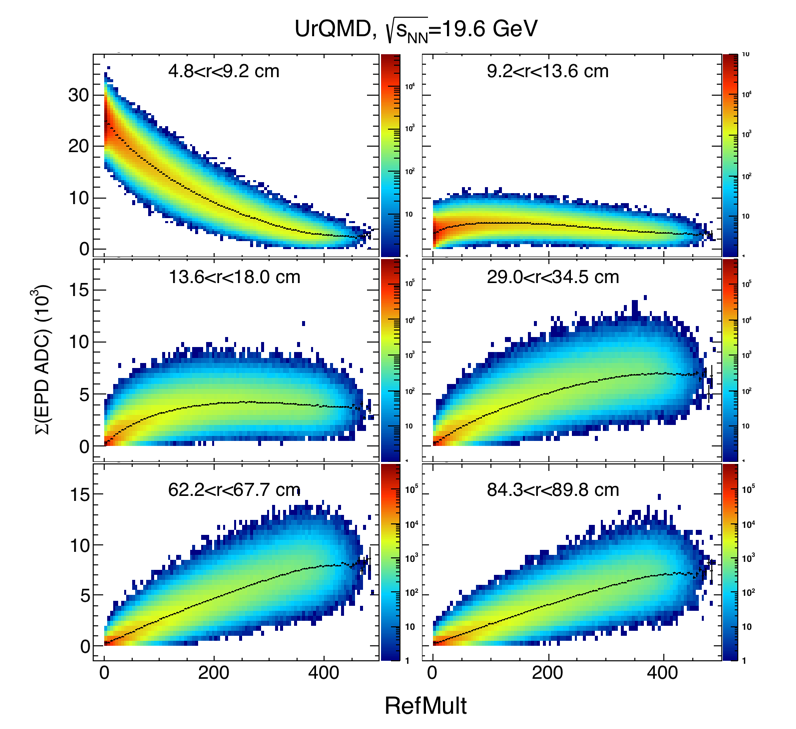}}
\end{center}
\caption{On the left is the sum of the ADC counts in six radial slices (out of 16) for $\sqrt{{s}_{NN}}$ = 7.7 GeV versus the reference multiplicity (RefMult) in the TPC.  On the right is the same distribution for 19.6 GeV.
}
\label{fig:UrQMDvEvR}
\end{figure}

In order to validate the simulations above with respect to the centrality resolution determined from the PHOBOS data above, collisions at two different energies ($\sqrt{{s}_{NN}} = 7.7 \textrm{~GeV~and~}\sqrt{{s}_{NN}} = 19.6$ GeV)  were simulated using UrQMD.  The STAR TPC multiplicity was simulated by summing the charged particle count of those particles with $|\eta|<0.5$ (RefMult) from a minimum-bias (MB) distribution.  The EPD has pseudo rapidity coverage of $\pm 2.1 < \eta < 5.1$, broken into 16 segments.  The particle counts were translated into ADC values using the parameterization of the prototype data that is discussed in Section \ref{sec_proto_data}.  The resulting parameterization of the ADC distribution was sampled for each hit, assuming that all charged particles leave a MIP in the EPD.  The total ADC count for a given radial segment was determined by summing the ADC values from all the hits within that segment, essentially convoluting the hit distribution with the data-driven ADC parameterization.  The correlation between the EPD ADC count versus the TPC RefMult for 6 different pseudo rapidity windows (out of the possible 16) is shown in Figure \ref{fig:UrQMDvEvR}.  As noted above, at low values of $r$, the distributions of the forward multiplicity, represented by the sum of the ADC, and the mid-rapidity multiplicity are anti-correlated.  However, the precise distributions and the value of $r$ where they switch to a correlated measure depend on the energy dependence.  This necessitates a small $\eta$ segmentation in order to have a similar centrality resolution at the different energies that will be recorded for BESII.

\begin{figure}[htbp]
\begin{center}
{
\mbox{\includegraphics[width=0.6\textwidth]{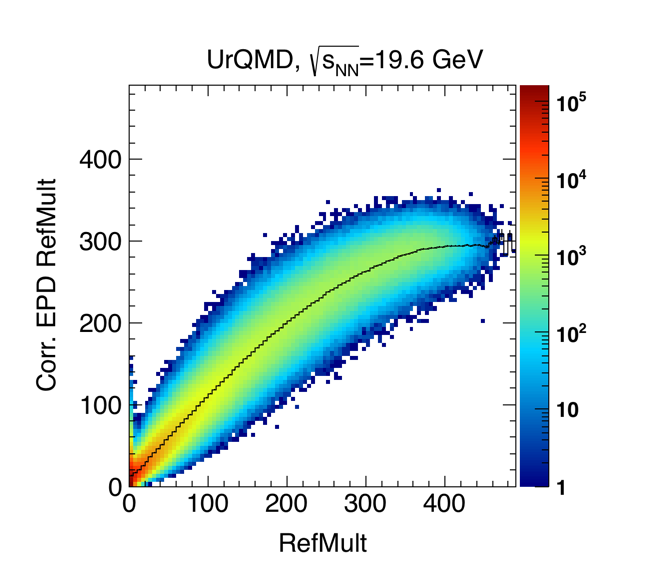}}
}
\end{center}
\caption{The corrected EPD RefMult versus the TPC RefMult for simulated UrQMD events at $\sqrt{{s}_{NN}} = 19.6$ GeV.
}
\label{fig:UrQMDEPDmultCorr}
\end{figure}

The plots in Figure \ref{fig:UrQMDvEvR} were used to combine the information into one "corrected" EPD multiplicity distribution.  This procedure would give us simple data driven way to determine the event centrality from the EPD ADC signals.  The corrected multiplicity is calculated event-by-event.  For each of the 16 pseudo rapidity bins, the corresponding 2D distribution is projected to the TPC RefMult at the calculated EPD ADC. The projected histogram, which is close to Gaussian, that results is sampled to get a statistically valid TPC RefMult value. Since some or the distributions show a better (anti-)correlations than others, the RMS of the projected histogram is used as an inverse weight to account for this effect.  The values from each of the 16 bins are added linearly with their related weight.  The average for that event is calculated, which gives a corrected EPD refmult value.  The corrected EPD RefMult versus the TPC RefMult for $\sqrt{{s}_{NN}}$ = 19.6 GeV is plotted in Figure \ref{fig:UrQMDEPDmultCorr} which shows a nice and relatively narrow correlation between the values. 

\begin{figure}[htbp]
\begin{center}
{
\mbox{\includegraphics[width=0.40\textwidth]{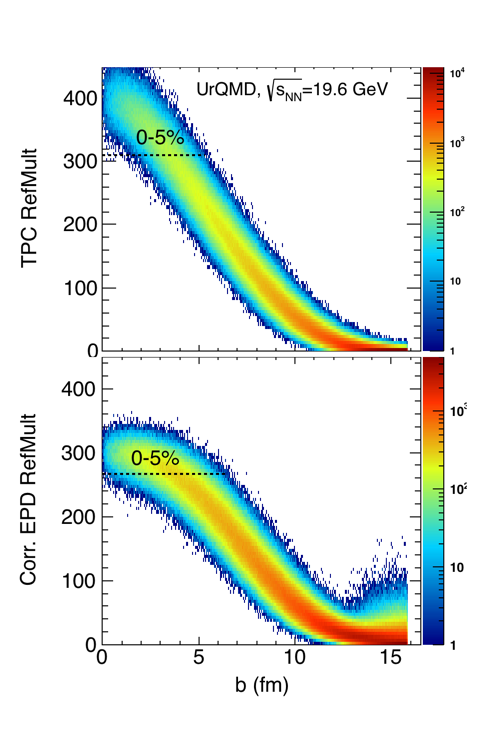}}
\mbox{\includegraphics[width=0.40\textwidth]{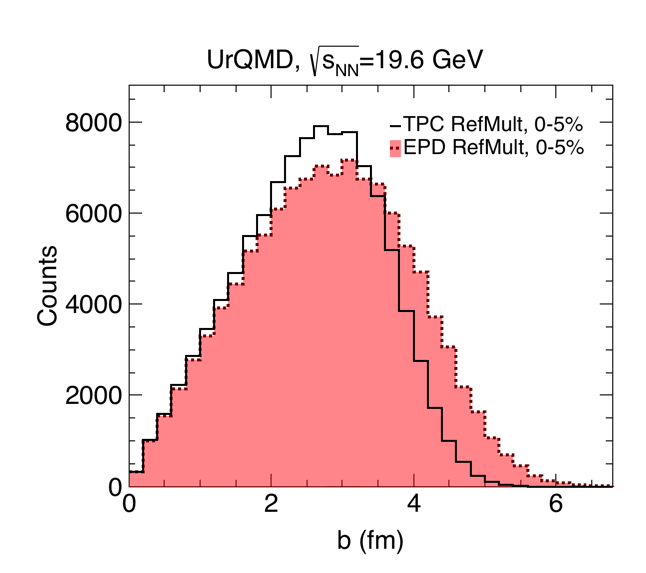}}
}
\end{center}
\caption{On the left are the distributions of multiplicity versus the impact parameter.  The top distribution is the TPC mid-rapidity multiplicity distribution.  The bottom is the EPD corrected multiplicity distribution.  The selection for the top 5\% of all events is shown by a dashed line in both cases.  On the right is the impact parameter distribution for these top 5\% of all events, with the EPD histogram in peach and the TPC histogram as an open frame.  The mean and sigma of the TPC distribution is 2.53 $\pm$ 0.99, and the mean and sigma of the EPD distribution is 2.79 $\pm$ 1.16, which is not significantly different.
}
\label{fig:UrQMDEPDmultCorr_centsel}
\end{figure}

The question is then how the resolution compares to the impact parameter distribution.  The resolution of the corrected multiplicity in the EPD is not very different from the  TPC, as shown in Figure \ref{fig:UrQMDEPDmultCorr_centsel}.  The most peripheral and most central events have some saturation, however the different between the impact parameter distributions after the top 5\% events are selected by multiplicity are nearly identical.  This is important as this means the centrality determination can be done entirely in the forward direction without any real loss in resolution, but also without the inclusion of autocorrelations.

\begin{figure}[ht]
\centering
\mbox{\includegraphics[width=0.45\textwidth]{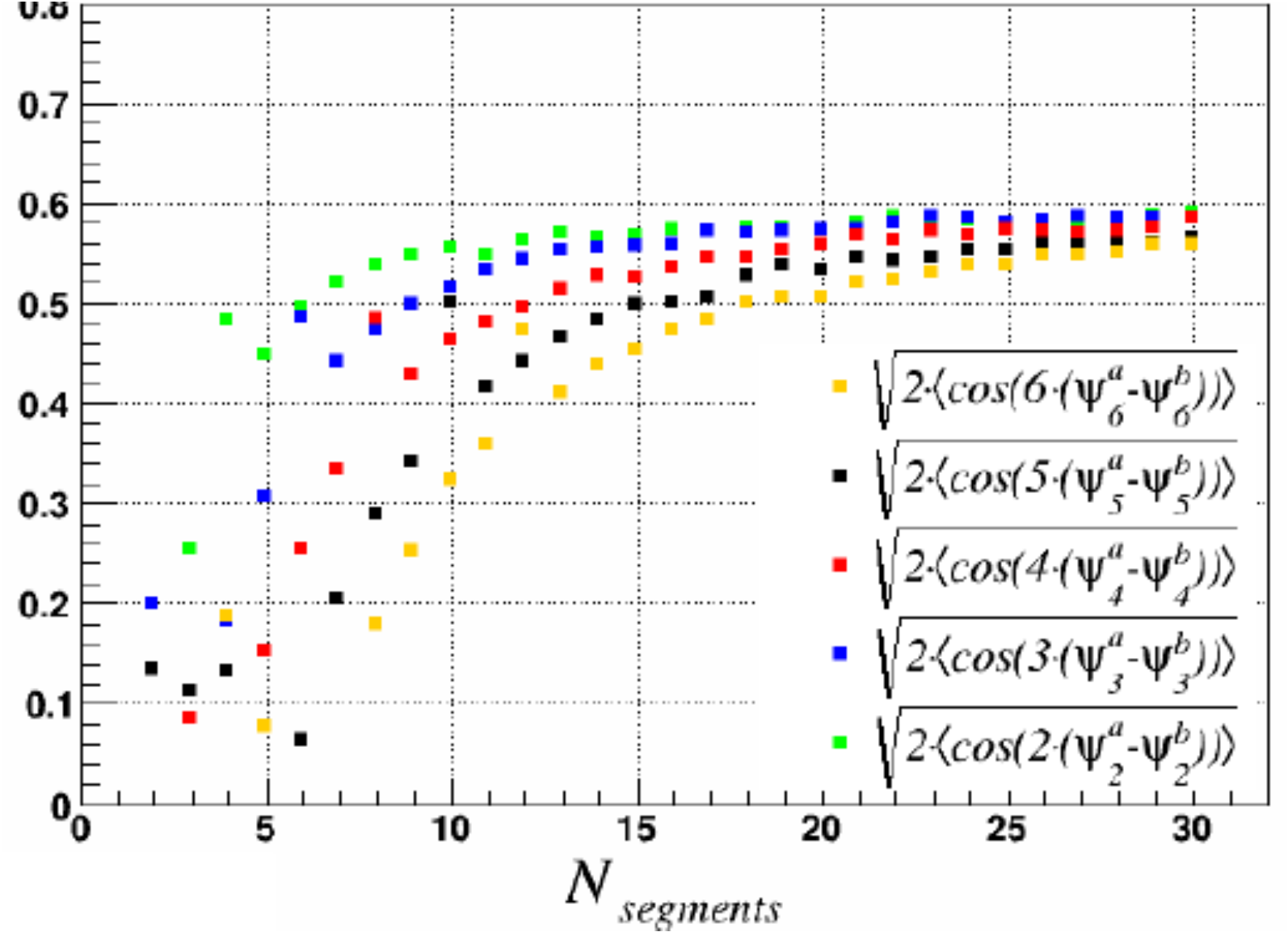}}
\caption{The event plane resolution for higher flow harmonics as a function of the azimuthal segmentation. At saturation takes place for harmonics $>$ 4 for larger than 20 azimuthal segments.
}
\label{fEP_higherorder}
\end{figure}
\begin{figure}[ht]
\centering
\mbox{\includegraphics[width=0.45\textwidth]{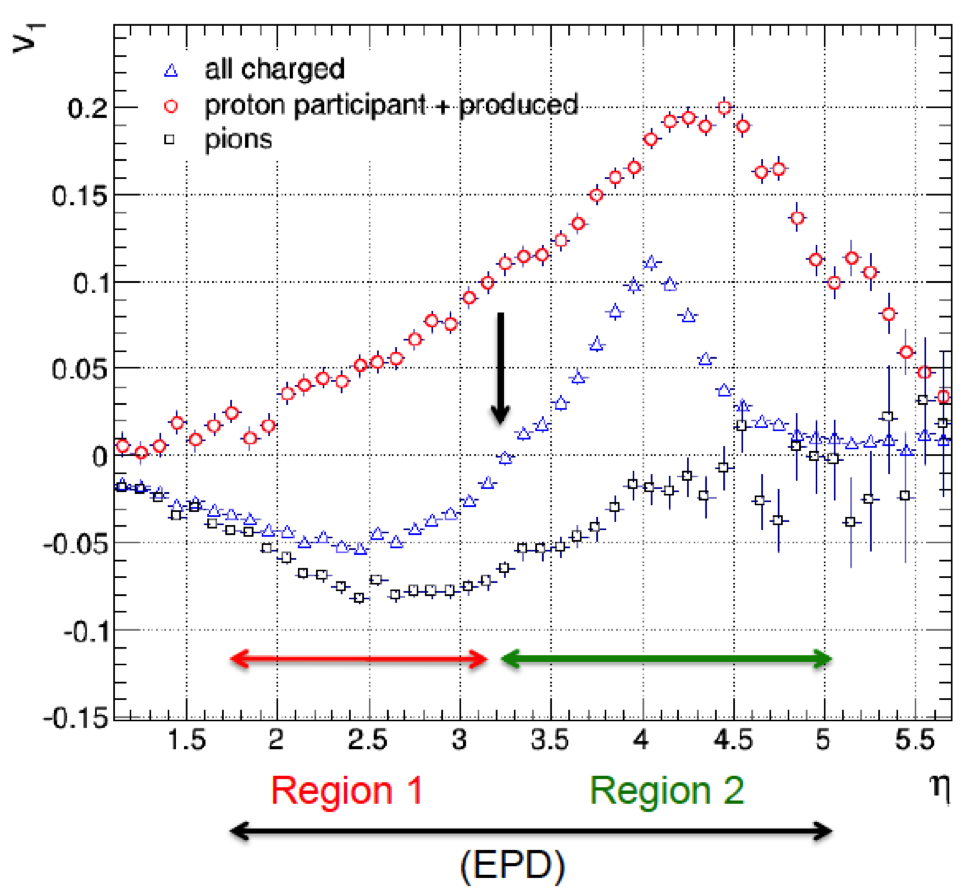}}
\mbox{\includegraphics[width=0.45\textwidth]{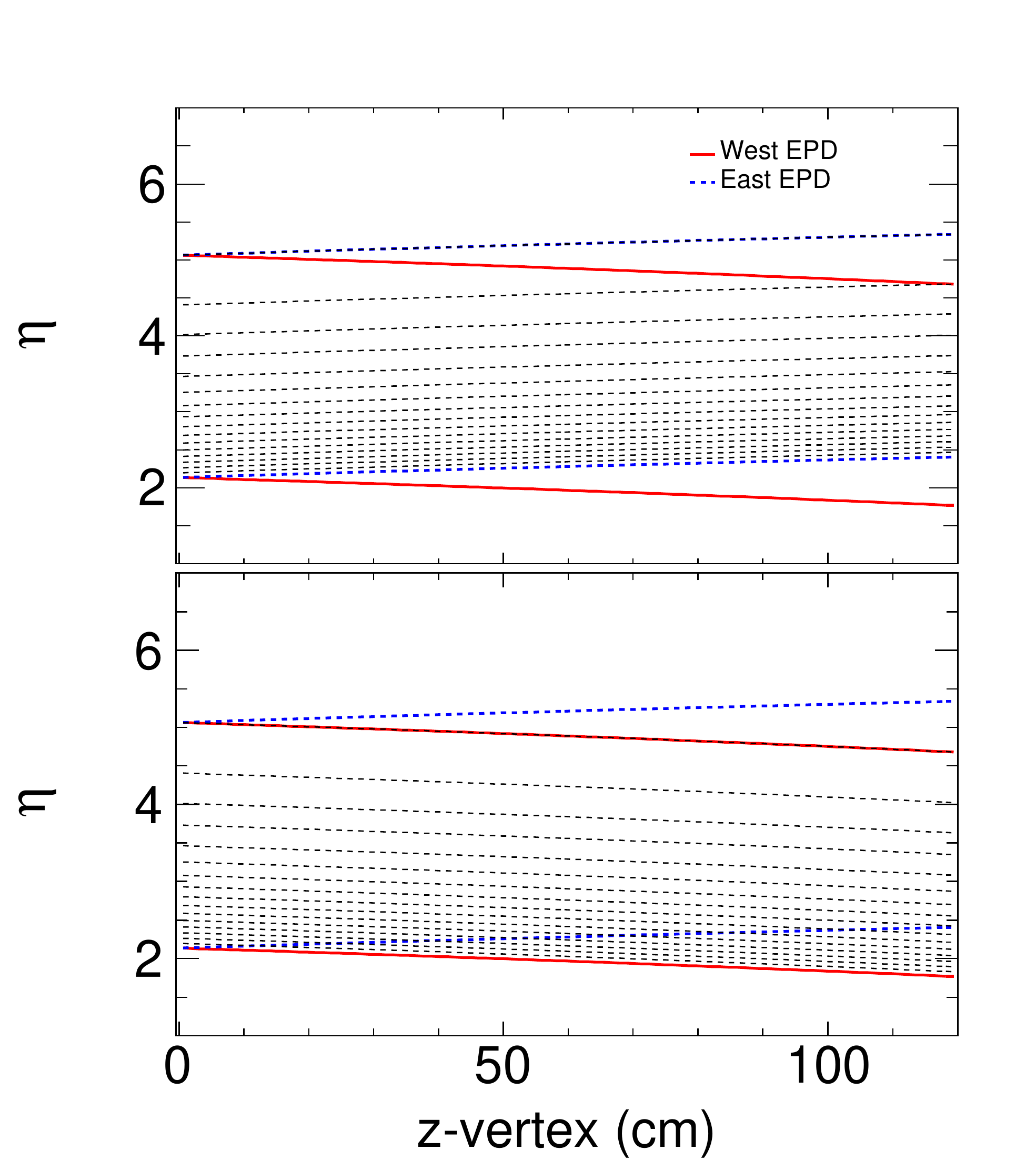}}
\caption{The top figure shows the v$_{1}$ signal from $\sqrt{s_{NN}}$ = 19.6 GeV UrQMD events as a function of $\eta$. The v$_{1}$ sign is changing for different particle species at different $\eta$.   The bottom figure shows the EPD acceptance in $\eta$ as a function of the z-vertex position with the dashed lines indicating the radial segments. It is split into upper/lower (EPD West segments/EPD East segments).}
\label{fEP_v1_eta}
\end{figure}

High azimuthal segmentation is needed for higher harmonics to improve the event plane resolution.  This will be important for many BES-II analyses.  As shown in Figure \ref{fEP_higherorder}, the number of segments needed for a good event plane resolution for ${v}_{5}$ or higher is at least 20.  High radial ($\eta$) segmentation helps keep the same acceptance for the East and West, regardless of where the collision vertex is located. Figure \ref{fEP_v1_eta} (bottom) shows how the $\eta$ acceptance is changing with the collisions z-vertex. Corrections for different z-vertex location can be made by having a large radial ($\eta$) segmentation. An application example is shown in the upper part of figure \ref{fEP_v1_eta}, which depicts the v$_{1}$ as a function of $\eta$ for $\sqrt{s_{NN}}$ = 19.6 GeV UrQMD events. Spectators and produced particles show varying v$_{1}$ signs as a function of $\eta$. By flipping the sign in different $\eta$ regions a better event plane resolution can be achieved. In addition to the z-vertex effect, the v$_{1}$ distributions are also changing significantly as a function of the collision energy, which makes it necessary to keep a large radial segmentation for adjustments.

\subsection{Trigger for BESII}
\label{sec_trigger}

The Beam-Beam Counter (BBC) had been used as the trigger detector for the first Beam Energy Scan.  The inner tiles of the BBC cover an area of $3.3 < \eta < 5.0$ and it has a timing resolution of 1 ns.  For the EPD to replace this BBC, it has to perform at least this well, and given the requirements of the BES-II program it will be to be an improvement.  The details of the trigger are discussed in Section \ref{sec_Triggering}, but in order to have high efficiency for peripheral events at least 12 radial segments are needed for triggering.  The EPD TACs need to be spread evenly in azimuth to reduce the systematic uncertainty.

\newpage

\section{EPD Design}
\label{sec_design}

\subsection{Prototype R\&D}
\label{sec_R_and_D}

We present here a comprehensive account of the operations conducted at LBNL concerning the R\&D of the proposed Event-Plane and Centrality Detector (EPD) at STAR. Briefly, the EPD serves to improve event-plane reconstruction for heavy-ion flow measurements and trigger detection for ``good'' events within a certain z-vertex range.

The design and goals of the EPD have evolved considerably over the course of the R\&D process, and each component of the final design, documented in what follows, was carefully chosen for optimal performance given various budget constraints. The fundamental characteristics of the EPD include radiation detection by plastic scintillating tiles and a redirection of the signal photons through wavelength shifting fibers. The signal is then coupled to clear optical fibers which are subsequently fed into Silicon Photomultipliers (SiPM) for signal amplification. The major design topics that were researched include optimizing the tile geometry, the fiber groove geometry, fiber coupling techniqes, and fiber polishing techniques. Throughout the research process, 3D-printing technology was utilized to quickly experiment with custom connector designs to be printed on-site at LBNL. \\

At the time of this writing, a full prototype sector has been installed at Brookhaven National Laboratory (BNL) for use in the STAR detector. Minor modifications have been implemented in the final design, contingent on the results from measurements using the aforementioned prototype as well as details of the installation and integration.  The design process and detector specifications will be covered in what follows.

\subsubsection{SiPMs}
\label{sec_SiPM}

Based on the physics requirements and the area to be covered, our baseline design uses a combination of scintillators and silicon photomultipliers (SiPM)~\cite{ref_SiPM} for the detector. This combination is the most promising technology choice. A few important characteristics of SiPMs are listed below, showing that SiPMs can replace standard photomultipliers and have in addition a few advantages:

\begin{itemize}
\item Time of Flight coincidence resolving time $\leq$ 250 ps
\item Gain on the order of $10^{6}$
\item Linear dependence of gain with voltage bias
\item Total quantum efficiency $\geq$ 20\% (wavelength dependent)
\item Cost on the order of \$20
\item Supply voltage $\sim$ 50V
\item Not sensitive to magnetic fields
\item SiPMs are small devices, allowing for compact designs 
\end{itemize}

Commercial SiPM technology as a replacement for standard photomultipliers (PMT) has been used for high energy experiments from 2005 to present day ~\cite{Kovalchuk:2005cp,Balagura:2005gh}. Tests show a similar or even better performance compared to standard PMTs. Many experiments are currently planning upgrades using SiPM technology, e.g. the CMS HCAL upgrade for the high luminosity runs~\cite{Anderson:2012zoa}. It is also planned to use SiPMs in the STAR EIC calorimeter and the FMS preshower detector, which is now under development~\cite{STAR_preshower}.

A part of the R\&D process was to check which SiPM's performance is sufficient for our specific purpose. This includes the following measurements:

\begin{itemize}
\item Efficiency for single MIP hits
\item Uniformity of pulse area and efficiency as a function of position
of hit on a scintillator
\item Pulse shapes (rise time and fall time) for MIPs
\item Gain vs. bias voltage
\item Timing resolution
\item Temperature stability
\end{itemize}

\subsubsection{Radiation Damage of SiPMs}
It was necessary to ensure that radiation damage does not affect our measurements within the two years of BESII running, as well as for any additional runs that utilize the EPD. It is estimated that the amount of radiation during BESII in collider mode is on the order of $3x10^{11}/cm^2$. The additional background radiation is still unknown. Based on neutron flux measurements in the STAR cave during the high luminosity p+p 510 GeV run~\cite{Fisyak:2013toa}, we can make an estimation for the expected radiation dose. The integrated number of neutrons for one run (100 days) with E$_{kin} \geq$ 100 keV close to the beampipe at a distance of 6.75 m is on the order of $10^{10}/cm^{2}$. This can be compared to detailed radiation hardness measurements of several SiPMs for the JLab hall B calorimeter, which have been recently performed~\cite{Qiang:2012zh}. Preliminary estimates indicate that we should be aware of possible radiation damage effects during the BES II run. We are investigating this issue in more detail by analyzing the prototype installed in RHIC in run 16.  A more comprehensive study will be done during our Run 17 commissioning run, however at the moment we see no evidence of radiation damage in the SiPM signals.

\subsubsection{Tile and Groove Geometry}
The final tile groove design was inspired by the Barrel Electromagnetic Calorimeter (BEMC) \cite{Beddo:2002zx}. Since the beginning of R\&D efforts, the sigma groove geometry for the WLS fiber seemed to be the most likely candidate for the final design, but other geometries were both designed and tested at LBNL. However, we found that the sigma geometry was capable of single photon detection and represented the simplest implementation. In addition, the small improvement in efficiency afforded by the other geometries was not found to be significant enough to justify complications that arose from, say, coupling multiple fibers to a single SiPM or increasing the risk of fiber damage associated with tightening the bending radius. In this respect, we stuck to the status quo, albeit with the groove closer to the center of the tile, as opposed to the BEMC choice of the grooves essentially at the tile edges [see figure to be inserted]. \\

In contrast with the BEMC, and due to unique spacial constraints set for the EPD placement at STAR, we do not feed the WLS fibers through the rear face of the tiles into an undulating channel. Rather, the fibers exit directly out the front face of each tile and are directed through a center channel between two sectors before being eventually organized through custom printed fiber collectors. This is because the SiPMs will be positioned near the large toroidal magnet in such a way as to shield them from radiation damage by fast neutrons. \\

All tile designs were made using the SolidWorks$^{\textrm{\textregistered}}$ software, elaborated on in the subsequent section, and sent to the LBNL machine shop for construction. The initial design had multiple tiles machined together, with a 90\% depth separation cut between each, as was done for the BEMC. This would constitute one ``sector'', in the shape of an elongated trapezoid, that would extend radially outward from the beampipe. Of course, since each tile must be optically isolated from any signals in other tiles, we experimented with filling the groove with epoxy and coating the tile sides with reflective paint. However, there would still be the possibility of signal contamination through the remaining 10\% of uncut tile holding the sector together. Ultimately, the final procedure involved breaking this 10\% remainder (not done for the BEMC) and painting and gluing it. This preserves the mechanical robustness of the sector, the primary reason for initially keeping the 10\% groove, while providing true optical separation between the tiles. \\

\subsubsection{Radiation Hardness of Optical Cement}
The materials for the EPD construction consist of 1 cm thick Eljen EJ-200 plastic scintillator, split into trapezoidal tiles with Kuraray Y-11(200) polystyrene Wavelength Shifting (WLS) fibers embedded and glued into grooves in the scintillator with Eljen EJ-500 optical cement.
Studies of single tiles indicated that in addition to the groove shape, gluing the wavelength shifting fiber (WLS) into the grooves in the scintillator with optical cement improved the optical contact.  However, this opened up the question of how the optical cement, along with the WLS fiber and the scintillator itself, respond to the radiation environment that is expected.  The optical cement had not been previously tested for radiation hardness; however there is documentation of a radiation test of a similar epoxy called Bicron 600 which shows significant radiation damage. These test results are not applicable to our detector, due to the fact that the measurement was conducted with a single dose of gamma radiation greatly exceeding our expected radiation limit\footnotemark. 

\footnotetext{Kirn, Thomas, et al. Absorption length, radiation hardness and ageing of different optical glues. No. CMS-NOTE-1999-003. CERN-CMS-NOTE-1999-003, 1999.}

\begin{figure}[H]
\centering
\includegraphics[width=0.5\textwidth]{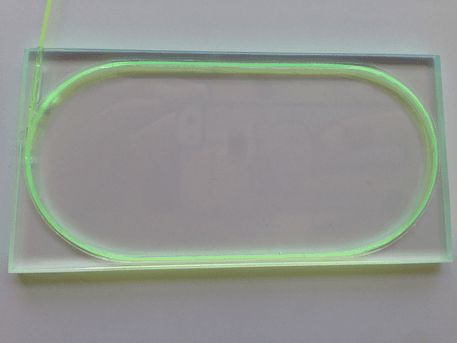}
\label{fig:tile}
\caption{12cm$\times$5.5cm$\times$0.5cm EJ-200 scintillator with a triple layer of Kuraray Y-11 WLS fibers glued with EJ-500 optical cement. This was one of the tiles used in the radiation hardness experiment.}
\end{figure}

\begin{figure}[H]
\centering
\includegraphics[width=0.6\textwidth]{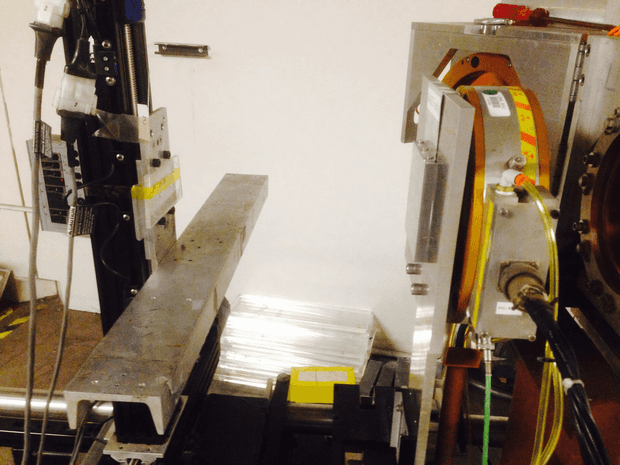}
\caption{\label{fig:setup} Each tile (left) was attached to a vertical bar directly in the beamline, roughly 25 cm from the beam window (right).}
\end{figure}

We estimate that the radiation dose for Beam Energy Scan Phase II (BES-II) will be a maximum of $3.0\times10^{11}$ particles/cm$^2$ per year. We designed and conducted an experiment at the 88-Inch Cyclotron at Lawrence Berkeley National Laboratory to test the effects of a range of radiation levels, exceeding the maximum expected level, on completed EPD tiles. We constructed and  assembled five identical tiles containing a fully polished WLS fiber looped in triple layer and held in by about 1.25 mL of EJ-500 epoxy, as shown in Figure \ref{fig:tile}. The arrangement of the tile with respect to the beamline is shown in Figure \ref{fig:setup}. Each tile was irradiated with a 50 MeV proton beam at one of five different levels of fluence, from $1.0\times10^{10}$ particles/cm$^2$ to $1.0\times10^{12}$ particles/cm$^2$. 

The irradiation effects were measured by recording signal amplitudes before and four days after radiation with a cosmic ray telescope setup consisting of two trigger detectors, with the test tile sandwiched in between. The tile with the greatest fluence (Tile 1) was measured again 3.5 weeks after irradiation, but no change was seen in the result. We measured the amplitude utilizing a trigger threshold of 6 mV for the two reference detectors, with typical signals on the order of 40 mV. For the analysis we removed bad events, which are defined as rapidly fluctuating signals. These cuts left us with an average of 200 measurements per tile, both before and after irradiation. 

The signal per tile was corrected for event-by-event fluctuations due to lower or higher energy deposits by dividing the tile signal by the average of the trigger detector signals. Finally, the efficiency of each tile before and after irradiation was calculated by dividing the average signal after irradiation by the average signal before irradiation. We found that the efficiency of the detector is 70\% when irradiated with $1.0\times10^{12}$ particles/cm$^2$. This level of radiation is about 3 times the maximum amount of radiation expected during BES-II. The irradiation data and calculated efficiency of each tile after radiation is listed in Table \ref{table:1}.

\begin{table}[h!]
\centering
\begin{tabu} to 1.0\textwidth { | X[c] || X[c] | X[c] | X[c] | X[c] | X[c] || }
 \hline
  & Tile 1 & Tile 2 & Tile 3 & Tile 4 & Tile 5\\
 \hline
 Flux (ions/cm$^{2}$s)  & $2.5\times10^8$ & $2.5\times10^8$ & $2\times10^8$ & $6.5\times10^7$ & $5\times10^7$ \\
 Fluence (ions/cm$^{2}$) & $1\times10^{12}$ & $5\times10^{11}$ & $1\times10^{11}$ & $5\times10^{10}$ & $1\times10^{10}$ \\
 Dose (kRad) & 300 & 150 & 30 & 15 & 3 \\
 Efficiency (\%) & 70 & 82 & 89 & 97 & 100\\
 \hline
\end{tabu}
\caption{Irradiation data per tile.}
\label{table:1}
\end{table}

\begin{figure}
\centering
\includegraphics[width=0.6\textwidth]{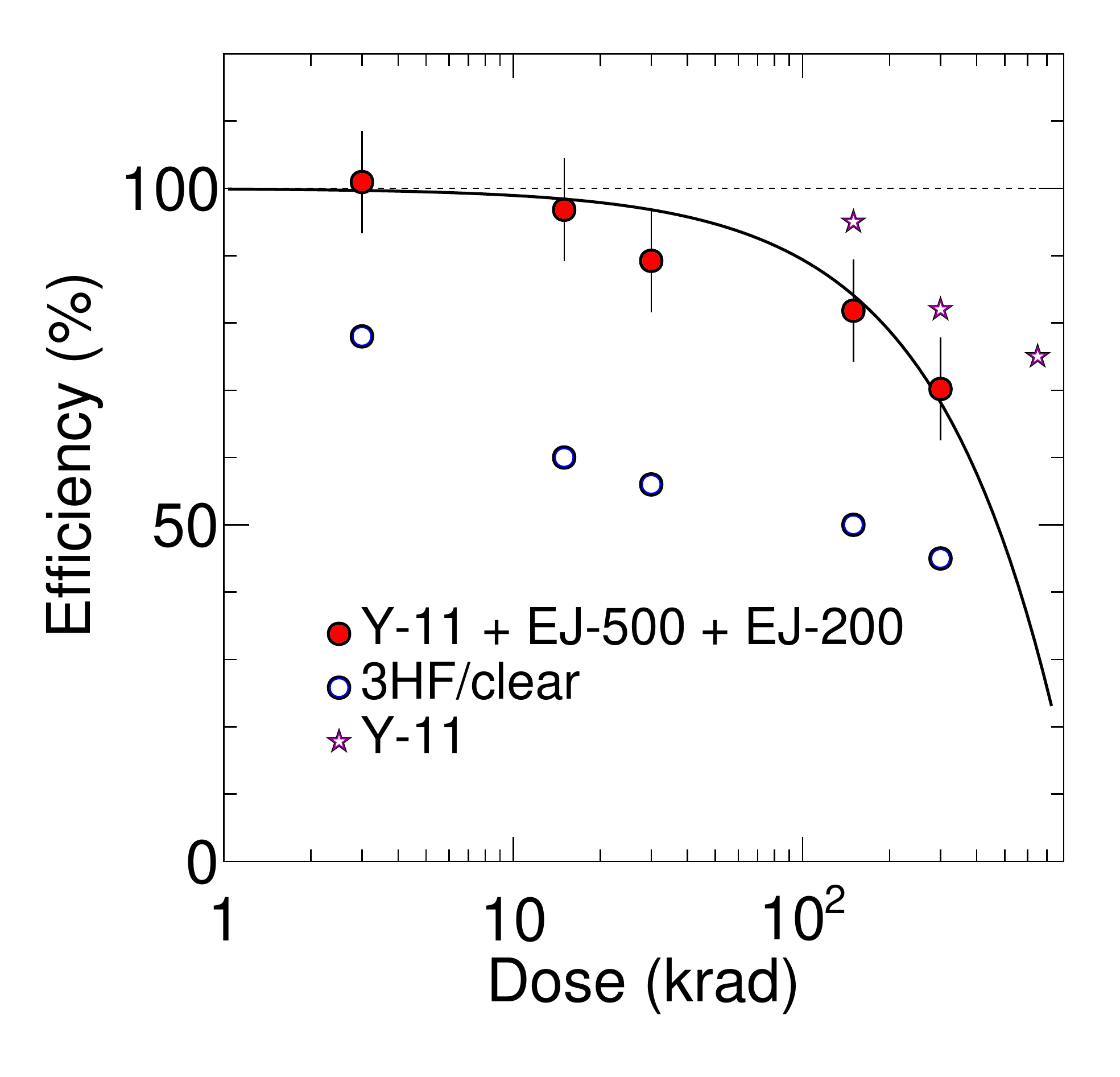}
\caption{\label{fig:eff} Measurement of the efficiency rate per dose of the assembled detector tiles compared with the reported efficiency rates of the Kuraray Y-11 WLS and 3HF clear optical fibers.}
\end{figure}

There have been studies conducted on the radiation hardness of the Kuraray WLS fibers that show agreement with our result of a 30\% loss in detector efficiency. One such study found that after a dose of 650 kRad the fibers had an efficiency of 70\% \footnote{Fleming, Jamie. Radiation Damage For The Hodoscope. 1st ed. 2012. Web. 26 Apr. 2016.}. For our experiment, we had $1.0\times10^{12}$ protons/cm$^2$ in our maximum irradiated tile, while the average energy deposited in the fibers during the experiment is calculated to be 19 $MeV\cdot{cm^{2}}/g$. This equates to a maximum dosage of 300 kRad in our WLS fibers, a factor of two less than the aforementioned study. A second study measured the attenuation length of Kuraray's 3HF clear optical fibers to decrease by about 50\% after an irradiation dosage of 300 kRad \footnote{K. Hara et al./Nucl. Instr. and Meth. in Phys. Res. A 411 (1998) 31—40}. The results of both studies are compared to our experiment results in Figure \ref{fig:eff}. Both studies report that while a small percentage of the radiation damage to the fibers is permanent, the fibers are able to partially recover from the damage after a rest time on the order of one to two months.

The results of our experiment show a 30\% efficiency reduction in our most highly irradiated tile. From the comparison of these results with the aforementioned studies we conclude that our measured efficiency is in agreement with the reduction in efficiency of WLS fibers that have been exposed to high dosages of radiation. This indicates that the WLS fibers account for the total radiation damage to the tiles. Consequently, we have determined that the optical epoxy suffers no noticeable impairment from the amount of radiation we applied during the experiment. Additionally, the 30\% reduction rate corresponds to the tile that received a radiation dosage around three times greater than that expected for BES-II. 

In conclusion, Eljen EJ-500 optical cement is considered radiation hard for our purposes and is satisfactory for use in the construction of the EPD. 

\subsubsection{Fiber Development}
We also tested to ensure that the SiPMs chosen for the detector worked well with the scintillator light wavelength. This included the installation of wavelength shifting (WLS) fibers, which depended on the specifications of the chosen SiPMs.  SiPM modules with dimensions of 3$\times$3 and 4$\times$4 mm from two companies, sensL and AdvanSiD, were tested with a scintillator. Different setups, with and without optical fibers, were investigated and tested in a two paddle cosmic ray setup. By triggering on one paddle, there was almost 100\% efficiency of seeing a count in the other paddle~\cite{cite_Matis}. \newline

\subsubsection{Tile Size}
\begin{figure}[htbp]
\begin{center}
{
\mbox{\includegraphics[width=0.25\textwidth]{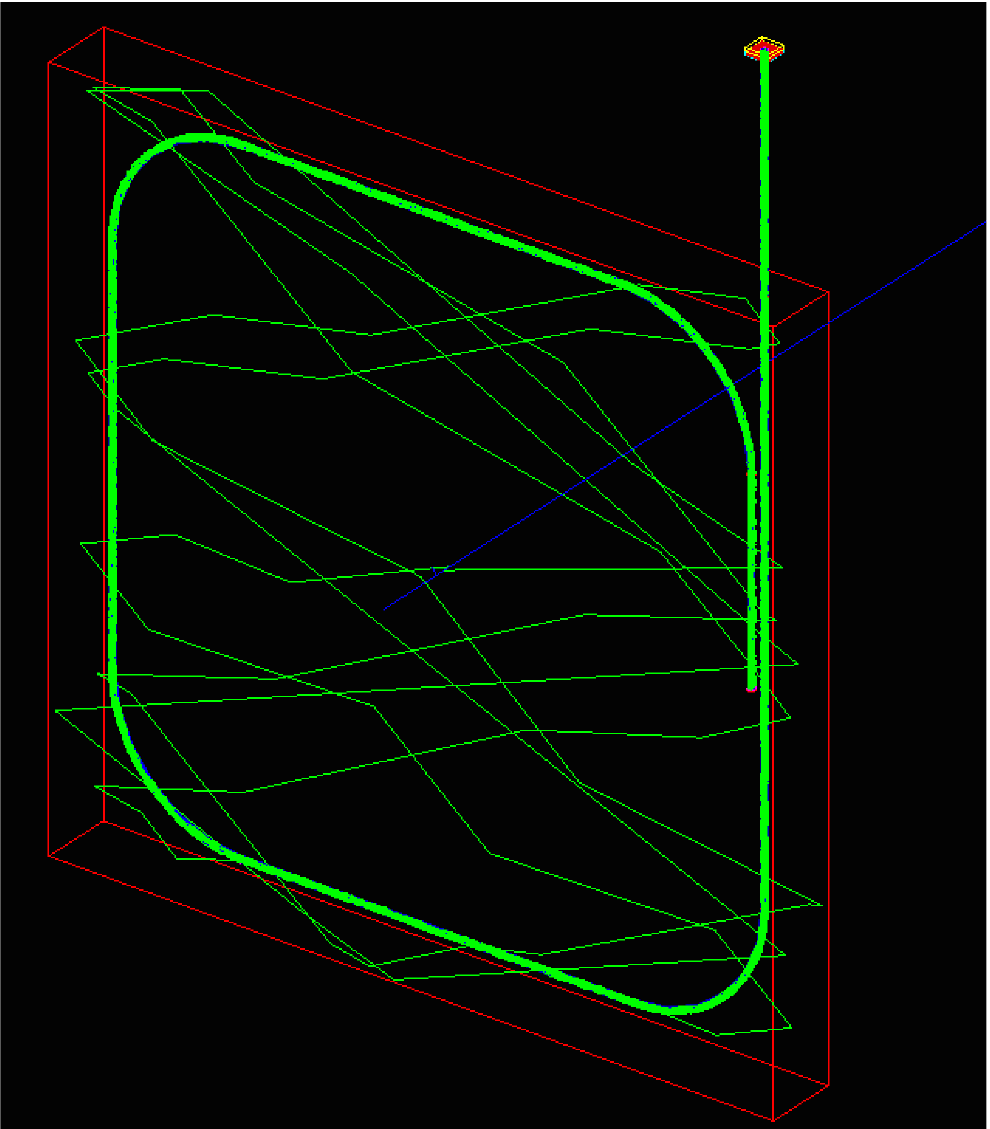}}
\mbox{\includegraphics[width=0.70\textwidth]{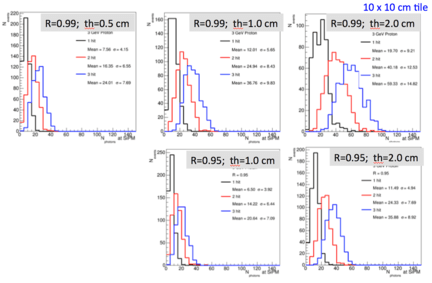}}
}
\end{center}
\caption{The left-most figure is the GEANT model for a single tile with photon tracks that result in a hit in the SiPM shown.  The right figures show the 1 hit, 2 hit and 3 hit distributions for a single tile for different tile thicknesses and different reflectivities.  This shows we we can not fully separate the different hits, however this indicates for all choices there is enough separation calculate the centrality and event plane via centrality weighting.
}
\label{fEPD_GEANT_Tile}
\end{figure}

\begin{figure*}[]
\centering
\resizebox{13cm}{!}{%
\includegraphics[bb = 50 400 550 800,clip,width=0.47\textwidth]{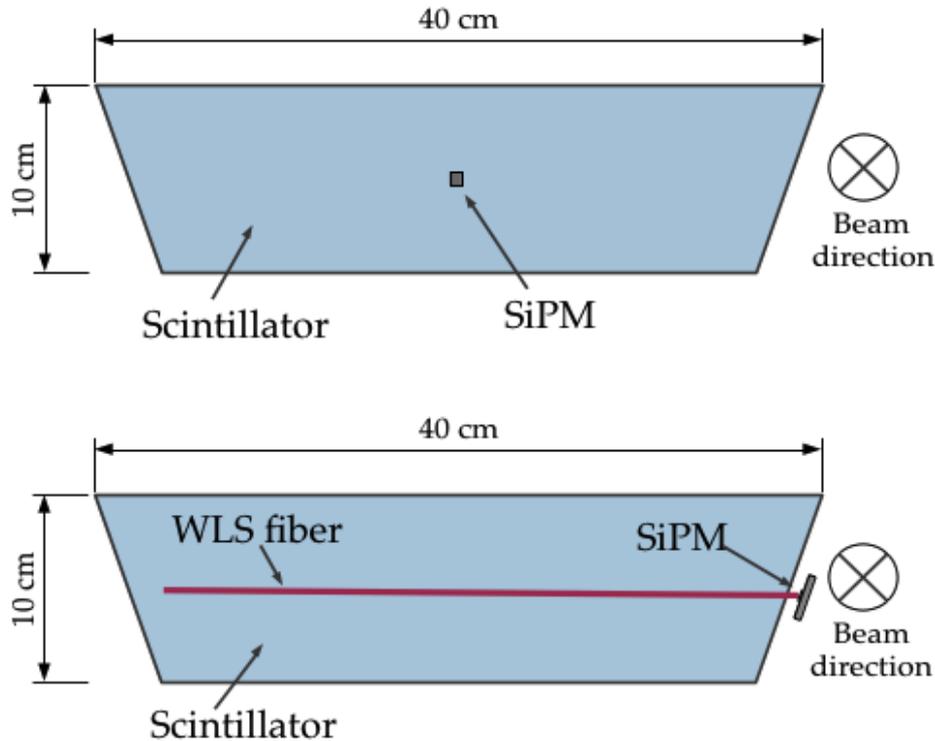}}
\caption{Drawing of a large scintillator tile and several possible arrangements of SiPMs, which have to be simulated and tested. The bottom drawing shows the additional installation of wavelength shifting (WLS) fibers, which may be needed.}
\label{fSci_tile}       
\end{figure*}

The geometrical configuration of the scintillators was developed utilizing results from simulations. The cross-sectional area of the scintillator was determined from the higher order event plane resolution, the hit probability and the resolution of the z position of the vertex. The amount of light collected was determined by the thickness of the scintillator. GEANT was used to simulate the light emission and collection in the scintillators and SiPMs. The results are shown in Figure \ref{fEPD_GEANT_Tile}. The simulation showed a 50\% increase in the number of photons when the thickness was increased from 0.5 cm to 1 cm, and an additional 60\% increase for a thickness of 2 cm. Based on the results of the simulations, the exact tile geometry was chosen and a series of experiments were performed including several arrangements. An example for trapezoidal tiles is shown in Fig.~\ref{fSci_tile} with two arrangements of WLS fiber and SiPMs which have proven to be less efficient than the sigma groove design we finally use. Emphasized in the figure is only the largest tile, which will be most problematic due to the small ratio of SiPM area to tile size, but similar tests have to be done for various tile sizes. These tests were in agreement with the simulation regarding the 50\% increase in the photon yield.  The tile thickness was chosen to be 1 cm, which is the same as the BBC. Results from the prototype installed in STAR for run 16 indicated that more light was not needed, and a thicker detector is more expensive, heavier and poses additional risk. 

We also found it necessary to develop and test polishing and wrapping procedures for the scintillators. For the WLS fibers a technique was developed to install them and to optimize the connection to the SiPMs.

The first tests were done with commercial hard- and software provided by the SiPM manufacturer. In this stage we also performed basic trigger tests with cosmics and a two or three tile setup. Once basic tests were done and the optimal SiPMs and scintillators were selected, we switched to STAR customized hardware and built the prototype. 

 In order to fully replace the BBC detectors we need to consider the following points:

\begin{itemize}
\item A hit time measurement is needed to determine the z-vertex position
\item An expansion of the existing readout system or a significant amount of pre-processing is needed 
\end{itemize}

We furthermore want to investigate whether it is possible to significantly improve the timing resolution with the EPD setup compared to the BBCs, which could lead to new applications, like out of time rejections. Those studies could end in a follow up proposal in order to develop the needed readout system.



\subsection{Major R\&D Results}

Over the course of our R\&D we have determined the best options for a number of vital elements of the construction of the EPD. These aspects and our conclusions are delineated in the following list.

\begin{itemize}
 \item During R\&D both 0.5 and 1.0 cm scintillator thicknesses were tested. The 1.0 cm thick tiles yielded about 35\% better signals than the 0.5 cm tiles. We conclude that a 1.0 cm thick detector is best for the signal quality of the EPD.
 \item We developed a method of polishing the cut scintillator edges. We found that the most satisfactory outcome came as a result of polishing with aluminum oxide powder with a grit size on the order of \SI{5}{\micro\metre}. For this reason we will use aluminum oxide powder in the scintillator polishing process of the EPD construction.
 \item Numerous design tests have shown a sigma shaped groove for the WLS fibers to be the superior design in terms of signal strength, tile coverage and connection economy. Other designs that proved to be inadequate in these areas include a straight fiber groove design as shown in figure \ref{fSci_tile} and direct connection of the SiPM to the scintillator. The sigma shape is the final fiber design for the EPD. 
 \item Looping the WLS fiber in a triple layer in the scintillator tiles produces a signal that is a factor of 2 times greater than if the fiber was only in a single layer. Such a design is still economical in terms of cost and labor. Consequently we have decided that triple layered fibers should be incorporated into the final EPD. 
 \item Throughout our R\&D we have used Eljen EJ-500 optical epoxy cement for the assembly of the prototype and test scintillator tiles. We have found this epoxy to be easy to handle and effective for our purpose. Before utilizing it to glue the fibers into place inside the scintillator grooves we run it through a centrifuge for mixing and degassing. Afterwards we place the epoxy in vacuum to draw out leftover air pockets and produce an optical cement free from bubbles. This process is efficient and allows for better quality signal collection by the WLS fibers. We will use EJ-500 optical cement for the construction of the EPD. 
 \item During the process of gluing the fibers into the scintillator grooves, we found that the best method for maintaining a clean scintillator surface free from epoxy residue is to cover the scintillator with Ultra-Slippery Tape Made with Teflon$^{\textrm{\textregistered}}$ PTFE from McMaster-Carr and then carefully cut the tape out from the grooves. This tape adheres well to the surface of the scintillator, cuts easily and leaves no residue when removed. This Ultra-Slippery Teflon Tape and method of masking the tiles will be used in the assembly of the EPD.
 \item The prototype utilized aluminum mylar to increase the inner reflectivity of the detector. Further R\&D has shown that Tyvek 1055B is a few percent better than aluminum mylar for this purpose and has the added benefit of not scratching the surface of the scintillator. Therefore the EPD will be wrapped in Tyvek 1055B.
\end{itemize}

\subsubsection{Computer Graphic Designs for the EPD}
 
Based on the simulations, described in section \ref{sec_sim}, it was determined that the optimal design of the detector would be two 24-sector disks on either side of the event plane. Each sector would extend radially from the beam pipe and touch side-to-side forming a complete icosikaihexagon. The prototype discussed in this paper is slightly larger than one of these 24 sectors. 

\begin{figure}[!h]
\centering
\includegraphics[width=0.3\textwidth]{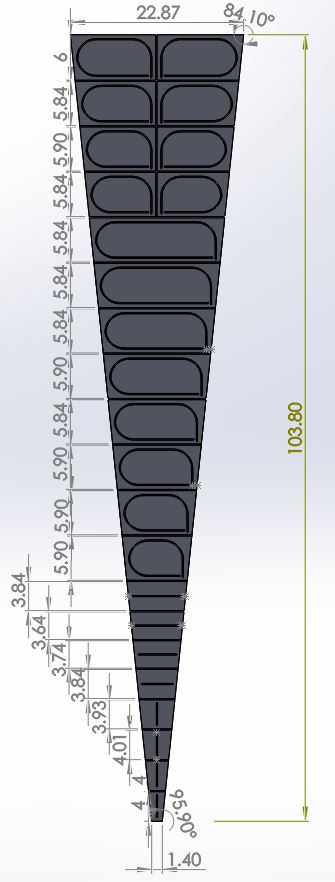}
\caption{\label{fig:PDC}SolidWorks model of the EPD prototype showing dimensions in centimeters.}
\end{figure}

The prototype was designed using a 3D computer-aided design software called SolidWorks$^{\textrm{\textregistered}}$  distributed by Dassault Syst\'emes. The model and the actual dimensions of the prototype in centimeters are pictured in figure \ref{fig:PDC}. In summary, the prototype is a narrow trapezoidal shape 103.8 cm long made from 1 cm thick plastic scintillator. It is 22.87 cm wide at the top and 1.4 cm wide at the bottom edge close to the beam pipe. The sector is divided into 24 tiles that are optically separated (see section \ref{sec_scintassembly}) and of varying widths ranging from four to six centimeters.  This design was longer than the required 90 cm, and consequently was installed into STAR horizontally below the beam-pipe.

\begin{figure}[!h]
\centering
\includegraphics[width=0.5\textwidth]{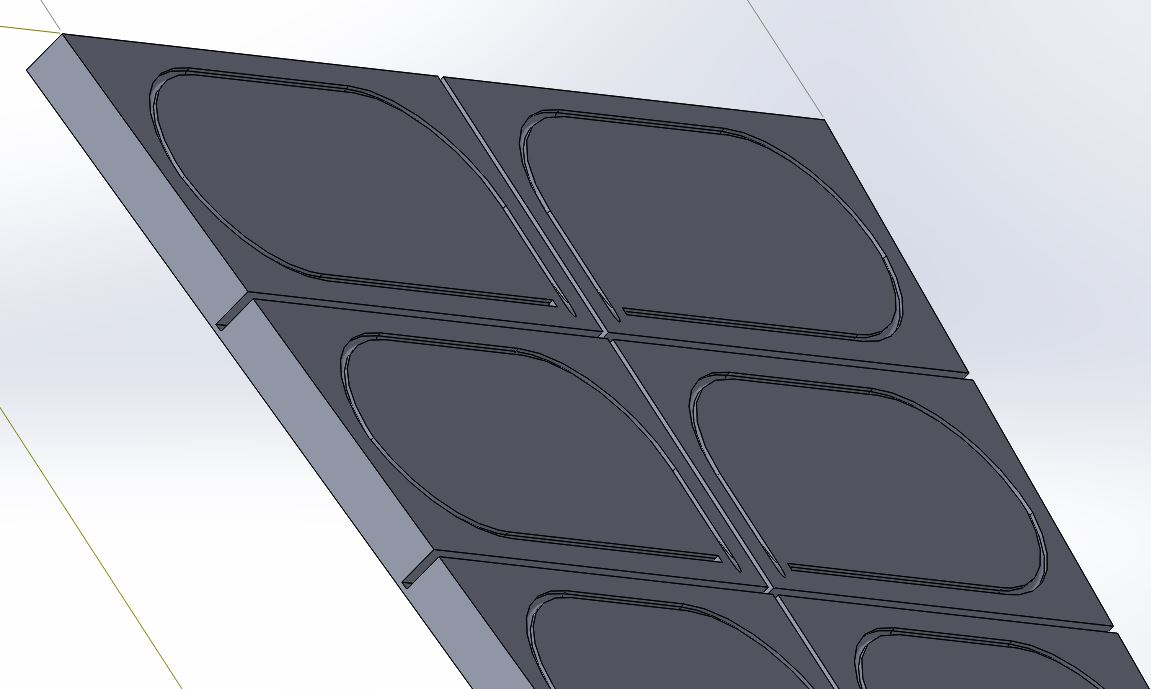}
\caption{\label{fig:PNZ}Closeup of the SolidWorks model of the EPD prototype showing the sigma shape and incline of the fiber grooves.}
\end{figure}

Each of the 24 tiles in the prototype has a 0.16 cm groove to hold the optical fibers used to collect event signals. In the larger tiles at the top of the prototype the grooves are sigma-shaped, with the fibers coming out of the scintillator at an angle at the bottom of each tile. This is illustrated in figure \ref{fig:PNZ}.

\begin{figure}[!h]
\centering
\includegraphics[width=0.4\textwidth]{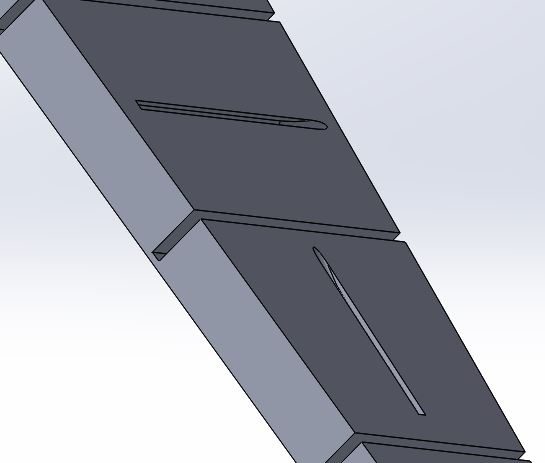}
\caption{\label{fig:PNBZ}Closeup of the SolidWorks model of the EPD prototype showing the shape and incline of the fiber grooves at the bottom of the detector.}
\end{figure}

The optical fibers used in the prototype have a maximum bending radius of 2.5 cm so the sigma-shaped grooves could not fit in all of the tiles. Therefore, straight line grooves, pictured in figure \ref{fig:PNBZ}, were employed in the smaller tiles at the bottom of the prototype.

\subsubsection{Scintillator Assembly}
\label{sec_scintassembly}

The prototype was cut in LBL's machine shop from three 30x30 cm 1 cm thick Eljen EJ-200 plastic scintillator tiles. The outside edges were polished in order to increase the reflectivity, which needs to be very high.  The groves between the tiles were cut 90\% down, then filled with epoxy in order to limit cross-talk.  The groves for the fibers were cut into the scintillator.

\subsubsection{Fiber Preparation}
\label{sec_fiberprep}

The fibers used in the scintillator tiles for signal collection are Kuraray Y-11(200) Wavelength Shifting Fibers (WLS). These fibers have an attenuation length of $\lambda =400$ cm. In order to transmit as much light as possible the fibers need to have polished ends cut perpendicular to the fibers' longitudinal axes. A six-step process as outlined below was utilized to achieve the desired quality of the fiber ends.
\begin{enumerate}
\item The required lengths of fiber were cut from the spool. This step leaves the ends of the fibers jagged, as shown in figure \ref{fig:fibers_polished} (top left).

\begin{figure}[htbp]
\begin{center}
{
\mbox{\includegraphics[width=0.48\textwidth]{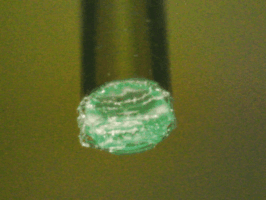}}
\mbox{\includegraphics[width=0.48\textwidth]{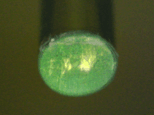}}
\mbox{\includegraphics[width=0.48\textwidth]{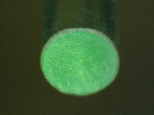}}
\mbox{\includegraphics[width=0.48\textwidth]{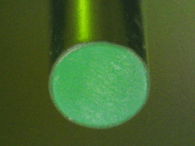}}
\mbox{\includegraphics[width=0.48\textwidth]{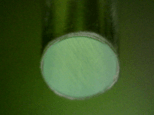}}
\mbox{\includegraphics[width=0.48\textwidth]{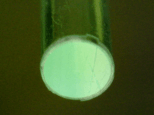}}
}
\end{center}
\caption{Kuraray Y-11(200) wavelength shifting fibers after different steps of cutting and polishing (see text for details).}
\label{fig:fibers_polished}
\end{figure}

\begin{figure}[!h]
\centering
\includegraphics[width=0.3\textwidth]{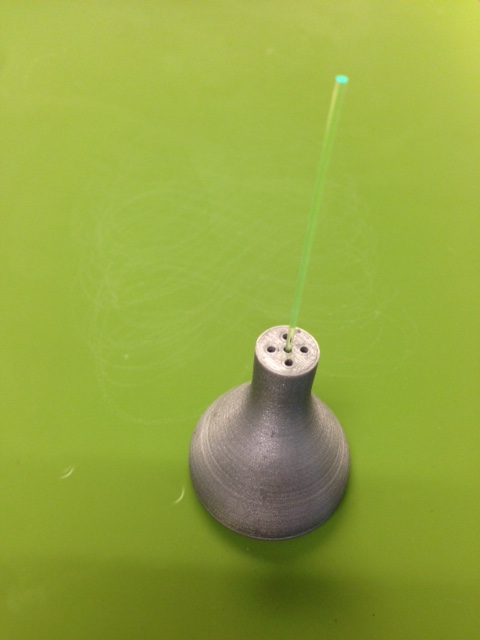}
\caption{\label{fig:green}Fiber holder and fiber on top of the green polishing sheet of the second lowest grit size.}
\end{figure}

\item \label{itm:step2} A razor fiber cutting tool was used to cut the ends of the fibers. When done correctly, the fiber ends were smoothly cut perpendicular to the longitudinal axis of the fiber with minimal fraying of the encasing cladding. However, the razor often left silver residue on the cut surface. These aspects are shown in figure \ref{fig:fibers_polished} (top right).

\item \label{itm:step3} Polishing began using a fiber polishing sheet with a grit of \SI{5}{\micro\metre}. These sheets are sold by Thorlabs, with the black \SI{5}{\micro\metre} grit size being the largest of four sizes. A plastic fiber holder was designed and fabricated with the 3D printer in order to maintain the fiber's perpendicular cut surface while polishing. The fibers were polished using a figure-eight pattern 30-40 times to achieve uniform quality in this step. The result of this step is shown in figure \ref{fig:fibers_polished} (middle left)

\item \label{itm:step4} The next highest grit size polishing sheet was then used on the fiber ends. The result of polishing with this grit size is shown in figure \ref{fig:fibers_polished} (middle right). Again, a uniform polish was achieved using a figure-eight pattern 30-40 times.

\item \label{itm:step5} The fiber ends were then polished in a figure-eight pattern with the second lowest grit 30-40 times. In this case the second lowest grit size is the green \SI{1}{\micro\metre} polishing sheet pictured in figure \ref{fig:green}. The result of this step is shown in figure \ref{fig:fibers_polished} (lower left).

\item \label{itm:step6} Finally, the ends of the fibers were polished in 30-40 figure-eights on the white sheet with the lowest grit size of \SI{0.3}{\micro\metre}. After this step the ends of the fibers had a smooth, mirror-like quality. In some cases this step revealed one or two deep scratches as shown in figure \ref{fig:fibers_polished} (lower right). To remedy this, steps \ref{itm:step3} through \ref{itm:step5} were repeated until a satisfactory quality polish was achieved. Also, if too much pressure was applied to the fiber while polishing the cladding often became frayed. In these cases it was often necessary to start over from step \ref{itm:step2}.
\end{enumerate}

\subsubsection{Coupling of the Fibers}
The signals collected by the WLS fibers in the scintillator tiles are read by Silicon Photomultipliers (SiPMs) located a short distance from the actual detector. In order to allow the signals to reach the SiPMs with as little loss as possible the WLS fibers are connected to clear optical fibers with an attenuation length of 10 meters. This necessitates two points of connection: fiber to fiber and fiber to SiPM. \\
\begin{figure}[htbp]
\begin{center}
{
\mbox{\includegraphics[width=0.47\textwidth]{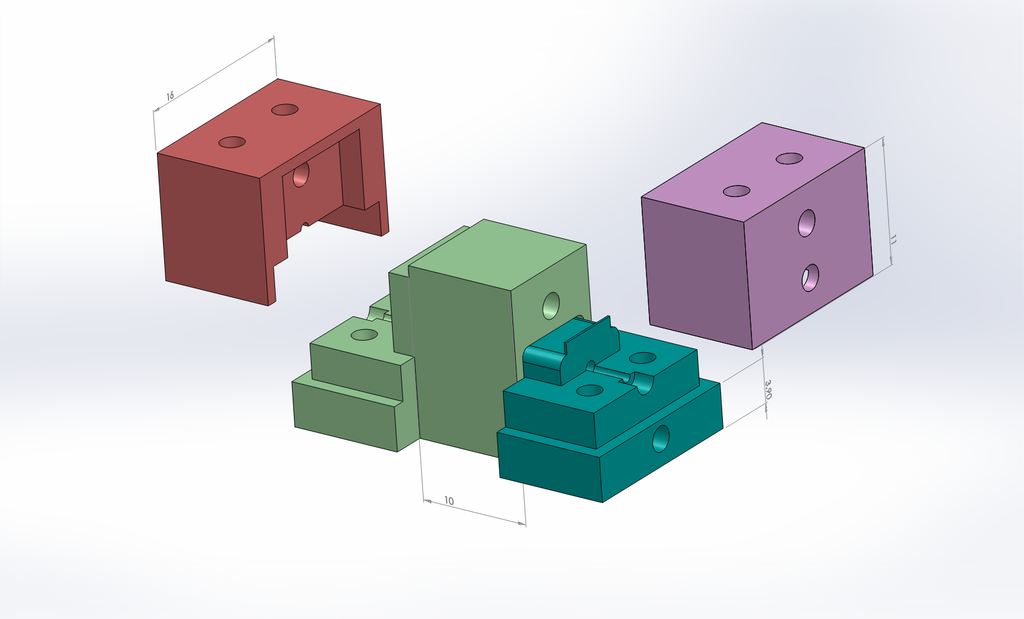}}
\mbox{\includegraphics[width=0.43\textwidth]{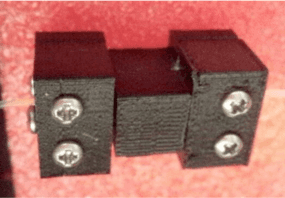}}
}
\end{center}
\caption{WLS to clear fiber connectors used for the EPD Prototype.  The three graphics on the left show the schematic for the three pieces that comprised the fiber connector.  On the right is a photo of the final assembly from the prototype.
}
\label{fig:PrototypeFiberConnectors} 
\end{figure}

 The connection between the WLS fibers and the clear optical fibers can result in appreciable signal loss if the fiber ends are misaligned. For this reason it was necessary to design a connector that is light-tight and correctly positions the axes of the fibers. The design was created in SolidWorks and built with the 3D printer. To increase the light-tight aspect of the connector the design was printed using black polylactic acid (PLA) plastic. The designs are shown in figure \ref{fig:PrototypeFiberConnectors}.  This design allowed the two fibers to butt up against each other.
  
 
 \begin{figure}[htbp]
\begin{center}
{
\mbox{\includegraphics[width=0.47\textwidth]{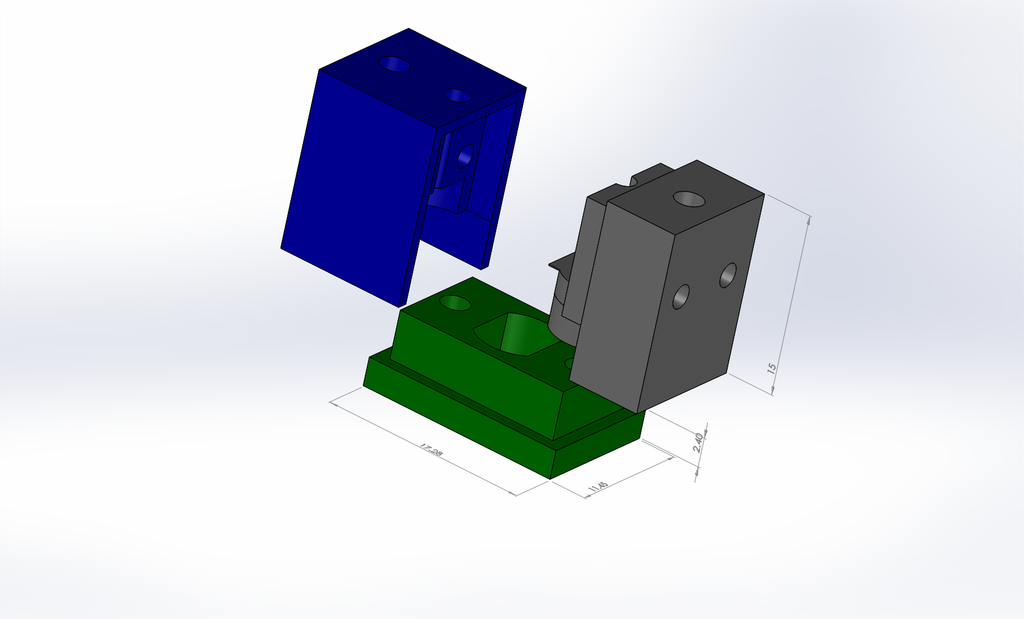}}

}
\end{center}
\caption{Fiber to SiPM connector for the Prototype.  Each connector only connected a single fiber and SiPM.
}
\label{fig:PrototypeSiPMConnectors} 
\end{figure}

\subsubsection{Fiber to SiPM Connection}
Similar to the fiber-fiber connector, the fiber-SiPM connector is also required to be light-tight to reduce signal noise. However, if the fiber extends too far into the connector it could easily break the avalanche photodiode (APD) array of the SiPM. These parameters were kept in mind during the design process of the connectors. 

To avoid breaking the APD, the insertion piece that holds the fiber in place was designed to extend into the SiPM holder by 1 mm, far enough to be light-tight while allowing room for the SiPM to fit without applying any pressure to the interface. The fiber is installed with its end flush with the end of the insertion piece, enabling the fiber to come into direct contact with the SiPM without breaking it. Once again black PLA plastic was used for the printing of the connectors to enhance their light-tight quality.  The schematic of the connector is shown in Figure \ref{fig:PrototypeSiPMConnectors}.

\subsubsection{Prototype Assembly}

\begin{figure}[htbp]
\begin{center}
{
\mbox{\includegraphics[width=0.65\textwidth]{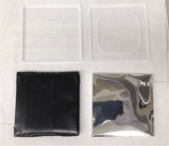}}
}
\end{center}
\caption{Test tiles for the EPD.  The upper right tile shows the sigma grove, the upper left shows a tile design that was discarded.  The lower right shows a tile wrapped in mylar, and the lower left shows a tile wrapped in mylar and then covered with black tape.
}
\label{fig:TilePrep}
\end{figure}

After polishing, the end of the WLS that will be embedded in the scintillator was painted with Eljen EJ-510 reflective paint in order to reflect back photons that were heading in the "wrong" direction.  Additionally, the edges of the scintillator itself was polished with aluminum oxide powder in order to increase its reflectivity. Then the fibers were glued into the grooves of the scintillator using Eljen EJ-500 optical cement for good optical contact.  In order to further ensure that signal photons were contained within the plastic scintillator, all tiles were tightly wrapped with aluminized mylar and then again with thick black paper. We found that it wasn't sufficient to simply take a piece of mylar and hand-wrap it around the tiles because the result isn't nearly tight enough. Instead, we drew a template cut for the mylar that would minimize the total surface area of the mylar while completely covering all faces of the tile. The mylar was then laid over the template and cut out with a box cutter for each tile. Any piece of the mylar that would be a fold on the tile was also perforated somewhat so as to ensure sharp edges all around. The result for test tiles is shown in Figure \ref{fig:TilePrep}, though for this black tape was used instead of black paper. We found that this was more than enough precaution to obtain a clean cosmic ray signal in the lab. The corresponding measurements can be found in the Testing section. \\

\begin{figure}[htbp]
\begin{center}
{
\mbox{\includegraphics[width=0.35\textwidth]{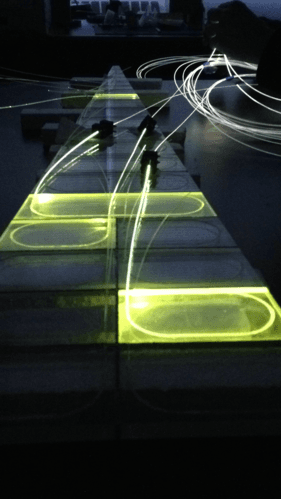}}
\mbox{\includegraphics[width=0.3\textwidth]{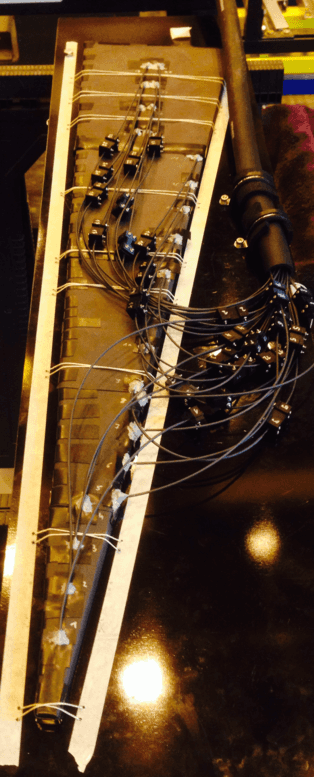}}
}
\end{center}
\caption{On the left is the light test done for the assembled prototype, where light was shined into the fiber optics.  On the right is the finished EPD prototype.
}
\label{fig:LightTest}
\end{figure}

After the EPD was assembled at LBNL it was sent to Brookhaven National Lab for installation.  At this time the shrink tube necessary for light-tightness was added to the WLS fibers that protruded from the EPD.  This was done via a heat gun.  Next the WLS and clear optical fibers were connected.  The 24 clear optical fibers were put into a single large black tube for light tightness.  The ends that protruded from this tube were also covered in shrink tubing.  The final assembly can be seen in Figure \ref{fig:LightTest}.

The far end of each clear optical fiber was connected to a SiPM.  These were then connected to the QT Boards that were borrowed from the FPS project for our use.  The electronics box was designed for Ultra Low Density (ULD) in order to facilitate any interventions that would be needed.

\begin{figure}[htbp]
\begin{center}
{
\mbox{\includegraphics[width=0.65\textwidth]{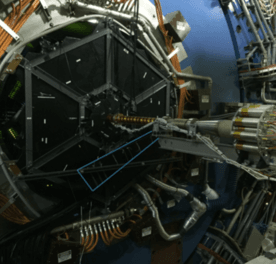}}
}
\end{center}
\caption{The installation of the EPD prototype in 2016.  The blue lines are drawn into the photograph in order to guide the eye.
}
\label{fig:PrototypeInSTAR}
\end{figure}

The EPD prototype was too long to be installed with the planned sector orientation.  It was installed underneath the beam in a mostly horizontal position.  This can be seen in Figure \ref{fig:PrototypeInSTAR}

\subsubsection{Prototype Data Analysis}
\label{sec_proto_data}
\begin{figure}[htbp]
\begin{center}
\mbox{\includegraphics[width=0.35\textwidth]{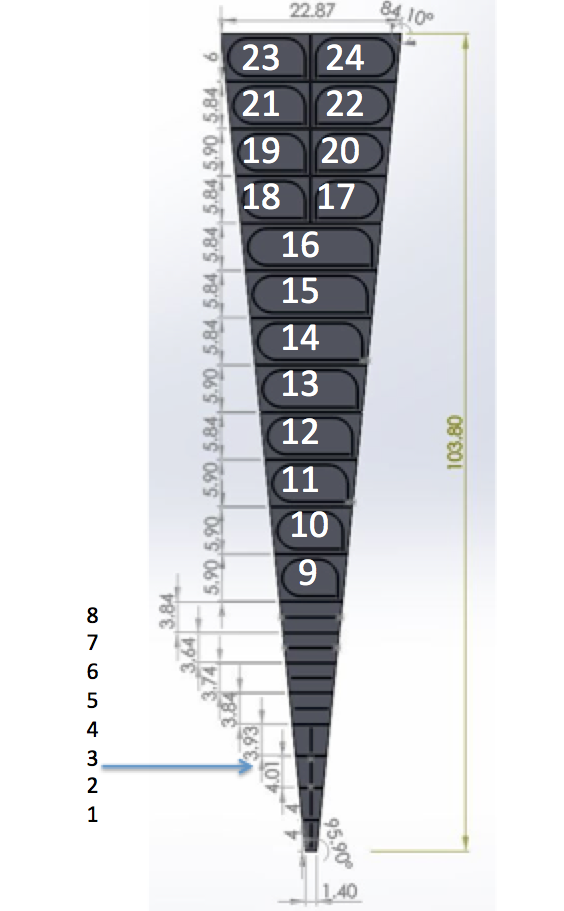}}
\mbox{\includegraphics[width=0.60\textwidth]{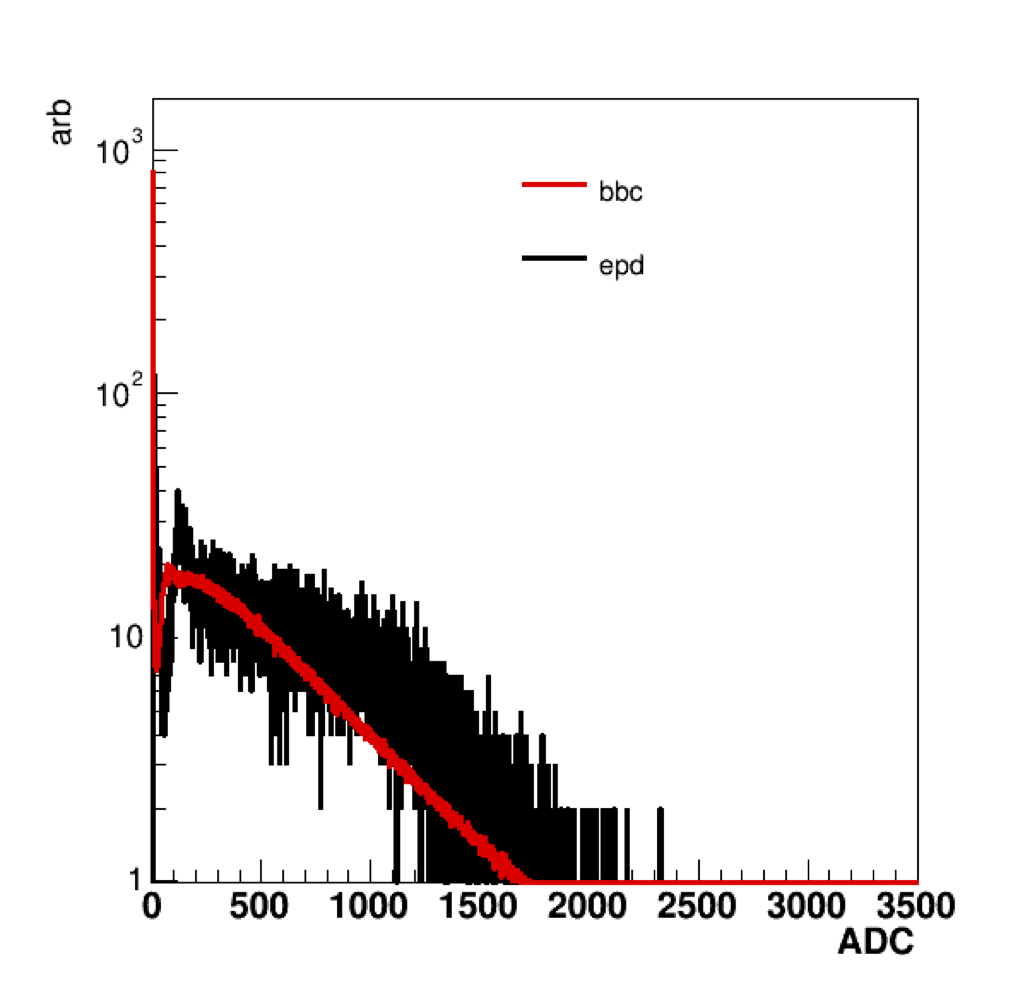}}
\end{center}
\caption{On the left is a diagram of the prototype EPD with the channel labels used during run 16.  On the right is the ADC distribution for one channel compared to the ADC distribution for the BBC for one channel.  They have a fairly similar distribution.}
\label{fig:EPD_channel_position}
\end{figure}

The prototype was installed prior to the start of run 16 with 24 channels and an additional 8 channels with timing information.  Two QT boards were used in the data collection process, and the map of the QT channels to EPD channels is shown in figure \ref{tab_QTChannelMap}.  The 8 channels with timing were chosen so that we could evaluate different size tiles and different fiber geometries.  A diagram of the EPD channels versus their position and tile geometry is shown in Figure \ref{fig:EPD_channel_position}.

\begin{table}[]
\centering
\caption{QT Channel to EPD Map}
\label{tab_QTChannelMap}
\begin{tabular}{|c|c|c|c|}
\hline
EPD Channel & QT TAC Card \#39 & QT TAC Card \#39  & QT Card \# 63 \\
 & (ADC) & (ADC Timing) & (ADC) \\
 \hline
1           & A-1                    & A-5                           &               \\
\hline
2           &                        &                               & A-1           \\
\hline
3           & A-2                    & A-6                           &               \\
\hline
4           &                        &                               & A-2           \\
\hline
5           & A-3                    & A-7                           &               \\
\hline
6           &                        &                               & A-3           \\
\hline
7           &                        &                               & A-4           \\
\hline
8           &                        &                               & A-5           \\
\hline
9           & A-4                    & A-8                           &               \\
\hline
10          &                        &                               & A-6           \\
\hline
11          & B-1                    & B-5                           &               \\
\hline
12          &                        &                               & A-7           \\
\hline
13          &                        &                               & A-8           \\
\hline
14          & B-2                    & B-6                           &               \\
\hline
15          &                        &                               & B-1           \\
\hline
16          & B-3                    & B-7                           &               \\
\hline
17          &                        &                               & B-2           \\
\hline
18          &                        &                               & B-3           \\
\hline
19          &                        &                               & B-4           \\
\hline
20          &                        &                               & B-5           \\
\hline
21          &                        &                               & B-6           \\
\hline
22          & B-4                    & B-8                           &               \\
\hline
23          &                        &                               & B-7           \\
\hline
24          &                        &                               & B-8       \\
\hline   
\end{tabular}
\end{table}

\begin{figure}[htbp]
\begin{center}
\mbox{\includegraphics[width=0.48\textwidth]{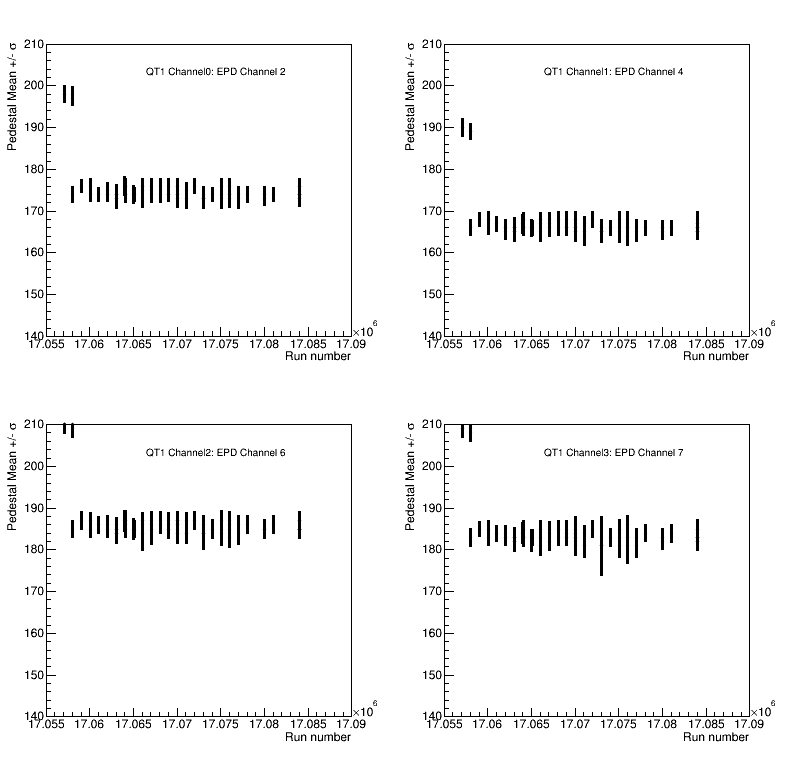}}
\mbox{\includegraphics[width=0.48\textwidth]{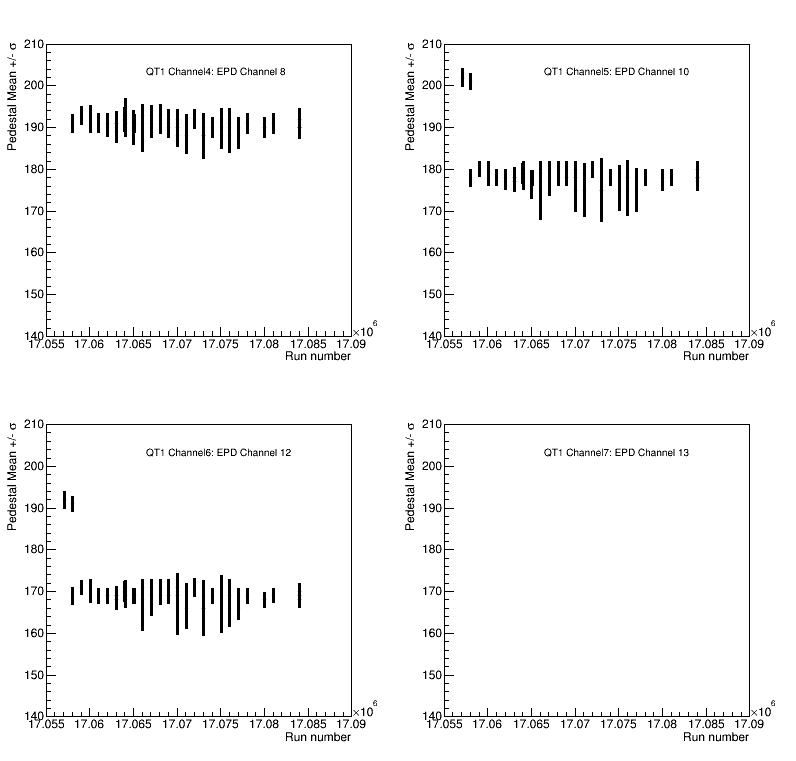}}
\mbox{\includegraphics[width=0.48\textwidth]{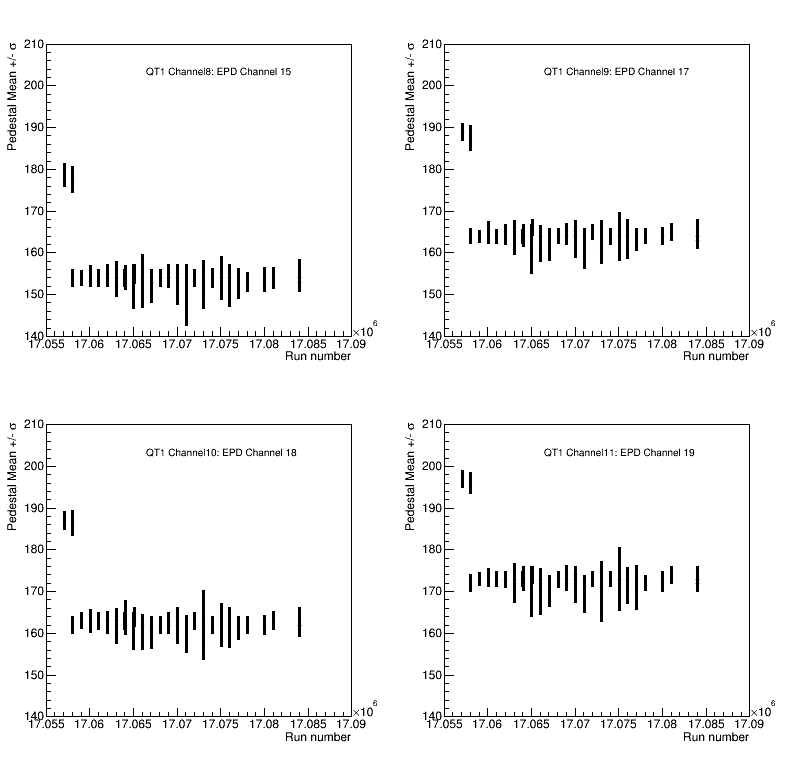}}
\mbox{\includegraphics[width=0.48\textwidth]{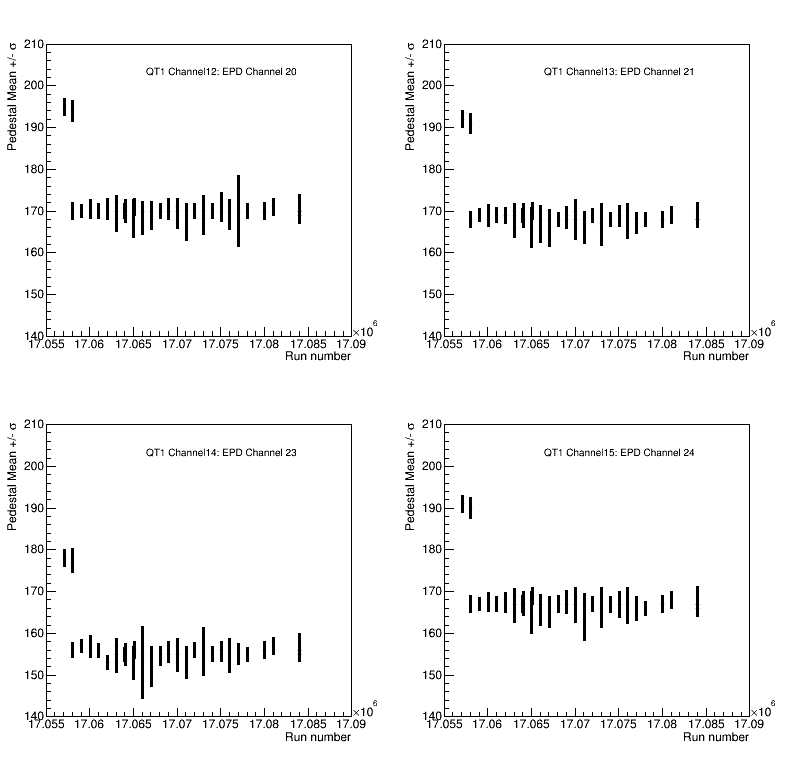}}

\end{center}
\caption{Pedestals versus run number for select channels.  During day 55 the pedestal voltages were adjusted.  During day 77, there was an issue with the EPD prototype that was not solved until day 111.
}
\label{fig:EPD_Pedestals}
\end{figure}

The EPD prototype started to record data on the 22nd day of the run.  At this point, the EPD was incorporated in the DAQ files so that the data is accessible to all.  Additionally, the EPD was included in the STAR pedestal runs, which allowed the STAR software frame to subtract the pedestals from the data.  For historical reasons, the EPD label within the STAR system for run 16 was FEQ.  Additionally, for future online development, access functions were added to StTriggerData, these will be developed into a complete package.   The pedestals for select channels during the run period to date are shown in Figure \ref{fig:EPD_Pedestals}. During day 55 the pedestal voltages were adjusted. During day 77, there was an issue with the EPD prototype that was not solved until day 111 (the beam was off during much of this time).

\begin{figure}[htbp]
\begin{center}
\mbox{\includegraphics[width=0.48\textwidth]{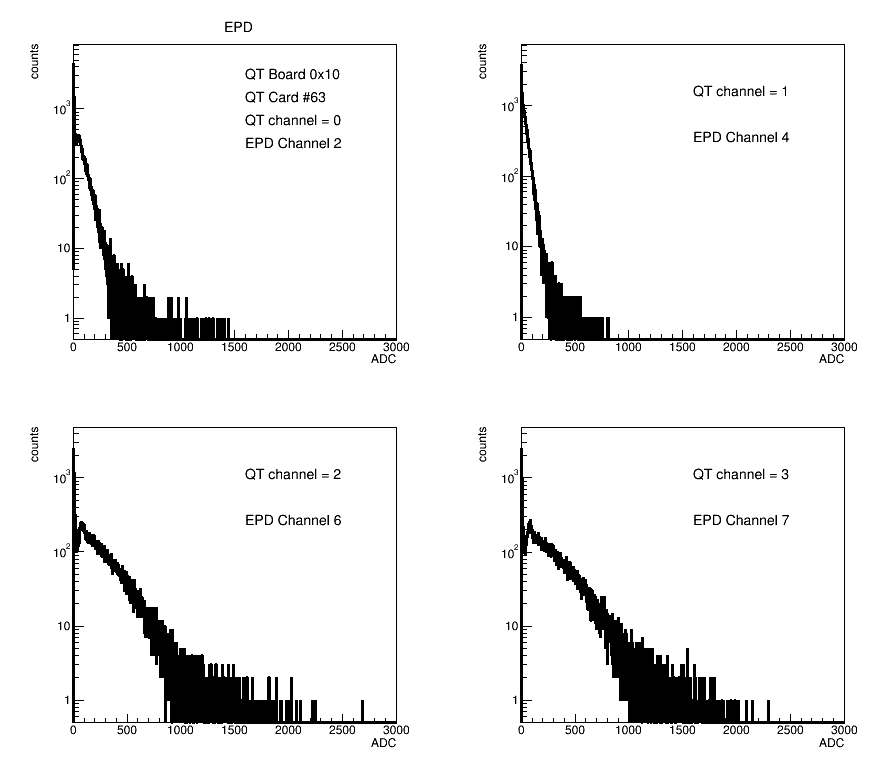}}
\mbox{\includegraphics[width=0.48\textwidth]{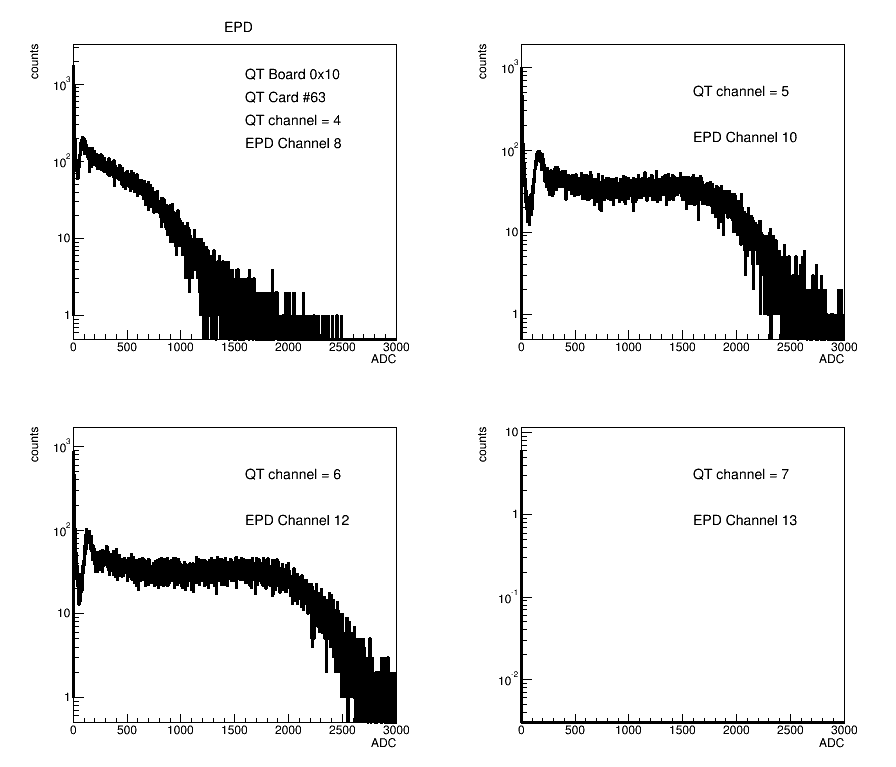}}
\end{center}
\caption{Data from 8 channels of the EPD.  There is no selection on trigger or centrality, so all events are represented here.  Channel 13 stopped working shortly into the run.}
\label{fig:EPD_1DData1}
\end{figure}

\begin{figure}[htbp]
\begin{center}
\mbox{\includegraphics[width=0.48\textwidth]{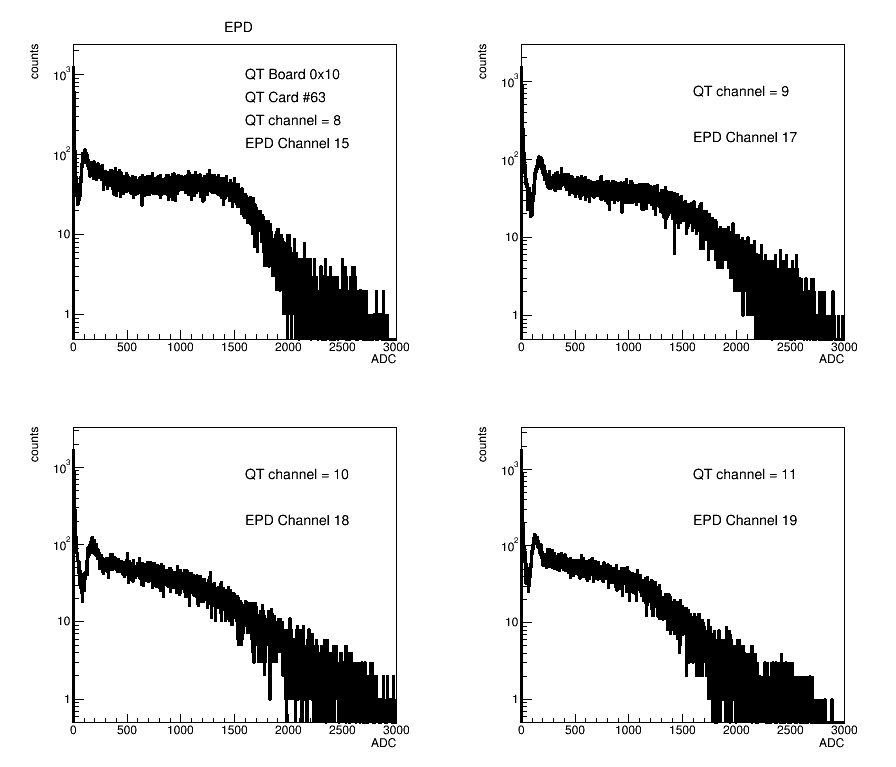}}
\mbox{\includegraphics[width=0.48\textwidth]{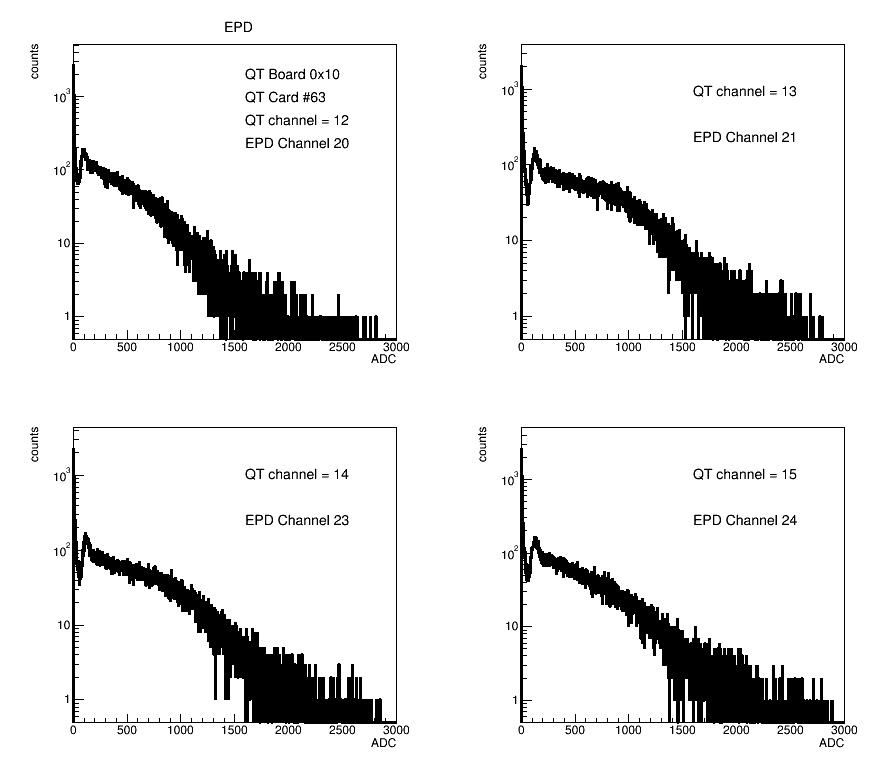}}
\end{center}
\caption{Data from 8 channels of the EPD.  There is no selection on trigger or centrality, so all events are represented here.}
\label{fig:EPD_1DData2}
\end{figure}

\begin{figure}[htbp]
\begin{center}
\mbox{\includegraphics[width=0.48\textwidth]{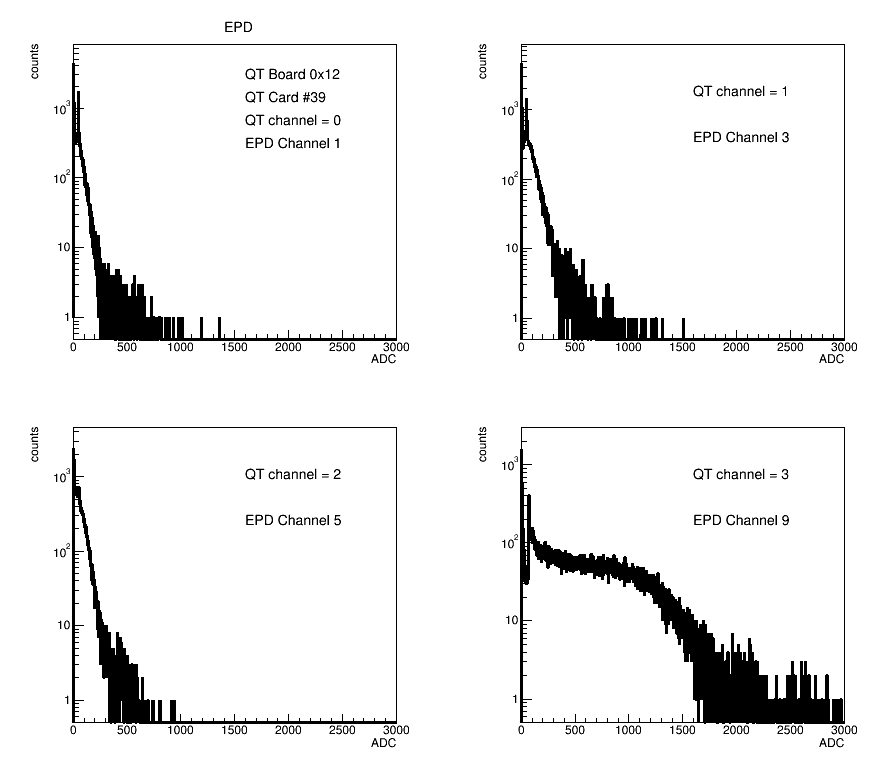}}
\mbox{\includegraphics[width=0.48\textwidth]{Figures/EPDEPD_QT0x12_ADCs2B.png}}
\end{center}
\caption{Data from 8 channels of the EPD.  There is no selection on trigger or centrality, so all events are represented here.  The four plots on the right show the ADC count from the timing output on the channels.}
\label{fig:EPD_1DData3}
\end{figure}

\begin{figure}[htbp]
\begin{center}
\mbox{\includegraphics[width=0.48\textwidth]{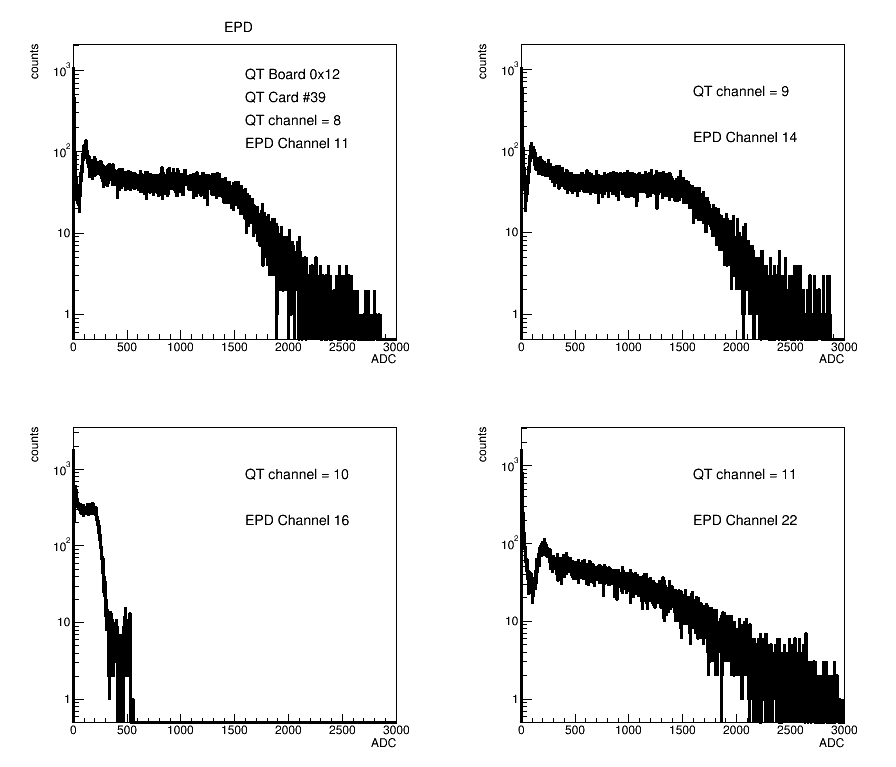}}
\mbox{\includegraphics[width=0.48\textwidth]{Figures/EPDEPD_QT0x12_ADCs2D.png}}
\end{center}
\caption{Data from 8 channels of the EPD.  There is no selection on trigger or centrality, so all events are represented here. The four plots on the right show the ADC count from the timing output on the channels.  You can see the beam structure in the lumps, especially in channel 13.}
\label{fig:EPD_1DData4}
\end{figure}

The ADC distribution from channel 14 is shown on the right side of Figure \ref{fig:EPD_channel_position}.  This is not overlapping data, however there were no trigger or centrality requirements for either measure.  The data for the ADC distributions for all channels are shown in Figures \ref{fig:EPD_1DData1},\ref{fig:EPD_1DData2},\ref{fig:EPD_1DData3}, and \ref{fig:EPD_1DData4} for the 24 channels and the 8 timing channels of the EPD prototype.  There is no centrality or trigger selection, which is indicated in the distributions as they are a convolution of 1, 2, 3 and more hits as well as other processes detailed below.

\begin{figure}[htbp]
\begin{center}
\mbox{\includegraphics[width=0.30\textwidth]{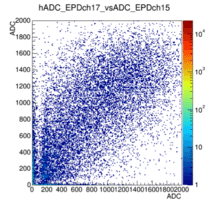}}
\mbox{\includegraphics[width=0.30\textwidth]{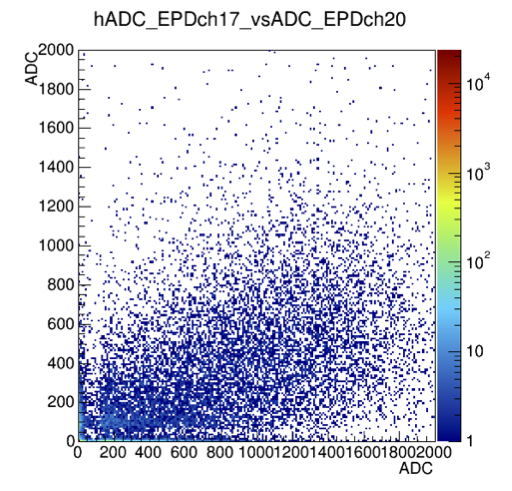}}
\mbox{\includegraphics[width=0.30\textwidth]{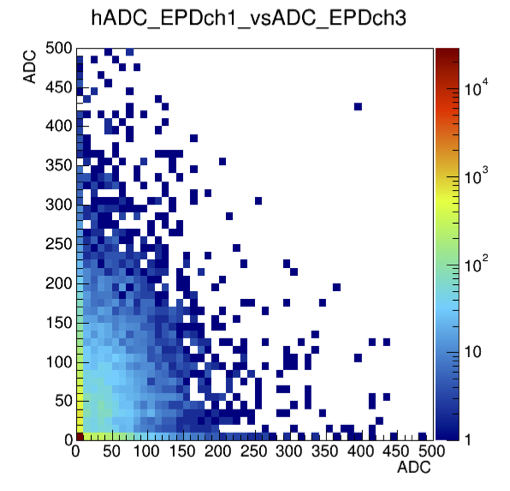}}
\end{center}
\caption{The ADC count in channel 17 versus 15 (large versus large) is shown on the right, the ADC count in channel 17 versus 20 (large versus medium) is shown in the middle and the ADC count in channel 1 versus 3 (small versus small) 
}
\label{fig:EPD_ADCcorr}
\end{figure}

To determine whether we were seeing a correlation of the ADC counts and the centrality of the collision, we compared the ADC values from different channels.  This can be seen in Figure \ref{fig:EPD_ADCcorr}, where the correlation is quite nice between the large tiles and less so for the smaller tiles.  It should be noted here that the small tiles have a single straight WLS fiber in them, a design that has been discarded for the final EPD design.

\begin{figure}[htbp]
\begin{center}
\mbox{\includegraphics[width=0.30\textwidth]{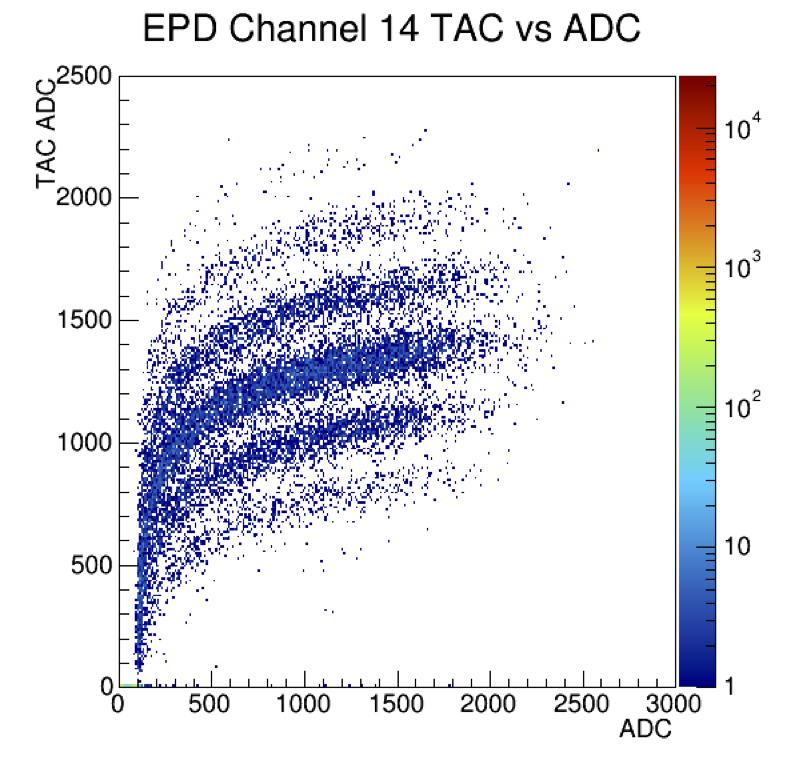}}
\mbox{\includegraphics[width=0.30\textwidth]{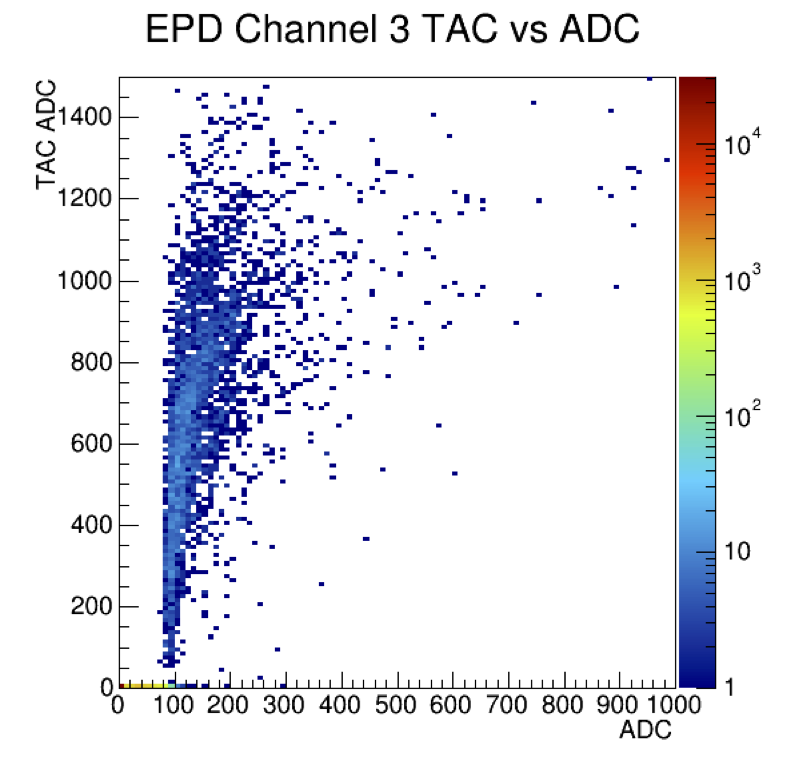}}
\end{center}
\caption{The left plot shows the relationship between the measured ADC and the TAC ADC for channel 14, which is a large tile.  The plot on the right shows the same distribution for channel 3, which is a small tile.
}
\label{fig:EPD_ADCTACcorr}
\end{figure}

We also looked at the channels that have timing capabilities and plotted the ADC of the signal versus the ADC of the timing signal, which is a measure of the time. This result is shown in Figure \ref{fig:EPD_ADCTACcorr}.

\begin{figure}[htbp]
\begin{center}
\mbox{\includegraphics[width=0.30\textwidth]{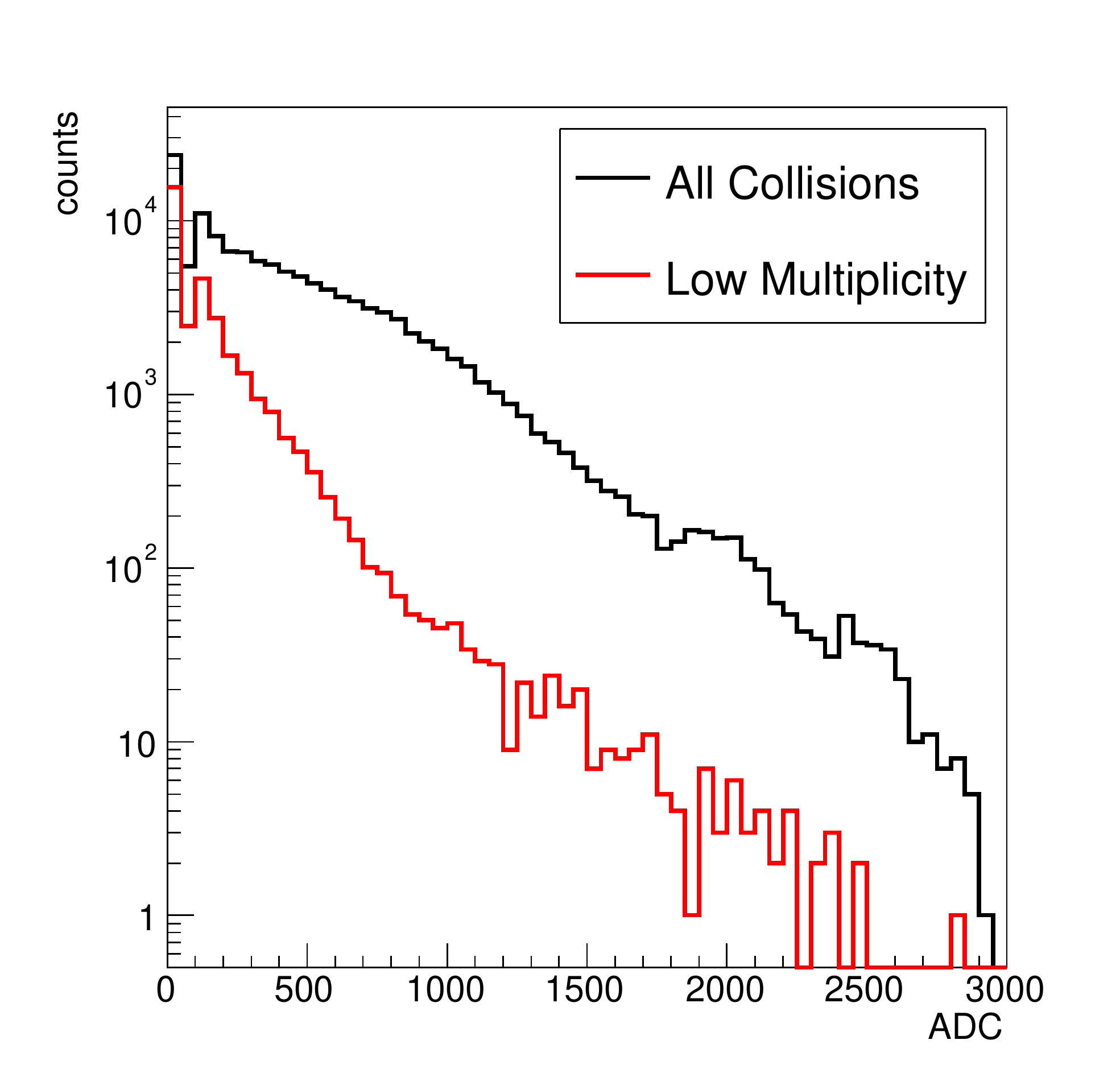}}

\end{center}
\caption{The ADC distribution in channel 24 for all collisions is in black, and for low multiplicity collisions is in red.
}
\label{fig:EPDChannel14_Periperal}
\end{figure}

In order to determine the number of photoelectrons that are produced for a single MIP it was necessary to select collisions where the probability of a double or triple hit is much less than in the ADC distributions shown above.  This was done by selecting events where the ADC count in the large EPD tiles was less than 500, enhancing the probability of selecting a peripheral event.   This result for EPD channel 24, which is a medium sized tile with a sigma groove fiber at the far end of the prototype is shown in Figure \ref{fig:EPDChannel14_Periperal}.

In order to characterize the distribution shown in Figure \ref{fig:EPDChannel14_Periperal}, we need to convolute a Landau distribution for the energy loss, with a Poisson distribution for the probability of measuring the number of photons, a Gaussian distribution to account for the electronic noise and finally an exponential or Gaussian function to account for the pedestal distribution.  This gives us 9 parameters including the overall normalization of the distribution, however, many of these parameters can be determined from physics or from knowledge of the STAR detector performance.

The Landau distribution can be characterized as: $$\cfrac{dP}{dE} = \textrm{Landau}(E,\textrm{MP},\textrm{WD})$$ where MP is the most probable value (the so called location parameter) and WD is the width (scale parameter).  To determine the probability P, we will have to integrate over dE.  The energy loss corresponds to the average number of photons, ${\bar{N}}_{\gamma}$ that we'll see.  This is determined via $${\bar{N}}_{\gamma} = A \times E$$ where $A$ = (energy required to produced a photon)$^{-1}\times$(Efficiency of capturing + transporting + detecting a photon).

Given ${\bar{N}}_{\gamma}$, the Poisson distribution will tell us the likelihood of measuring a number of photons ${N}_{\gamma}$ which is:
$$\cfrac{dP}{d{N}_{\gamma}} = \textrm{Poisson}({N}_{\gamma}, {\bar{N}}_{\gamma})$$
As before, we will need to integrate to determine P, though this time over ${N}_{\gamma}$.

The nominal ADC measured for a given number of photoelectrons is: ADC$_{nom}$ = Gain$\times {N}_{\gamma}$ + Offset.  The gain should be roughly 3, based on previous experience from STAR electronics.  The electronics noise is then characterized as:
$$\cfrac{dP}{d\textrm{ADC}} = \textrm{Gaussian(ADC,ADC}_{nom},\sigma)$$
where $\sigma$ is the with of the QT ADC noise.

Lastly, the pedestal can be characterized as:
$$\cfrac{dP}{d\textrm{ADC}} = {A}_{ped}{e}^{-\textrm{ADC}/{\tau}_{ped}}$$
where ${A}_{ped}$ is the amplitude and roughly proportional to the number zero hit events and ${\tau}_{per}$ is the width.

Combining these together, the resulting equation is:

$$P(\textrm{ADC}) = \textrm{Norm}[\textrm{PED}(\textrm{ADC},{A}_{ped},{\tau}_{ped}) + $$
$$\textrm{~~~}\int^{\infty}_{0} dE \cfrac{dP}{dE}(E,MP,WD)  \int^{\infty}_{0} d{N}_{\gamma} \cfrac{dP}{d{N}_{\gamma}}({N}_{\gamma},AE)\textrm{exp}\biggr(\cfrac{{[\textrm{ADC}-(\textrm{Gain}\times{N}_{\gamma} + \textrm{Offset}]}^{2}}{2{\sigma}^{2}}\biggr)]$$

Unfortunately in the distribution shown in Figure \ref{fig:EPDChannel14_Periperal} we will have 2-hit and 3-hit events. This needs to be accounted for in the equation, adding two more parameters: $\cfrac{{N}_{2-hit}}{{N}_{1-hit}}$ and $\cfrac{{N}_{3-hit}}{{N}_{1-hit}}$.

The 11 parameters of the distribution are (Note, values set by data were fit by eye):

\begin{enumerate}
  \item ${A}_{ped}$ - Set by the data
  \item ${\tau}_{per}$ - Set by the data
  \item MP$_{\textrm{Landau}}$ = 1
  \item WD$_{\textrm{Landau}}$ = 0.2
  \item ${N}_{\gamma}/$unit energy 
  \item Gain$_{\textrm{ADC}}$ = 3
  \item Offset$_{\textrm{ADC}}$ = 0
  \item Noise$_{\textrm{ADC}}$ = 1 - 3
  \item Norm - Set by the data
  \item TurnOnOneHit = 1 (Needed to plot 1-hit, 2-hit, 3-hit contributions separately)
  \item $\cfrac{{N}_{2-hit}}{{N}_{1-hit}}$ = 10\%
  \item $\cfrac{{N}_{3-hit}}{{N}_{1-hit}}$ = 5\%
\end{enumerate}

The energy scale is in arbitrary units (which matters for parameters 2 and 3), but what matters is the ratio WD/MP, which should be between 0.18 - 0.22 for a 1 cm thick scintillator.  Setting MP = 1 means that E = 1 is the most probable value of the energy deposited by 1 MIP.  This means that MP for 2 MIPS is 2MP and for 3 it is 3MP and so on.  WD will be the same for all distributions.

\begin{figure}[htbp]
\begin{center}
\mbox{\includegraphics[width=0.76\textwidth]{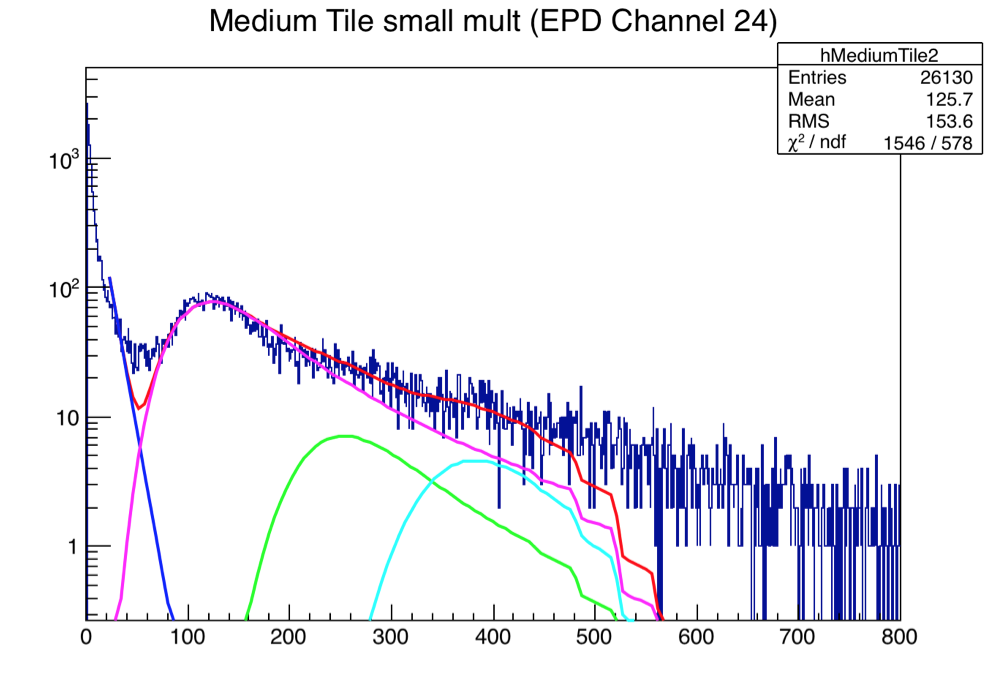}}
\end{center}
\caption{The ADC spectrum for channel 24, which had the following fit parameters not set above: ${p}_{1} = 0.05$, ${p}_{2} = 10$, ${p}_{5} = 42$, ${p}_{9} = 2.4e4$.
}
\label{fig:Channel24_ADCSpectrumFit}
\end{figure}

Parameter t, the average number of photoelectrons that we detect for an MIP, is what we are most interested in finding out.  This was measured to be around 40 in the lab, which is in good agreement with the fit value of 42 which was determined by a fit of the ADC spectrum from channel 42 as shown in Figure \ref{fig:Channel24_ADCSpectrumFit}.

The ADC structures are large, on the order of 50, so parameter 8 is irrelevant.

The relationship between these parameters and the structures seen in the distribution are:

The positions of the peaks is determined by: $\biggr(\cfrac{{N}_{\gamma}}{\textrm{MIP}}\biggr)\times {\textrm{Gain}}_{\textrm{ADC}}$

The widths of the peaks is determined by: $\biggr(\cfrac{{N}_{\gamma}}{\textrm{MIP}}\biggr)\times {\textrm{Gain}}_{\textrm{ADC}},$ WD $\times$ Gain$_{ADC}$, Noise$_{\textrm{ADC}}$.  (Though the last term is irrelevant).  

The first and second terms are related to the energy-loss-fluctuations and the finite ${N}_{\gamma}$ statistics respectively.

The fit for channel 24, a medium sized tile with a sigma grove, is shown in Figure \ref{fig:Channel24_ADCSpectrumFit}.  This fit yielded 42 photons per MIP, in good agreement with the lab studies that have been done.

\subsubsection{Number of photons per MIP and SiPM saturation}

\begin{figure}[htbp]
\begin{center}
\mbox{\includegraphics[width=0.76\textwidth]{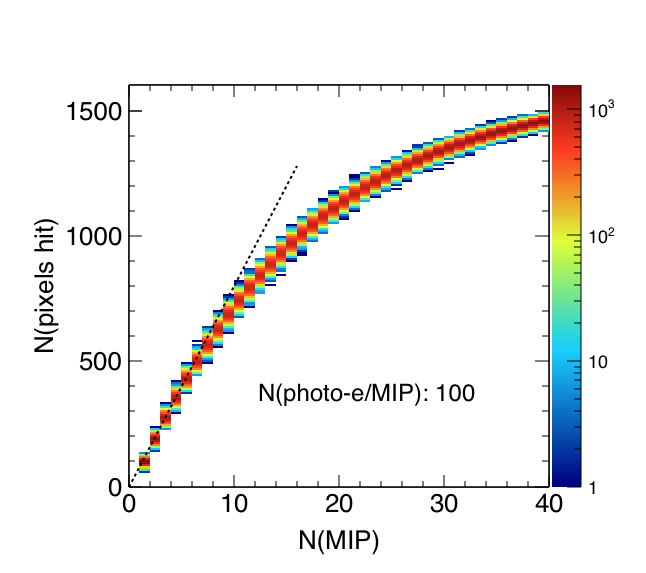}}
\end{center}
\caption{Number of hit SiPM pixels in clear fiber acceptance as a function of MIPs. A non-linear response is expected after about 10 MIPs.
}
\label{fig:SiPM_saturation}
\end{figure}

\begin{figure}[htbp]
\begin{center}
\mbox{\includegraphics[width=0.76\textwidth]{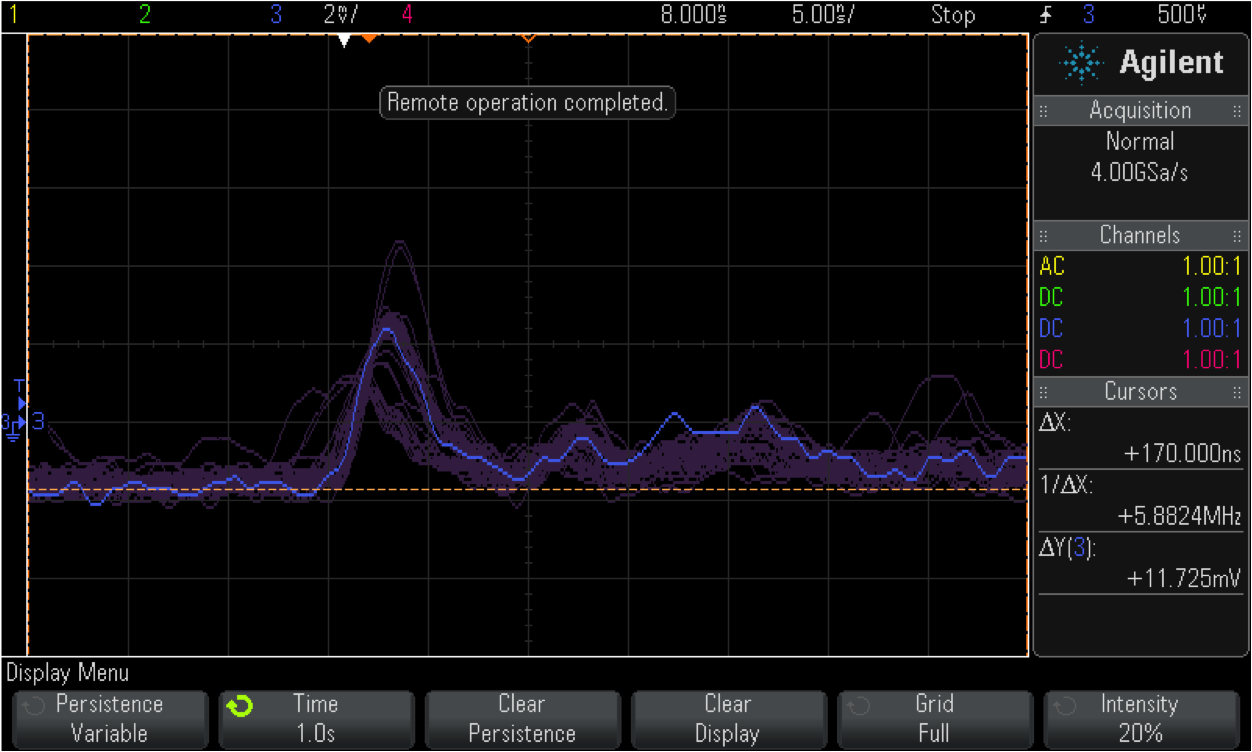}}
\mbox{\includegraphics[width=0.76\textwidth]{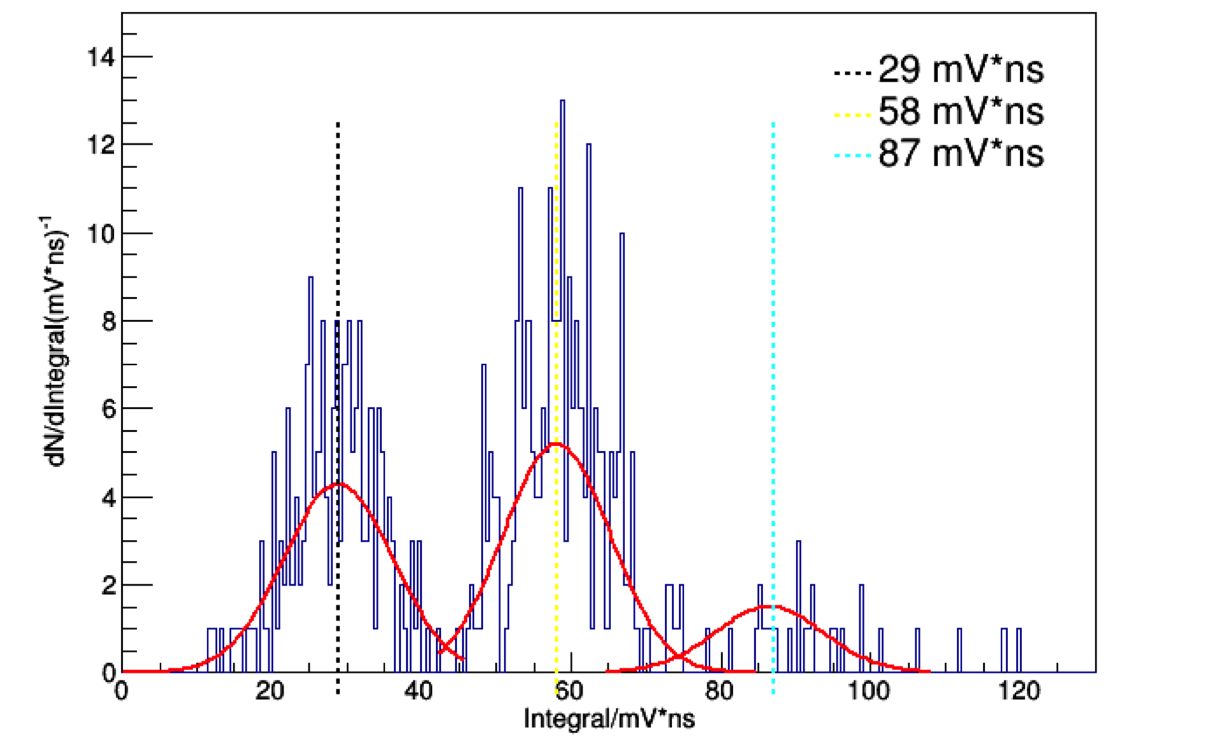}}
\end{center}
\caption{Top: Dark noise of the SiPMs, showing single, double and triple pixel signals. Bottom: Integrated dark noise signals, showing the peaks
for single, double and triple pixels.
}
\label{fig:SiPM_signals}
\end{figure}

Signals of bare SiPMs, not connected to any detector, were measured in the lab to get the energy information for single fired pixels. Fig.\ref{fig:SiPM_signals} shows the 
dark noise signals on the scope (top) and the integrated signals (bottom). The integrated double and triple signals scale as expected. This gives an estimate for the signals of 
single, double and triple photons. We divided the average signal of a triple WLS layer, 1cm scintillator tile design by the integrated single dark noise signal and get
~250 photoelectrons/MIP.
The prototype has only a single WLS fiber loop, which reduced the light output by about a factor of 2. The coupling to clear fibers for the prototype has an efficiency of 0.6.
The attenuation due to WLS fibers (0.4 m attenuation length) and clear fibers (10 m attenuation length) is 0.5. All factors together result in an estimated number of 40 photoelectrons/MIP for the prototype, which is in agreement with the extracted number from section \ref{sec_proto_data}.
For the EPD we are going to use the S13360-1325PE type SiPMs with a pixel pitch of 25$\mu$m and clear fibers with 1.15 mm diameter. That results in about 1600 pixels on the fiber area. The coupling efficiency is assumed to be at least 80\%, the attenuation for WLS and clear fiber is 0.5, which results in about 100 photoelectrons/MIP. That means at maximum 16 MIPs can be detected in the extreme case. Figure \ref{fig:SiPM_saturation} shows a MC calculation of the SiPM saturation as a function of MIPs hitting the scintillator. A non-linear response is expected after about 10 MIPs.

\subsubsection{Timing Resolution}

To determine the timing resolution of the EPD we measured an assembled test tile with a cosmic ray telescope setup using two larger scintillators for triggering. The top trigger scintillator, channel 1, was a 10 cm x 10 cm x 1 cm tile with a Kuraray Wavelength Shifting (WLS) fiber embedded in a spiral shape. The bottom trigger tile was the same size as the top with a sigma shaped embedded WLS fiber serving as channel 2. The middle tile measured as channel 3 was identical to a medium-sized tile from a sector of the EPD: a 1 cm thick trapezoid with a WLS fiber embedded in a triple layer sigma shape. The measurement was taken with Hamamatsu C12332-01 Multi-Pixel Photon Counter (MPPC) power supply boards and a 2.5 GHz oscilloscope.

During the measurement the trigger level for all three tiles was set to 6.0 mV and 1,000 events were recorded.  Any channel voltage larger than 6.0 mV was considered the start of a good signal and the time was recorded for the three readout channels as $T_{1}$, $T_{2}$ and $T_{3}$ respectively.

\begin{figure}
\centering
\includegraphics[width=0.8\textwidth]{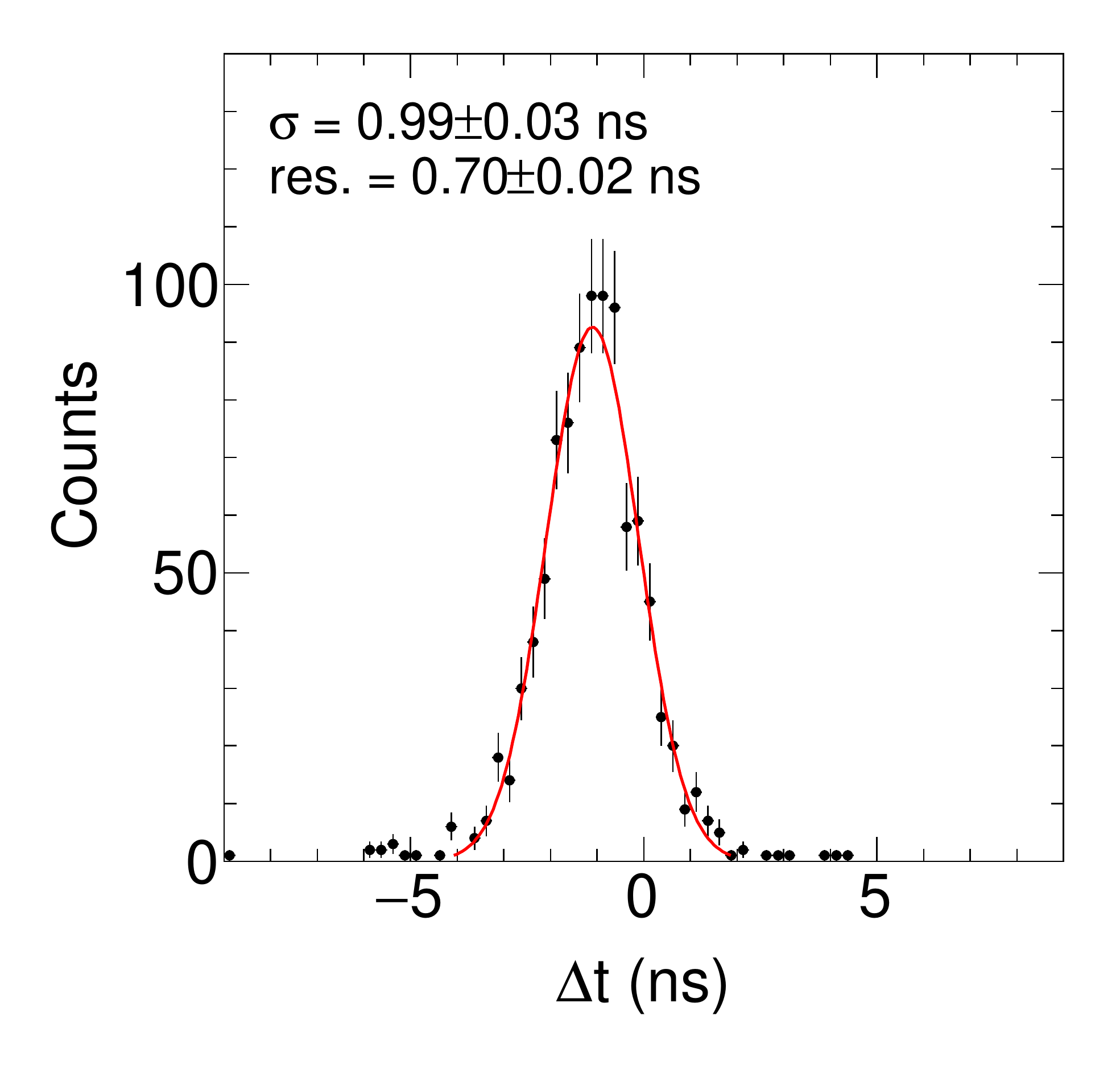}
\caption{\label{fig:timingres}The gaussian distribution of the time difference between the test tile and trigger 1 ($T_{1}-T_{3}$) from which the timing resolution of the EPD is calculated.}
\end{figure}

The distribution of the time difference is shown in figure \ref{fig:timingres}. The time difference distributions for $T_{1}-T_{3}$ and $T_{2}-T_{3}$ are similar and therefore only the distribution for $T_{1}-T_{3}$ is shown. The distributions were fit with a gaussian and the sigma value was recorded as 0.99 ns. For the trigger reference we calculated the distribution for $T_{1}-T_{2}$, which was similar to the other distributions. 

The timing resolution is calculated as $\sigma / \sqrt{2} = 700$ ps for the triple layer fiber test tile. A second set of measurements was taken with the same setup and parameters for a test tile with only a single layer WLS fiber in an embedded sigma shape. For this second measurement $\sigma = 0.95\pm 0.02$ ns and the timing resolution is calculated to be 670 ps. This shows that there is no significant difference between the single layered tile and the triple layered tile in terms of timing resolution. The timing resolution for the BBC detector is on the order of 1.0 ns, which is significantly higher than the projected resolution of the EPD. In conclusion, the EPD can successfully replace the BBC as a trigger detector.

\subsection{Scintillator Construction}
The final design for the EPD consists of two disks (wheels) made of 1.2 cm thick Eljen EJ-200 plastic scintillator with an inner radius of 4.5 cm and an outer radius of 90 cm.  Each disk will be composed of 24 sectors, for a total of 48 sectors.  A super sector will consist of two sectors, for a total of 12 super sectors per wheel.  A single piece of plastic scintillator can be used to create two super-sectors.  Each super sector will be divided into 31 separate tiles.  The inner-most tile, extending from a radius of 4.5 to 8.5 cm, will span the entire super sector.   All subsequent tiles will only span a sector.

In order to ensure that the tiles are optically separated, 1.65 mm wide grooves will be milled down 50\% to create the individual tiles.  This groove will be painted and filled with glue.  The super-sector will then be turned over, and the procedure will be repeated on the other side, which will result in tiles that are completely optically separated.  A channel for the wave-length shifting fibers will be milled down the center of the super-sector.  This channel will start at a depth of 0.16 cm directly after the first channel and smoothly ramp to 0.6 cm at the end of the scintillator.  The width of the channel will start at 0.15 cm and increase by 0.30 cm for each set of tiles it passes until it reaches a width of 1.65 cm a distance of 29.4 cm from the tip.

The grooves for the WLS fibers will be in the configuration we have called the "sigma groove".  This maximizes the light collection by increasing the amount of fiber contained within an individual tile.  The fiber will be coiled around three times in each tile, so the groove will be 1.6mm wide and 3.6mm deep.

\subsection{Fiber preparation}
The EPD uses two different optical fibers:  Kuraray Y-11(200) polystyrene Wavelength Shifting (WLS) and Kurray 3HF clear optical fiber.  We will use 1.0 mm WLS fibers, which is the same as the EPD prototype.  The clear fibers will be increased to 1.15 mm in order to facilitate the fiber-to-fiber connection.  The WLS fiber has a maximum bending radius of 2.5 cm, which was a design consideration in the tile size.  At Lehigh University the fibers will be glued into their connectors and then cut using a diamond knife.  After which, the polishing procedure outlined in Section \ref{sec_fiberprep} will be followed.  The ends of the WLS fibers that will be embedded into the scintillator will be painted with  painted with Eljen: EJ-510 reflective paint to redirect the photons into the correct direction. 

\subsection{Fiber Connector}
\label{sec_fiberconnector}
\begin{figure}[htbp]
\begin{center}
{
\mbox{\includegraphics[width=0.30\textwidth]{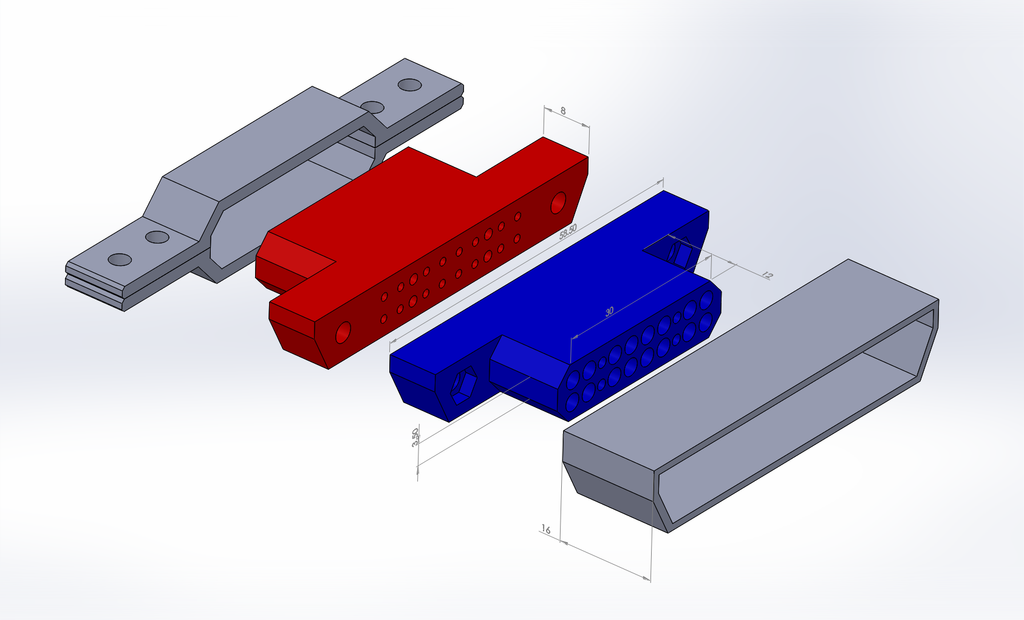}}
\mbox{\includegraphics[width=0.30\textwidth]{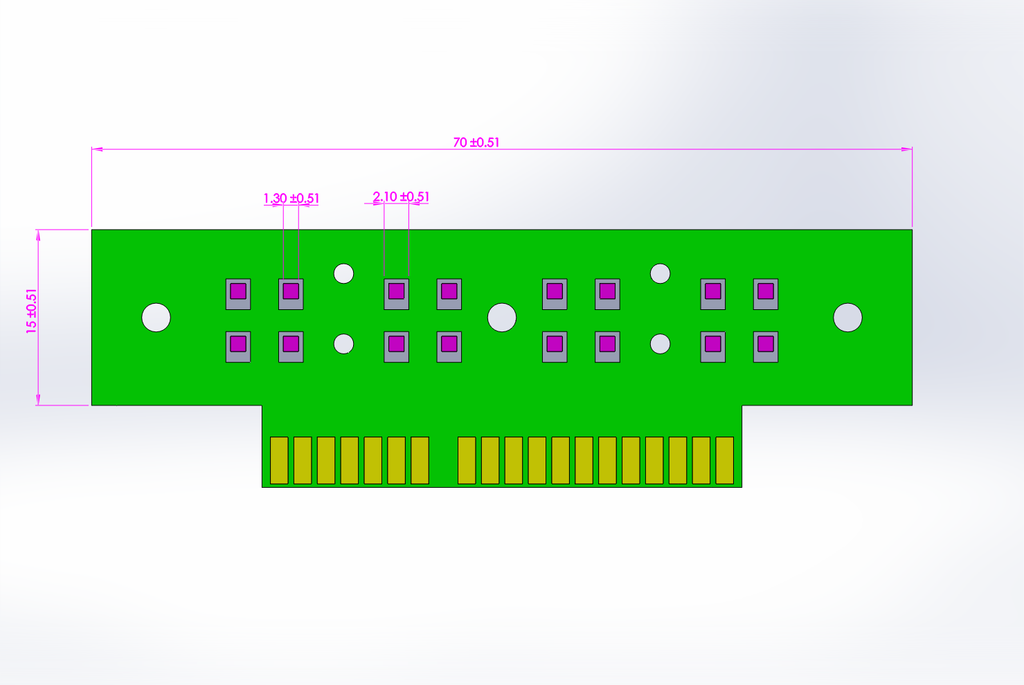}}
\mbox{\includegraphics[width=0.30\textwidth]{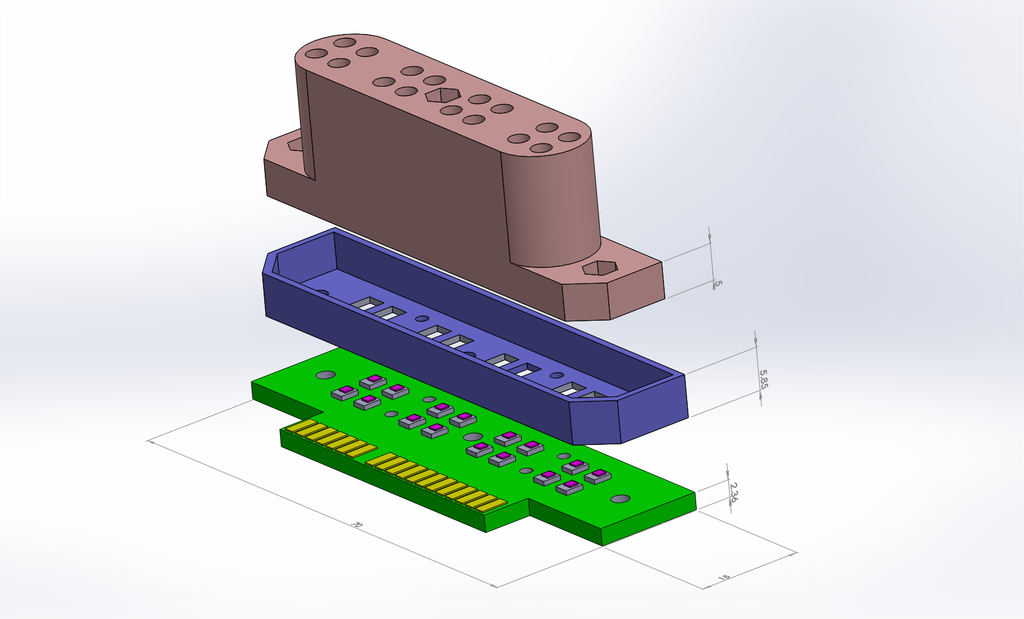}}
}
\end{center}
\caption{On the left is the fiber-to-fiber connector, including the light tight sleeve.  The middle figure shows the SiPM board, which will contain 16 SiPMs.  The bottom part of the board will attach to the FEE board that will be used to read out the signals.  The right figure shows the SiPM to fiber connector.
}
\label{fig:EPDConnectors} 
\end{figure}

There are two types of connectors that are needed for the the EPD detector.  One will connect the WLS fibers coming from the scintillator to the clear optical fibers that take the signal to the read-out boards.  This can be seen on the left in Figure \ref{fig:EPDConnectors}.  These connectors will hold 16 fibers and will be printed from polylactic acid (PLA) at LBNL.  Each super-sector will require 2 fiber-to-fiber connectors.  In addition to the fiber holder, there is a sleeve that slides down over the connection to ensure light-tightness.  The fiber-to-SiPM connector shown on the right of Figure \ref{fig:EPDConnectors}  will include the SiPM board shown in the middle of Figure \ref{fig:EPDConnectors}.

 \begin{figure}[htbp]
\begin{center}
{
\mbox{\includegraphics[width=0.45\textwidth]{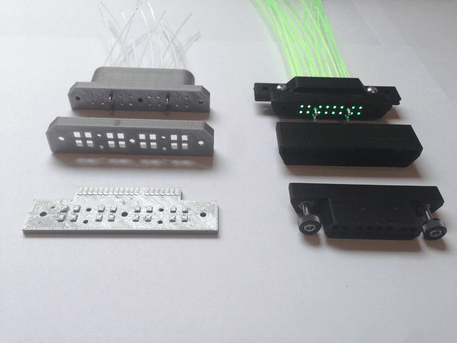}}
\mbox{\includegraphics[width=0.45\textwidth]{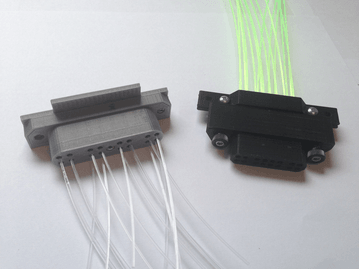}}
}
\end{center}
\caption{Prototypes of the fiber-to-fiber and fiber-to-SiPM board connectors.
}
\label{fig:EPDConnectorsReal} 
\end{figure}

\begin{figure}[htbp]
\begin{center}
{
\mbox{\includegraphics[width=0.9\textwidth]{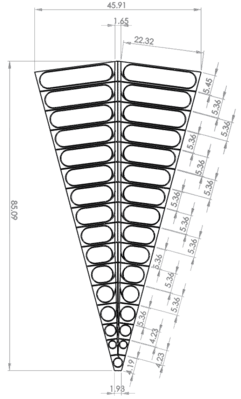}}
}
\end{center}
\caption{Schematic drawing of the super-sector with dimensions in cm.
}
\end{figure}

\begin{figure}[htbp]
\begin{center}
\includegraphics[width=0.48\textwidth]{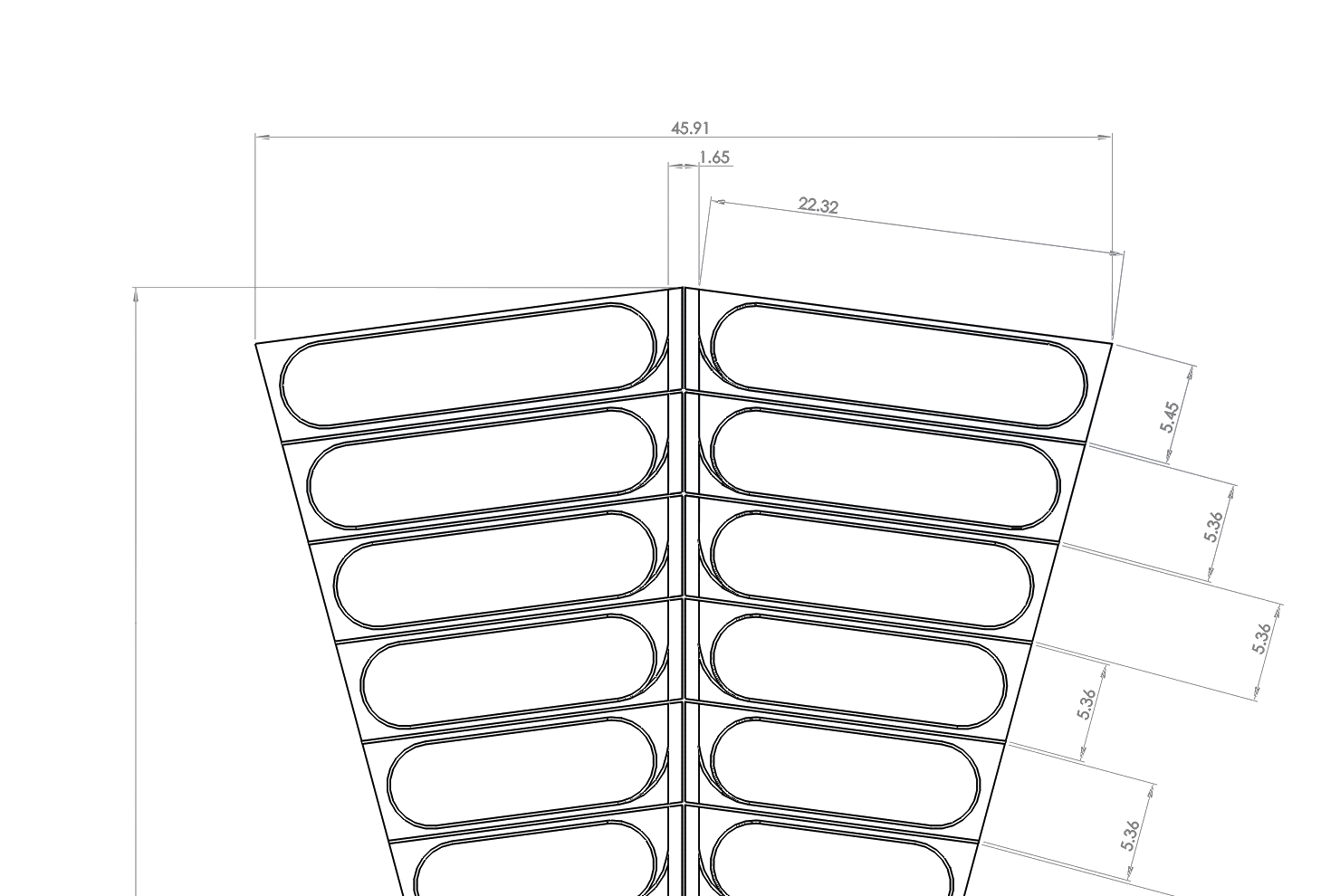}
\includegraphics[width=0.48\textwidth]{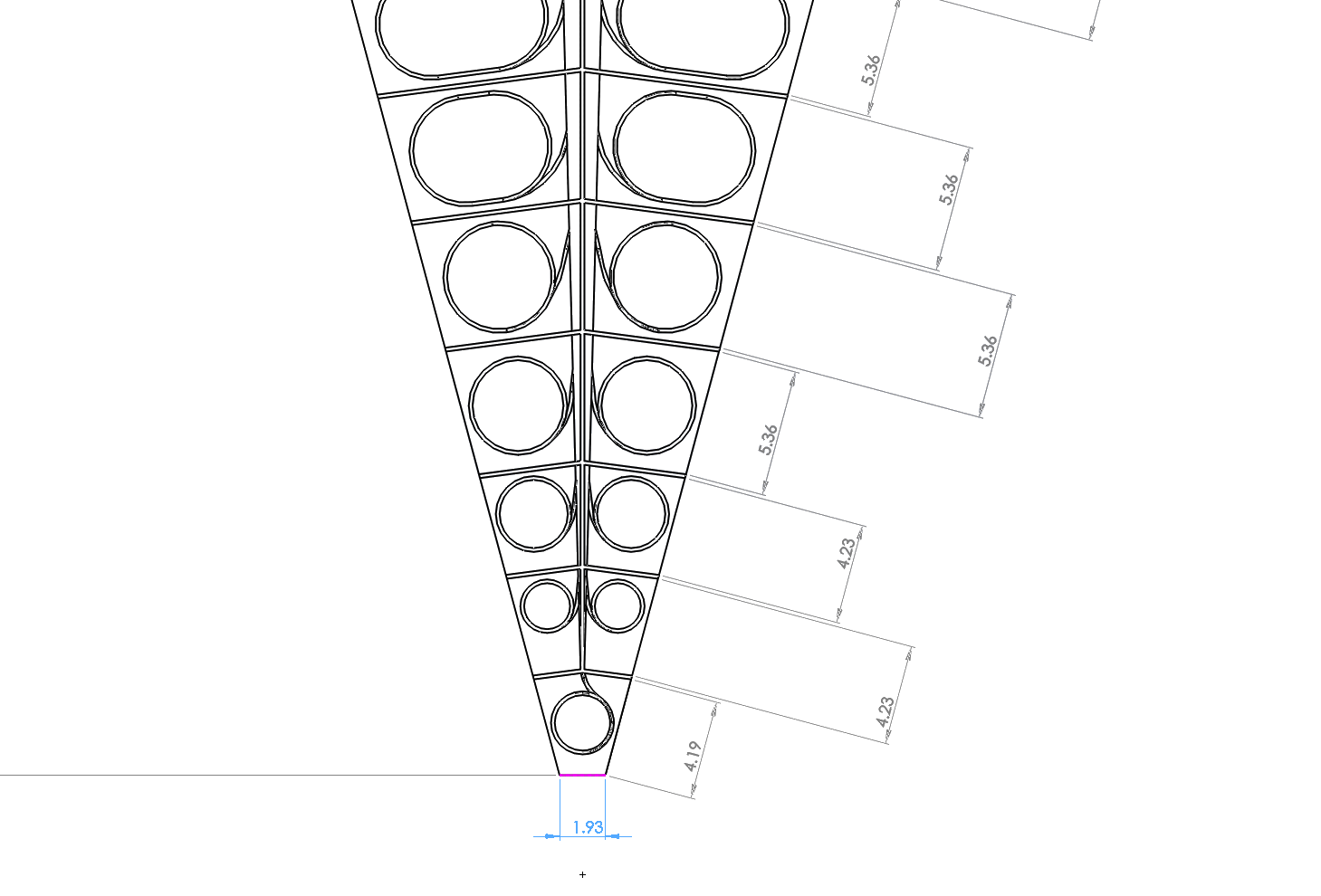}
\end{center}
\caption{Detailed schematic drawings of the upper and lower parts of the super-sector with dimensions in cm.
}
\end{figure}

\subsection{Super-Sector Final preparation}
The edges of the scintillator will be polished using aluminum oxide powder.  Then the polished WSL fibers will be glued into the super sectors using EJ-500 optical cement.  At this point, the entire super sector assembly will be wrapped in Tyvek 1055B, which has a lower probability of scratching the scintillator surface than the aluminized mylar used in the prototype.  Lastly, the entire assembly will be wrapped in Light-Absorbing Black-Out Paper from McMaster.

\subsection{Electronics}
\label{sec_electronics}

\begin{figure}[htbp]
\begin{center}
{
\mbox{\includegraphics[width=0.5\textwidth]{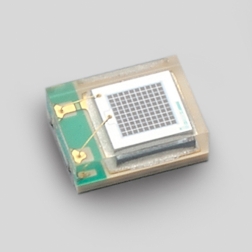}}
}
\end{center}
\caption{A 1.3x1.3mm Hamamatsu S13360-1325PE MPPC (SiPM) as it will be used for the EPD.
}
\label{fig:SiPM}
\end{figure}

Signals will be read out by 1.3x1.3mm Hamamatsu S13360-1325PE MPPC (SiPM), see Fig. \ref{fig:SiPM}, which will be attached to the optical fibers coupled to the wavelength shifting fibers from the scintillator.  The SiPMs will be connected to a FEE board developed for the forward calorimeter upgrade and for the FPS by Gerard Visser. The FEE board requires a single multi-drop flat cable to bring in LV (5V and -90V) as well as slow control bus, and it will have a SMB connector for the signal output. The signals will be readout by existing QT boards as a part of STAR trigger system. 
Gerard will design an adapter to interface between the twisted pair cables and the QT boards to make differential output possible from the FEE cards.

\subsection{Mechanical Support}
\label{sec_mechanical}

\begin{figure}[htbp]
\begin{center}
{
\mbox{\includegraphics[width=0.65\textwidth]{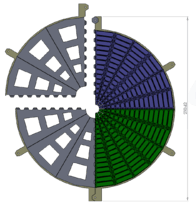}}
}
\end{center}
\caption{Schematic diagram of the mechanical support structure for mounting the EPD as viewed from the nominal collision point.  Two quadrants are shown fully populated with three EPD supersectors each (green and blue). 
One quadrant (upper left) is shifted for clarification.
}
\label{fig:MechanicalSupport}
\end{figure}

The mechanical support structure for mounting the EPD in STAR IR is shown in
Figure \ref{fig:MechanicalSupport}. The frame consists of four interlocking quadrants milled from 3/8 inch
thick fiberglass-reinforced epoxy laminate sheets, FR-4, by BNL Central Machine
Shop. This material is stiff enough to support three EPD supersectors in each
quadrant, and the resin binder in FR-4 is especially flame resistant. To reduce the
material budget in front of the VPD, the EPD frame will not extend beyond the VPD
acceptance, constraining the inner radius of the frame. Cutouts in the EPD frame
will reduce its weight without compromising the overall structural integrity. The
EPD support structure is designed to attach to the same six mounting studs on the
STAR magnet pole tip as the BBC detector array that it will replace during BES II.
Modularity (easy removal and reinstallation) of an EPD quadrant fully loaded with
supersectors, along with reproducibility of the mounting position over the full BBC
acceptance were major considerations in the design done by LBNL and BNL.

\subsection{Triggering for BESII}
\label{sec_Triggering}

The EPD will replace the BBCs as the principle trigger for the BES-II program. Therefore,
it is important to consider the requirements for functionality that this places on the
system. The BES-II program will need a minimum bias trigger in order to meet several of
the physics goals. The physics goals that specifically require a minimum bias trigger
are high $p_T$ hadron suppression, directed flow, and the correlations necessary to study
the chiral magnetic effect. In addition, the trigger must be able to resolve Au+Au collider
events from backround events. Finally, the EPD will be necessary as a trigger for the
fixed-target program.

\begin{figure}[htbp]
\begin{center}
{
\mbox{\includegraphics[width=0.65\textwidth]{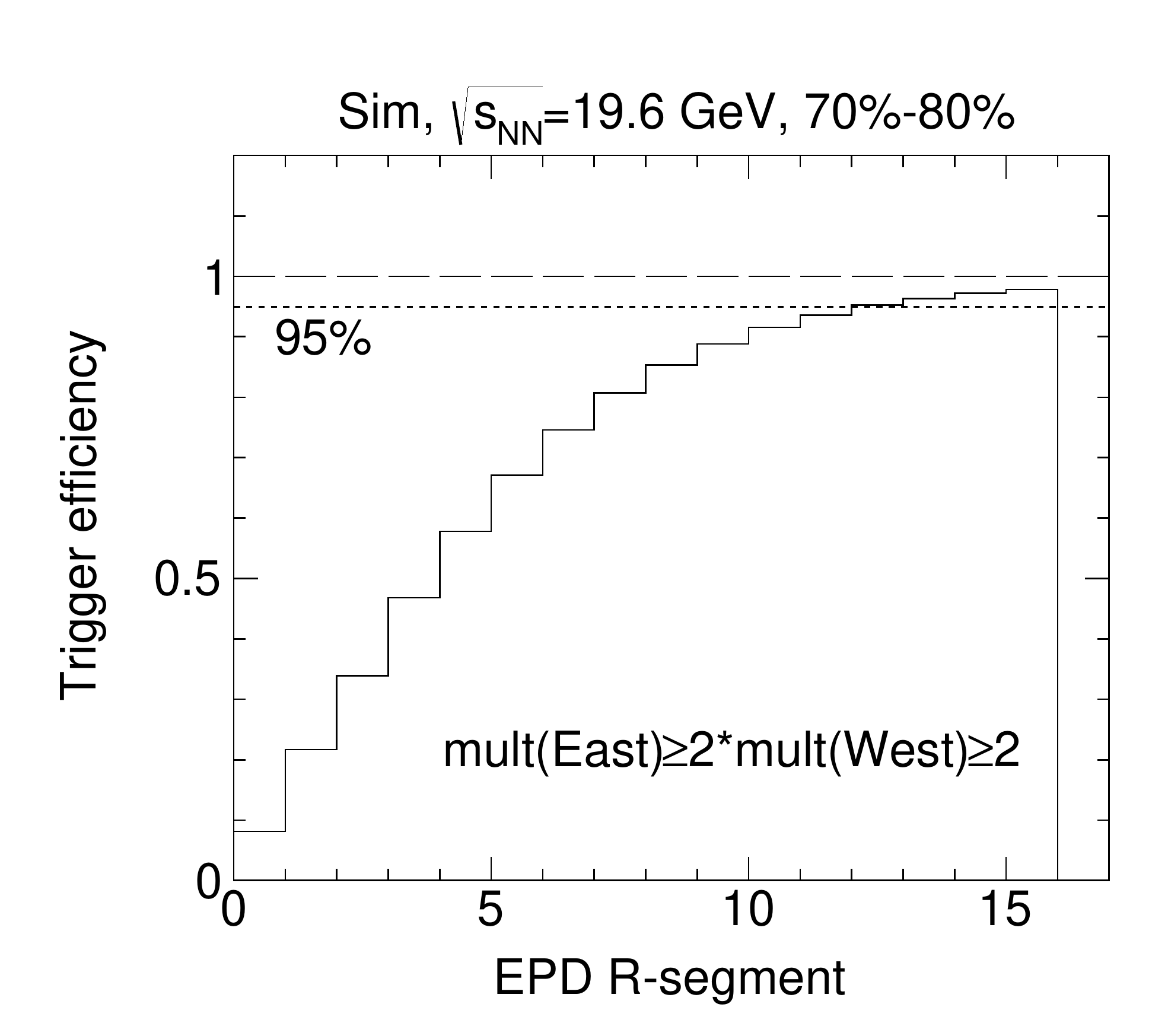}}
}
\end{center}
\caption{Trigger efficiency for peripheral collisions at $\sqrt{{s}_{\textrm{NN}}} = 19.6$ GeV versus the R segment of the EPD, with a trigger requirement of at least two his in both the east and the west.
}
\label{fig:EPDTriggerEff}
\end{figure}

The minimum bias trigger requirement places constraints on the solid angle coverage
and the rise time and timing resolution of the EPD system. The EPD will cover the
psuedorapidity range from $2.1 < |\eta| < 5.1$ . For the energies proposed for the BES-II 
program, the beam rapidities will range from $y = 2.1$ to $y = 3.05$, however rapidity
and psuedorapidity are quite different for protons. Due to their fermi momentum of the 
spectator protons and fragments will spread only a few tenths of a degree off the beam 
axis and will therefore not be incident on the inner tiles of the EPD, which extend in to 0.7
degrees. Therefore, the trigger forward protons and pions from the interaction region. The
solid angle coverage of the EPD implemented in the trigger system must be sufficently large 
to provide a highly efficent trigger for all beam energies and centralities. Using the
published $dN/d\eta$ distributions from PHOBOS for 19.6 GeV Au+Au collisions, we can
estimate the trigger efficiency for peripheral events as a function of the the number
of EPD radial segments included at the trigger level. For 95\%
efficiency for peripheral events, data from at least 12 radial segments would be needed 
(see Fig. \ref{fig:EPDTriggerEff}). For this study, we assumed at least two hits on both the 
East and West EPD. 

\begin{figure}[htbp]
\begin{center}
{
\mbox{\includegraphics[width=0.80\textwidth]{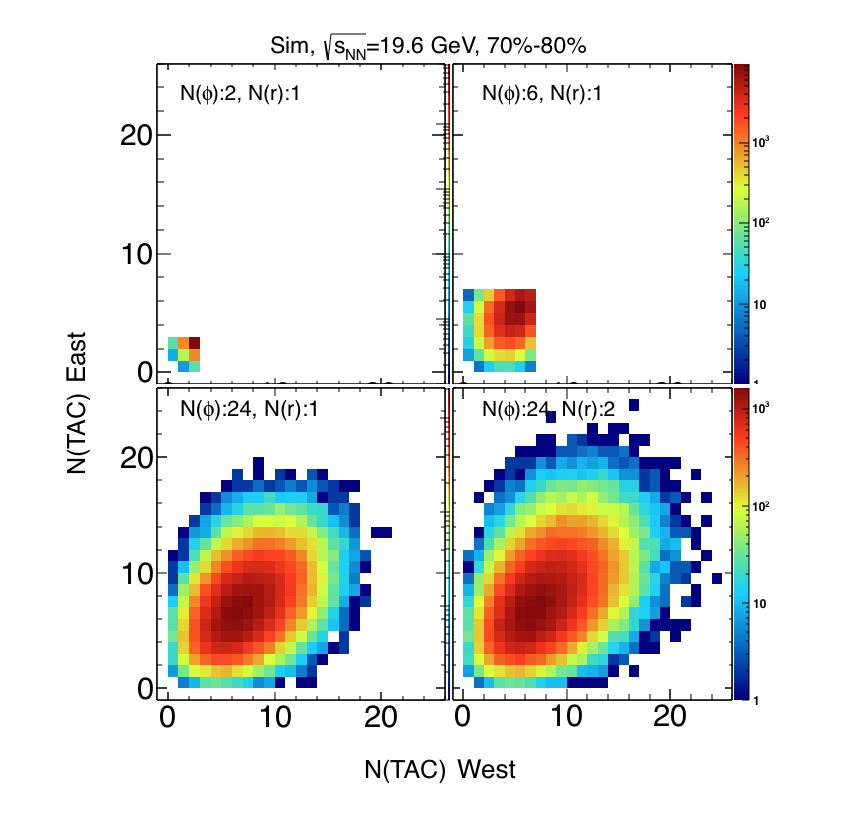}}
}
\end{center}
\caption{Number of fired TAC channels on the East side versus West EPD side for four different configurations. The input
for the simulation are $\sqrt{s_{NN}}$ = 19.6 GeV peripheral events.
}
\label{fig:EPDTriggerTAC}
\end{figure}

\begin{figure}[htbp]
\begin{center}
{
\mbox{\includegraphics[width=0.65\textwidth]{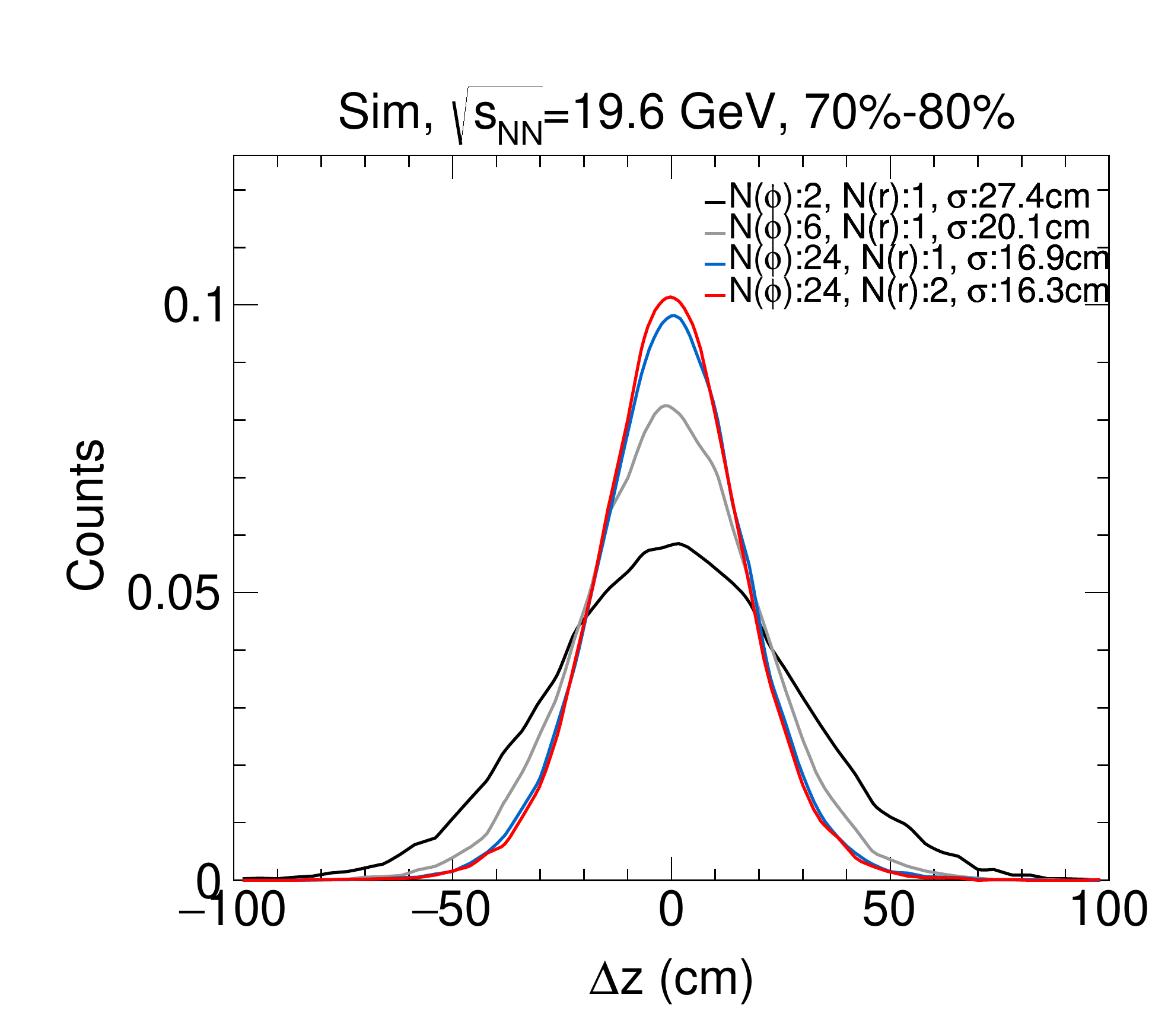}}
\mbox{\includegraphics[width=0.65\textwidth]{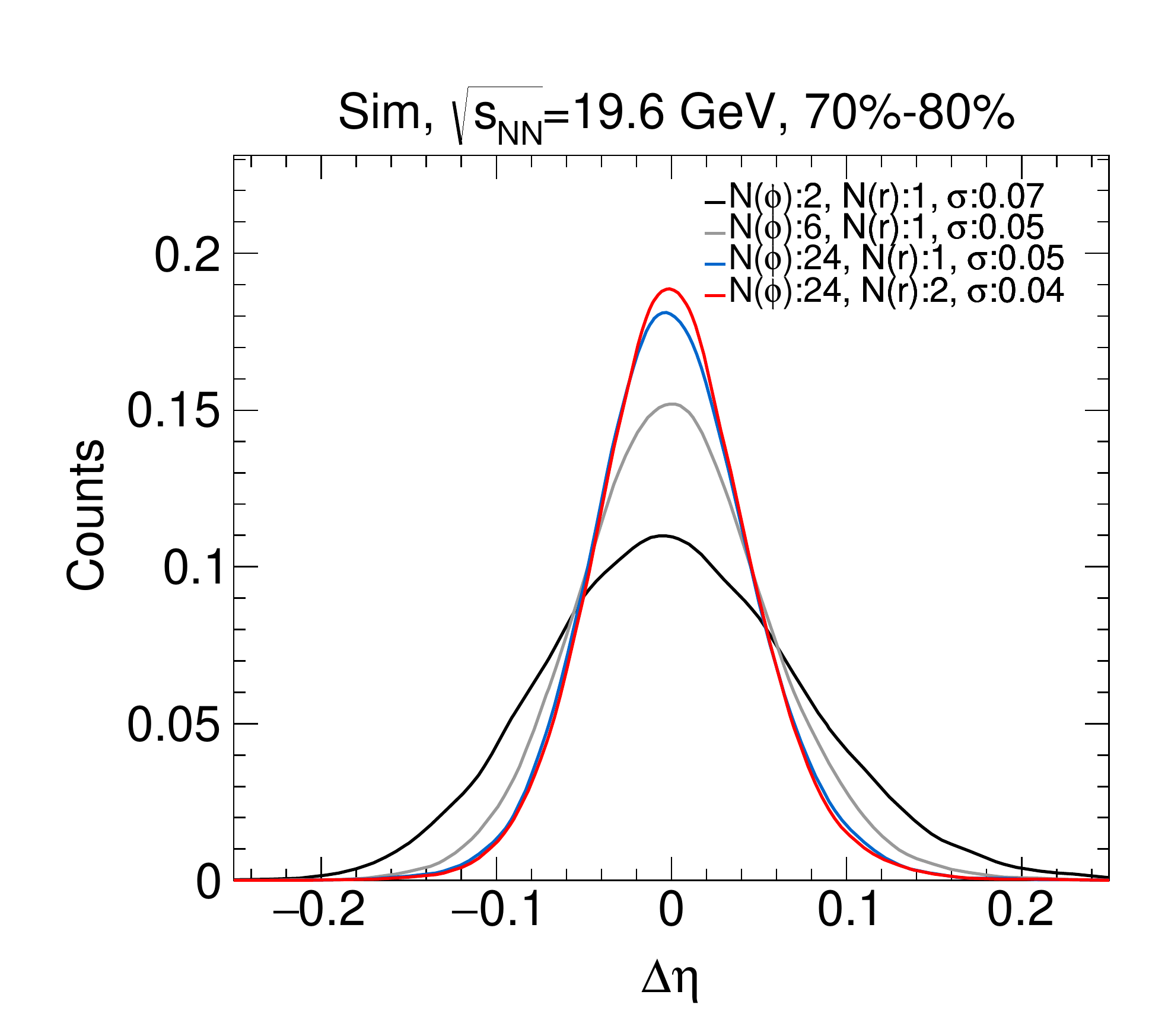}}
}
\end{center}
\caption{z-vertex and $\eta$ resolution for four different configuration of the EPD TAC channels. The input
for the simulation are $\sqrt{s_{NN}}$ = 19.6 GeV peripheral events.
}
\label{fig:EPDTriggerRes}
\end{figure}

In order to maximize the probability of multiple timing signals, the  EPD TACs should be 
distributed azimuthally. Each EPD slat should have a single TAC signal. The rise time and 
timing resolution are set by the coincidence window of the trigger. The width of
this window is set by the longitudinal extent of the bunches. For
BES-II, CAS have proposed to increase the luminosity by stretching out the bunches in
$z$. This will have the effect of distributing the interaction vertices between $
-1.0 m < V_z < +1.0 m$. A vertex offset of one meter East (West) will create a timing offset
of 6 (-6) ns, requiring at least a 12 ns coincidence window. A 24 ns coincideince window would
allow interactions across the entire volume of the TPC. Thus the
timing requirements for the EPD are simply that the rise time and resolution be small
compare to 12 (or 24) ns. The 2016 tests have demonstrated a 700 ps timing resolution, which
is sufficent for an efficient minimum bias trigger.

More stringent requirements for the EPD system are imposed by the need to reject
background events during the BES-II program. During the BES runs in 2010 and 2011,
most of the triggers which were recorded were background collisions between beam
halo and portions of the beam pipe. This was not a problem for BES-I, because the
DAQ system had enough bandwidth to record all triggers and the background could be
efficiently rejected offline. However this will no longer be the case with the increased 
luminosities expected for BES-II. For the 7.7 GeV colliding system, during BES-I, we 
recorded good events at an average rate of 6 Hz. At the start of each twenty minute 
fill, background triggers were being recorded at a rate of 75 Hz. Generally, a single 
collision between one ion in the beam halo and the material of the beam pipe would 
not fire the trigger, as it requires an East and West coincidence. Therefore, one 
component of the background comes from a random coincidence between one East-going
and one West-going background collision. It may have been possible for a single background
event at $V_z > 5$ m to fire both the East and West BBCs (assuming the
trigger coincidence window was larger than 24 ns). Further analysis of the BES-I data is
needed to determine the relative contributions of these two trigger mechanisms; however,
as a worst case scenario, we will assume all the background to come from random
East-West coincidences. For random coincidences, the background rate should increase as the
square of the lumininosity. The expected luminosity increase for 7.7 GeV in
BES-II is a factor of four, therefore we should be prepared for a background trigger rate of
$ 4^2 \times 75 = 1200$ Hz. This is still no a problem for DAQ1000.
The challenge for the trigger and DAQ systems of STAR come with the higher BES-II
beam energies. For 19.6 GeV, we recorded good events at an average rate of 100 Hz. At the start
of each forty minute fill, we were recording background events at a rate of about 250 Hz. The 
luminosity increase predicted for BES-II is a factor of fifteen, which yields a good event rate of 
1500 Hz, and a background rate of 56 kHz! Without serious background rejection, we may 
be seriously challenged to run at the higher BES-II energies. Figure \ref{fig:EPDTriggerTAC} is showing
for $\sqrt{s_{NN}}$ = 19.6 GeV peripheral collisions the number of fired East EPD TACs vs the number
of fired West EPD TACs for four different configurations of the TACs. The upper two plots show a setup
with 2 and 4 TACs, splitted along the azimuth, the lower two plots shows 24 TACs, one per super-sector.
The lower right figure shows and additional splitting alone the radial directions between the 7th and 8th
radial segment to keep some flexibility for the various energies and possible collision vertices along z. 
The z-vertex and $\eta$ resolution can be calculated based on those simulations. The results are shown
in figure \ref{fig:EPDTriggerRes}. A clear improvement of the resolutions is achieved from 2,4 azimuthal 
segments to 24. The additional split along the radial direction does not improve the resolution further,
mainly due the low particle multiplicity for peripheral collision. The additional split is going to improve the 
resolution  for more central collision and might be useful for further trigger conditions to be developed
by the Beam Energy Scan focus group. For Run17 we suggest to have 12 TAC channels, 4 per super-sector.

The background events are coming from collisions from beam halo ions with the beam pipe 
material, and therefore they are all in essence fixed-target events with a low $Z$ target nucleus. 
We have been studying the nature of these collisions using the BES-I data. The background 
events look qualitatively different from the good collider events; they have a lower multiplicity and 
they are very one-sided. Therefore, a trigger algorithm using multiplicity information as a function of
pseudorapidity would be most efficient at rejecting background. Good collider events at the center
of the detector will produce a hit pattern symmetric around $z=0$. There will be
some asymmetry introduced by offset vertices, which will not be uncommon with the
long bunches proposed for BES-II, however this can be corrected for using the timing
difference between the East and West EPD. For background events, center-of-mass
rapidity is shifted by 1.05 units for the 7.7 GeV collider system, and
1.5 units for the 19.6 GeV system. Because the East-West EPD timing difference
should come from random coincidences, it should not be related to the vertex
location. Therefore, a requirement on the balance between the East and West
multiplicity distributions, corrected for the vertex position slew, could
effectively reject the background. To best characterize the $\eta$ distribution
at the trigger level, ADCs for the EPD should be distributed radially with
sixteen channels each for the East and West sides.

To extend the physics reach of the BES-II program, STAR has installed an internal
fixed gold target. Special fixed-target runs will be taken, and these will need a customized 
trigger. For the fixed-target test runs in 2015, a coincidence between the BBC-East 
and a TOF multiplicity signal, along with a BBC-west veto, was developed
to provide a clean central trigger for Au+Au events. The principal short-coming of
this trigger was the inability to provide a Level-1 minimum bias trigger for FXT events.
The challenge is in distinguishing FXT Au+Au events from collisions beam between gold
ion in the beam halo and the aluminum beampipe. The TOF multiplicity cut was used to
suppress the lower multiplicity Au+Al background. The EPD upgrade will have improved
$\eta$ segmentation, and this combined with TOF multiplicity information could
better reject background events.

\newpage



\section{Management Plan}
\label{sec_management}

\begin{figure}[htbp]
\begin{center}
{
\mbox{\includegraphics[width=0.8\textwidth]{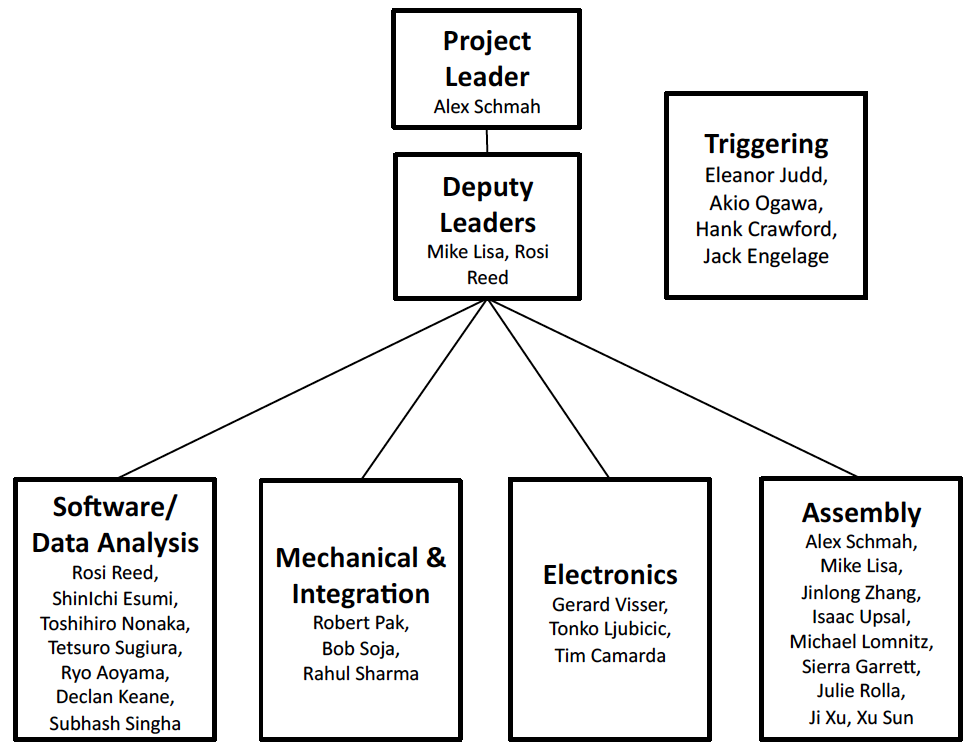}}
}
\end{center}
\caption{The EPD organizational structure. 
}
\label{fig:Organization}
\end{figure}

The list of Institutions, people and tasks is listed in Table \ref{tab:Institutions}
\begin{table}[]
\centering
\caption{People and Institutions}
\label{tab:Institutions}
\begin{tabular}{ll}
\textbf{Institution}            & \textbf{People}                                                                     \\
OSU                    & Mike Lisa, Isaac Upsal, Julie Rolla                                        \\
LBNL                   & Alex Schmah, Michael Lomnitz, Sierra Garrett, \\
                            & Jinlong Zhang, Grazyna Odyniec, Ji Xu \\
Lehigh University      & Rosi Reed                                                                  \\
UC Davis               & Daniel Cebra, Kathryn Meehan                                               \\
CCNU                   & Xu Sun                                                                     \\
Stony Brook University & Roy Lacey, Benjamin Schweid                                                \\
BNL                    & Robert Pak, STSG (Bob Soja, Rahul Sharma)                                  \\
Kent State University  & Declan Keane, Subhash Singha                                               \\
Indiana University     & Gerard Visser                                                              \\
Tsukuba                & ShinIchi Esumi, Toshihiro Nonaka, Tesuro Sugiura, Ryo Aoyama                              
\end{tabular}
\end{table}

\section{Integration with STAR}
\label{Integration}

\subsection{Mechanical Support}

\begin{figure}[htbp]
\begin{center}
{
\mbox{\includegraphics[width=0.65\textwidth]{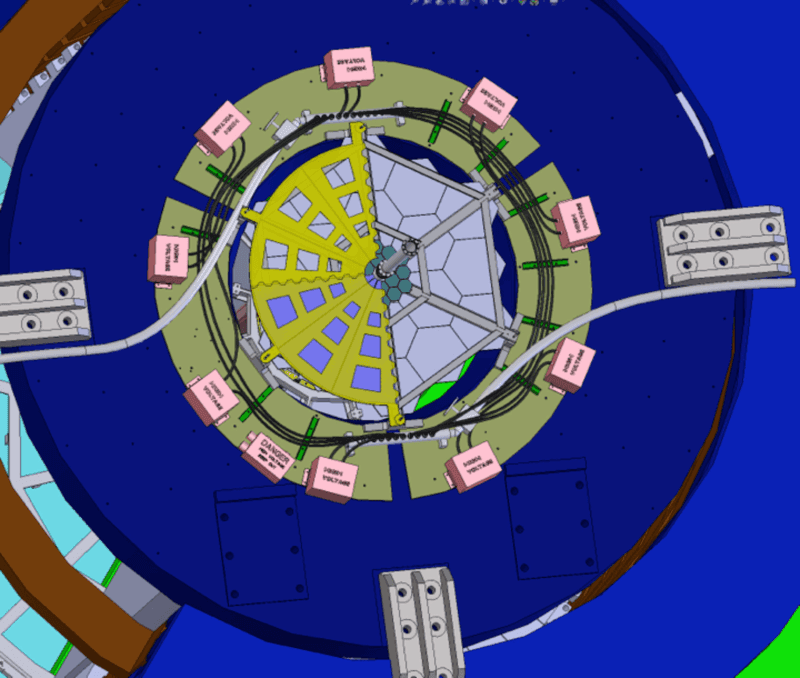}}
}
\end{center}
\caption{Schematic diagram for one half of the EPD support structure
mounted on the east STAR magnet pole tip behind the BCC array (honeycomb
pattern) as viewed looking toward the nominal collision point. A section of the
beam pipe is shown through the center and one quadrant of the EPD frame
(between 6 and 9 oÕclock) is fully populated with three supersectors.}
\label{fig:ELDinSTAR}
\end{figure}

A staged approach for EPD installation is planned, in which one fully operational
quadrant of the EPD will be mounted behind the full BBC array in RHIC Run 17.  The
location will be on the east side of STAR under the beam pipe, i.e, between 6 and 9
oÕclock, as shown in Figure \ref{fig:ELDinSTAR}  The inner tip of the three supersectors will
cantilever off the end of the support structure in the acceptance of the VPD. The
fiber-to- fiber connectors (see Section \ref{sec_fiberconnector}) extend beyond the outer edge of the
support frame to facilitate easy disconnect/reconnect. The strain-relieved fibers
will run to light-tight boxes (not shown in Fig. \ref{fig:ELDinSTAR}) containing the SiPMs mounted at
the 3 and 9 oÕclock positions on the magnet pole tip. For the EPD engineering run in
2017, coaxial RG-58 cables will carry the 96 signals from the SiPM FEE cards in the
light-tight box at the 9 oÕclock position to QT modules in electronics racks under the
east staircase in STAR IR. The racks also contain the electronics for EPD slow
controls, i.e., TUFF box. Installation of full EPD arrays on both sides of STAR will
occur prior to BES II (the west side is essentially a mirror image of the east side),
with sufficient time for commissioning the detector.

\subsection{Software Integration}
\label{sec_softwareIntegration}
The analysis software for the prototype EPD has been developed, however this will be developed and integrated
in the STAR software framework for the run 17 install of the 1/4 wheel.  In addition, monitoring plots and other important
information for the shift crew will be developed.  We will also develop the offline analysis software so that it is integrated into
the STAR software framework for run 17, in order to do a full commissioning run.  This will allow any additional develop for run 18 to be completed, and the results from the commissioning run to be analyzed.


\section{Risk Analysis}
\label{Risk}

One aspect of the risk is the weight of the detector. We can perform some quick density * volume calculations to approximate the total weight of the EPD. The EJ-200 1 cm thick plastic scintillator has a density of 1.02 g/cc \cite{ej200}. Using these values and the aforementioned dimensions of the EPD, the weight of the scintillator plastic for each super sector is 2.6 kg.  This means for each wheel the weight of the plastic alone would be 31 kg.  The addition of the backing structure brings that up to a bit over 43 kg for an entire wheel.   However, to mitigate the risks posed by this weight each wheel will be broken into 4 quadrants which weigh close to 10 kg each. 

The fibers and electronics represent another risk.  The current strategy is to feed the WLS fibers through the back of the scintillating tiles and to place the box containing the SiPMs safely behind the STAR magnet some distance from the beampipe to avoid damage from fast neutrons and other radiation. As mentioned earlier, the WLS fibers are not directly coupled to the SiPMs but are instead coupled to clear optical fibers soon after protruding from the plastic scintillators. We created a custom-designed connector in SolidWorks which is built at LBNL with the group's 3D-printer. The connector was designed to be as light-tight as possible while also firmly holding the fibers in close contact for minimal signal loss. 

\subsection*{Safety}
The safety risks of the EPD are low. No gases are used in the structure of the detector and voltages of 90 V or less will be employed in its operation. Only approved low voltage power supplies and previously used and tested electronics will be installed.



\newpage
\section{Cost and Schedule}
\label{sec_costnschedule}
\subsection{Cost}

Table \ref{tab:costs} shows the costs in U.S.D. of the EPD project for Run17 (only 1/8 installation) and the total
costs for Run18. Costs for QT-board are not included yet since it is unclear at the moment which option (see below) 
is finally selected. Costs for power supplied, crates, DSM and TAC cards are only estimated due to similar restrictions.

\begin{tabular}{  l c c } \hline
{\bf Component} & 24 {\bf super-sectors} & {\bf 3 super-sectors} \\ \hline
 SiPM &  \$ 13,200.00 & \$ 1,845.00 \\
 Scintillator  & \$ 13,560.00 & \$ 3,300.00 \\
 Wrapping material  & \$ 3,000.00 & \$ 400.00 \\
 Wavelength shifting fibers  & \$ 5,900.00 & \$ 2,860.00 \\
 Clear fibers  & \$ 17,050.00 & \$ 2,200.00  \\ 
 Scintillator workshop/labor & \$ 28,000.00 & \$ 3,000.00\\
 Fiber polishing workshop/labor & \$ 10,000.0 &  \$ 2,000 \\
 Connectors  & \$ 3,000.00 &  \$ 400.00\\
 Mechanical/support structure  & \$ 30,000.00 &  \$ 3,750.00 \\
 FEE card design and integration & \$ 20,000.00 & \$ 20,000.0 \\
 FEE cards(SiPM + contr. board) & \$ 30,000.00 &  \$ 5,000.00 \\
 TAC cards + DSM boards  & \$ 20,000.00 & \$ 0.00\\
 Power supplies, crates, cables  &  \$ 50,000.00 &  \$ 2,000.00\\ \hline
 {\bf Total} & {\bf 243,710.00} & {\bf 46,755.00}
\label{tab:costs}
 \end{tabular}

The following sections provide a simple breakdown of the sources of the estimated costs.

\section*{Silicon Photomultipliers}
Hamamatsu SiPM's, series S13360-1325PE, were quoted on 04/29/2016 for two scenarios (1/8 and full EPD):
\begin{itemize}
\item 100 SiPM's + 10 extra for 1/8 design \\

\begin{tabular}{| c | c | c |} \hline
 Price per SiPM (U.S.D.)& Total Cost (U.S.D.) \\ \hline
 18.45 &  1,845.00 \\ \hline
 \end{tabular}
 
 \item 850 SiPM's + 150 extra for full design \\
 
\begin{tabular}{| c | c | c |} \hline
 Price per SiPM (U.S.D.)& Total Cost (U.S.D.) \\ \hline
 13.21 & 13,210.00 \\ \hline
 \end{tabular} 
 \end{itemize}

 \section*{Scintillator}
 Eljen EJ-200 Scintillator Plastic (100cm x 65cm x 1.2cm), quote received on 04/11/16: \\
 
 \begin{tabular}{| c | c | c |} \hline
 No. of pieces & Cost per piece(U.S.D.) & Cost(U.S.D.) \\ \hline
 2 & 1148 & 2,296 \\ \hline
 4 & 997 & 3,988 \\ \hline
 15 & 904 & 13,560 \\ \hline
  \end{tabular}
  
  \section*{Wavelength shifting fibers}
  Kuraray Y-11 1.0mmD, 500m BSJ multi-cladding, round, spool, S-type. Quote received on 6/17/2015, each super-sector includes 39 m of wavelength shifting fiber with a cost per meter of \$5.71\\
  
  \begin{tabular}{ | c | c  | c |}\hline
  Number of supersectors & Total length + 10\% allowance & Total cost (U.S.D.) \\ \hline
  3 & 130 & 743 \\ \hline
  24 & 1030 & 5,881 \\ 
  \hline
  \end{tabular}
  
  \section*{Clear Optical fibers}
  Kuaray Clear-PS 1.15mmD, BJ, round, spool fiber. Quote received 03/14/2016. Estimate 4 m of fiber per channel implies a total of $\sim$ 3.1km for full EPD, however fibers are available only in 5 km spools:\\
  
  \begin{tabular}{ | c | c | c | } \hline
  Quantity & Price per m(U.S.D.) & Total \\ \hline
  5000 & 3.41 & 17,050 \\ \hline
  \end{tabular}
  
  \section*{Scintillator workshop/labor}
  Scintillator plastics will most likely be machined at shops in OSU. Estimate received on 05/03/2016 \$35.00 per hour implies $\sim$ \$1000 per super-sector: \\
  
  \begin{tabular}{ | c | c | c | }\hline
  Super-sectors & Cost per super-sector(U.S.D.) & Total cost (U.S.D.) \\ \hline
  3 &  1000.00 & 3,000.00\\ \hline
  28 & 1000.00 & 28,000.00\\ \hline
  \end{tabular}
  
 \section*{Mechanical/support structure}
 3,750 (U.S.D.) per quadrant, including labor.
 
  \section*{Fiber polishing workshop/labor}
  \$43 per hour at Lehigh university. \$10,000.00 for all super-sectors.
 
  \section*{Connectors}
  To be printed at LBNL, the estimate includes both Fiber-Fiber and SiPM-Fiber connectors, total cost of \$ 3,000.00.
  
  \section*{TAC card + DSM boards}
  Cost estimated at roughly \$ 20,000.00
  
  \section*{FEE cards (SiPM + controller card)}
 Costs are estimated so far.
  
  \section*{Power supplies, crates, cables}
  Estimated at roughly \$ 50,000.00

\subsection{QT-boards}
\label{sec_QT}

Possible readout schemes for the EPD are listed below:

\begin{itemize}
\item EPD is mainly using QT-boards from the FMS and FPS (FMS and FPS don't operate during/after 2018) \\
\item EPD is using the existing QT-board technology but new boards need to be build (FMS and FPS operate during/after 2018).
We would need about 30 QT boards, each is about \$1500, which makes in total about \$45,000.00. \\
\item The new DEP boards are used
\end{itemize}

\subsection{Schedule}
\label{sec_schedule}
\begin{figure}[htbp]
\begin{center}
{
\mbox{\includegraphics[width=0.95\textwidth]{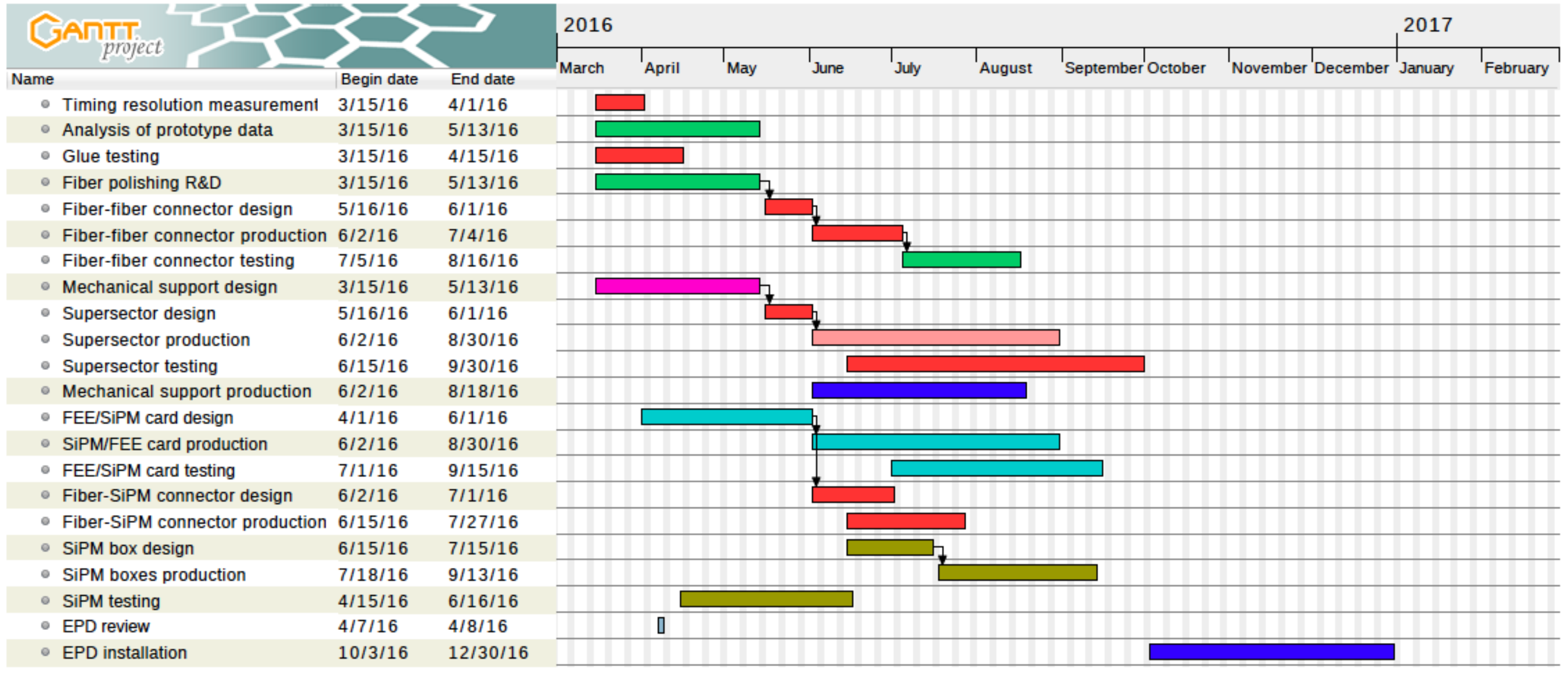}}
}
\end{center}
\caption{Schedule for the 1/8 detector build.}
\label{fig:schedule1_8}
\end{figure}

The time chart view of the project schedule for the 1/8 detector is shown in shown in Figure \ref{fig:schedule1_8}.

The plan for Run17 operations is to commission the EPD:

\begin{itemize}
\item 1/8th of the detector should be installed on one side of STAR \\
\item New supersector design \\
\item  New FEE boards are tested \\
\item One super-sector is equipped with the new DEP board \\
\item New coupling of fibers and fibers to SiPMs \\
\item New mechanical support structure is installed \\
\item  Online/Offline check of data/correlations as done for prototype \\
\item Online software for prototype will be expanded, the whole system
  is similar to FPS \\
\item Offline: ADC/TDC values will be treated the same as for BBC
  (is already working for prototype data) \\
\item  Calibration: Pedestal runs as for BBC/FPS, already done for prototype
\end{itemize}

The entire EPD will be completed by October 2017, which will allow it to be installed for run 18 and leaves a few months of contingency.  The EPD production will continue after the completion of the 1/8 detector build in order to meet this deadline and avoid the loss of key personnel.
 


\section{Acknowledgments}
We would like to thank Elke Aschenauer, John Campbell, Oleg Eyser, Leo Greiner, Patrick Huck, Peter Jacobs, Eleanor Judd, Bill Llope (for providing the UrQMD simulations), Howard Matis, Akio Ogawa, Tim Camarda, Art Poskanzer, Hans-Georg Ritter, Ernst Sichtermann, Jim Thomas, Oleg Tsai, and Paul Sorensen for their help and for important discussions.
\label{sec_acknow}



\end{document}